\documentclass[preprint,12pt]{elsarticle}
\usepackage[textwidth=480pt]{geometry}
\usepackage{subcaption}
\usepackage{graphicx}
\usepackage{gacaps}
\usepackage{ga3dplot}

\usepackage{amssymb}
\usepackage{amsmath}

\usepackage{tabularx}	% Defines tabularx environment
\usepackage{ar}
\usepackage{bm}
\usepackage{printlen}
\usepackage{tikz,pgfplots}
\usetikzlibrary{decorations.markings,arrows.meta,calc,shapes,intersections,backgrounds,patterns,bending}
\usepackage{placeins}

\usepackage{algpseudocode}
\let\ig\includegraphics
\let\tw\textwidth

\newcommand{\myls}[3]{\lineSymbol[#3]{#1}{#2}{.8ex}{.8pt}{}{#2}}

\newcommand{\ly}[2]{\lineSymbol[#1]{none}{#2}{.8ex}{.5pt}{#2}{#2}}
\newcommand{\lyy}[2]{\lineSymbol[#1]{none}{#2}{.8ex}{.8pt}{#2}{#2}}

\definecolor{C0}{HTML}{1F77B4} 
\definecolor{C1}{HTML}{FF7F0E}
\definecolor{C2}{HTML}{2CA02C}
\definecolor{C3}{HTML}{D62728}
\definecolor{C4}{HTML}{9467BD}
\definecolor{C5}{HTML}{8C564B}
\definecolor{C6}{HTML}{E377C2}
\definecolor{C7}{HTML}{7F7F7F}
\definecolor{C8}{HTML}{BCBD22}
\definecolor{C9}{HTML}{17BECF}

\newcolumntype{L}[1]{>{\raggedright\let\newline\\\arraybackslash\hspace{0pt}}m{#1}}
\newcolumntype{C}[1]{>{\centering\let\newline\\\arraybackslash\hspace{0pt}}m{#1}}
\newcolumntype{R}[1]{>{\raggedleft\let\newline\\\arraybackslash\hspace{0pt}}m{#1}}

\tikzset{
     >={Latex[length=.2cm]}
}
\tikzset{
    mylab/.style={label={[xshift=.03\tw,yshift=+.1em]below:\footnotesize{(#1)}}}
}
\tikzset{
    mylab1/.style n args={3}{label={[xshift=#2,yshift=#3]below:\footnotesize{(#1)}}}
}

\let\vvec\mathbf

\newcommand\uv{\vvec{u}}
\newcommand\fv{\vvec{f}}
\newcommand\qv{\vvec{q}}
\newcommand\qdv{\dot{\qv}}
\newcommand\qddv{\ddot{\qv}}
\newcommand\Hmat{\mathsf{H}}
\newcommand\Cmat{\mathsf{c}}
\providecommand\T{\mathsf{T}}
\providecommand\qvf{\qv_u}
\providecommand\qdvf{\dot{\qv}_u}
\providecommand\qddvf{\ddot{\qv}_u}
\providecommand\qvi{\qv_p}

\providecommand\qddvi{\ddot{\qv}_p}

\newcommand\HmatD[1]{\Hmat_{#1}}
\newcommand\CmatD[1]{\Cmat_{#1}}

\newcommand\Bdy\Gamma
\newcommand\xB{\vvec{X}}
\newcommand\Fld{\Omega_f}
\newcommand\NB{N_B}
\newcommand\Ndof{N_{\textit{dof}}}
\newcommand\Ninv{N_{\textit{p}}}

\newcommand\Fbdy{\mathcal{F}}
\newcommand\Mbdy{\mathcal{M}}
\newcommand\Gbdy{\mathcal{G}}
\newcommand\Nbdy{\mathcal{N}}

\newcommand\vsp{\vvec{\hat{v}}}

\providecommand\rhof{\rho}
\providecommand\Bpre{k}

\providecommand{\eqcite}[1]{eq.~\eqref{#1}}

\journal{Journal of Fluids and Structures}

\begin{document}

\begin{frontmatter}



\title{Fluid-structure interaction of multi-body systems:\\ Methodology and applications} 


\author[uc3m,mit]{G Arranz\corref{cor1}}
\ead{garranz@mit.edu}
\author[uc3m]{C Mart\'{i}nez-Muriel}
\author[uc3m]{O Flores}
\author[uc3m]{M Garc\'{i}a-Villalba}

\address[uc3m]{Bioengineering and Aerospace Eng. Dept., Univ. Carlos III de Madrid, Spain}
\address[mit]{Department of Aeronautics and Astronautics, Massachusetts Institute of Technology, Cambridge, USA}

\cortext[cor1]{Corresponding author}

\begin{abstract}

We present a method for computing fluid-structure interaction problems for multi-body systems. 
The fluid flow equations are solved using a fractional-step method with the immersed boundary method proposed by Uhlmann [J. Comput Phys. 209 (2005) 448].
The equations of the rigid bodies are solved using recursive algorithms proposed by Felis [Auton. Robot 41 (2017) 495].
The two systems of equations are weakly coupled, so that the resulting method is cost-effective.
The accuracy of the method is demonstrated by comparison with two-
and three-dimensional cases from the literature: the flapping of a flexible airfoil, the self-propulsion of a plunging flexible plate, and the flapping of a flag 
in a free stream.
As an illustration of the capabilities of the proposed method, 
two three-dimensional bio-inspired applications are presented: 
an extension to three dimensions of the plunging flexible plate and 
a simple model of spider ballooning. 
\end{abstract}




\end{frontmatter}

% \linenumbers

% main text

%%%%%%%%%%%%%%%%%%%%%%%%%%%%%%%%%%%%%%%%%%%%%%%%%%%%%%%%%%%%%%%%%%%%%%%%%%%%%%%%
%% INTRODUCTION  
%%%%%%%%%%%%%%%%%%%%%%%%%%%%%%%%%%%%%%%%%%%%%%%%%%%%%%%%%%%%%%%%%%%%%%%%%%%%%%%%

\section{Introduction \label{sec:intro}}

The understanding of the mechanisms of biological motion, 
such as insect flight, fish swimming, bacteria swarming 
or seed dispersal by wind, has been shown to be important 
for scientific and engineering applications.
Clear examples are the recent developments in micro-air vehicles \cite{decroon2009,rithcher2011,keennon2012}
or swimming robots \cite{triantafyllou1995,hirata2000,yu2004}.
However, despite these advances, there are still gaps in our knowledge of the physics underlying the motion of biological systems.
This restricts artificial systems from achieving the performance of  biological systems.
%
%In this framework, direct numerical simulations have proven to be an effective tool to study these kind of problems.
 
%In particular, one of the shortcomings associated to the modelling of biological motion is that it is  computationally challenging.
%
Filling these gaps in our knowledge is challenging. 
On one hand, performing controlled experiments with biological systems is difficult, and so is the interpretation of the results of those experiments. 
On the other hand, the detailed simulation of these systems can become computationally expensive.
Note that, from a physical point of view, the biological motion of insects/fish/bacteria/seeds can be described as a fluid-structure interaction (FSI) problem, 
in which one or more deformable bodies are immersed in a fluid.
The dynamics of the bodies is a direct result of their hydrodynamic interaction with
the surrounding fluid, which is driven by their deformation (active or passive).
As a consequence, the resulting problem is a highly non-linear problem consisting of the coupling between the equations of the 
fluid motion and the equations of motion of the bodies.
%%
%In this regard, this kind of FSI problem are computationally challenging compared to fluid-structure problems in which the kinematics of the structure are prescribed.
%%
%\textcolor{red}{not only coupling but also the dynamic equations difficult for complex bodies.}
%
%An additional aspect to be considered is the geometrical variability of the bodies found in applications:
Moreover, these bodies are usually geometrically complex: 
they may have mobile appendages (e.g., the wings of flying animals or robots, fins of aquatic swimmers, etc.), and/or they may be deformable and subject to large deformations. 
This geometrical variability/complexity poses additional problems when modelling this kind of problems, since 
the fluid-solid interface changes with time and 
the equations of the body dynamics can become complex to derive and to solve.

In computational fluid dynamics, solid-fluid interfaces can be represented with 
two families of procedures that differ in the spatial discretization of the fluid near the solid:
conforming mesh methods and non-conforming mesh methods \citep{deng2013}.
In the former, the interface condition is treated as a physical boundary, 
which requires the definition of a boundary fitted mesh.
As a consequence, the grid must be adjusted due to the movement of the bodies as time evolves, 
entailing an increase in the computational cost.
%
% First of all, for ALE, re-meshing is not a requirement. The mesh will definitely move, but it (i.e. number of nodes and connectivity) can remain the same. Re-meshing is only needed when the mesh quality deteriorates significantly for large deformations. For the amount of deformation mentioned in the last two 3d cases, it's very likely that no re-meshing will be needed when solved with ALE. For immersed boundary methods, depending on the flavor, a physical boundary does exist in some cases, such as in the sharp-interface IBM.
%
The Arbitrary Eulerian-Lagrangian (ALE) method in \citet{donea2017} is an example of conforming mesh methods.
On the other hand, in non-conforming mesh methods, a physical boundary between the bodies and the fluid does not exist, 
but constraints are imposed to the fluid equations such that interface conditions are fulfilled at the boundaries.
As a consequence, adjusting the mesh fluid domain is not required, thus reducing the algorithm 
complexity and the computational cost.
Among the different non-conforming mesh method, such as level-set, volume-of-fluid or phase-field
methods \citep{tschisgale2020},
%immersed methods such as the 
immersed boundary methods (IBM) have proven to be 
a very useful tool to reproduce the arbitrary motion of solids immersed in a fluid \citep{mittal2005,griffith2020,uhlmann2005,pinelli2010,breugem2012,kempe2012}.

The literature shows multitude of examples of application of IBMs to problems similar to those of biological motion.
These examples focus mainly on deformable bodies of a particular topology, 
such as filaments \citep{bhalla2013,wiens2015,tschisgale2020}; 
membranes \citep{detullio2016,kim2009}; or finite 
volumetric structures \citep{zhang2004,tian2014}.
In these methods, the deformation of the bodies is computed by means of different methods (i.e., classical continuum mechanics, networks of point-masses and springs, etc.), which may lead to algorithms of varying complexity. 
In some situations, the dynamics of the biological system allow modelling a complex, deformable body
as a set of rigid bodies connected among them by kinematic constraints 
(see for example \citet{liu2009,suzuki2015}).
Under this approach, the dynamics of the bodies are represented as a system of non-linear differential equations
which is influenced by the inertia properties of the rigid bodies and the connections among them.
Note that there is a large variability of possible system configurations so that, in general, the system of non-linear differential equations needs to be re-derived every time the configuration is modified
(i.e., adding a body, such as a tail, or changing the type of links between bodies).
%
%Using for example Lagrangian mechanics may lead to tedious derivation of such equations.
%
In this regard, robotic algorithms stand as an outstanding choice to compute the dynamics of a complex
system of rigid bodies with a semi-automatic procedure  \cite{featherstone2014}.

Most of the works which study the FSI of multi-body system derive the equations of motion
specifically for the problem under consideration  \citep{zhang2013,arora2018,suzuki2019,yao2019}.
But a few of them propose strategies that allow solving generic multi-body system interacting with a surrounding flow.
For example, \citet{wang2015} reported a vorticity-based immersed boundary projection method with a strong coupling between the fluid Navier-Stokes equation and the multi-body dynamics equations. 
The coupling is implemented using Lagrange multipliers (i.e., strong coupling), and the paper provides a detailed explanation on how to obtain the equations for the system of rigid bodies for an array of linked planar plates (i.e., only rotations between bodies are allowed). 
More recently, \citet{li2018} coupled a multi-body algorithm with the existing finite volume method of 
the commercial software, Ansys, to solve the flow around the bodies.
Again, only rotations where allowed between the bodies conforming the multi-body system.
The time coupling between both algorithm was accomplished in a staggered fashion, leading to a weak coupling
of both systems.
Finally, \citet{bernier2019} recently developed an algorithm combining a 2D vortex particle method coupled with a multi-body solver, using the projection and penalization techniques of \citet{gazzola2011}. 

In this article, we present a methodology to compute the dynamics of multi-body systems immersed in an
incompressible Newtonian fluid using a partitioned (non monolithic) approach.
The flow equations are solved by means of Direct Numerical Simulation (DNS) where the presence of the bodies
is modelled using the IBM proposed by \citet{uhlmann2005}.
The dynamics of the multi-body systems is solved using recursive dynamic algorithms
in reduced coordinates developed by \citet{felis2017}. 
As a result, the proposed methodology allows the definition of different multi-body 
system configurations with no modification of the main algorithm. 
That includes the number of bodies of the system and its topology, 
as well as the properties of the links (spherical, revolute or prismatic joints, 
or any combination of them) and their equations (free movement, springs, dampers, etc.).
Secondly, the coupling between the flow equation and the dynamics equations is very simple,
yet robust enough to study a wide range of bio-inspired locomotion problems.

The structure of the paper is as follows: the methodology 
is presented in \S~\ref{sec:met}; 
a validation of the methodology with existing cases reported in the literature is found in \S~\ref{sec:val};
in \S~\ref{sec:res}, two illustrative problems solved using this methodology are presented;
and, finally, the major conclusions of this study are gathered in \S~\ref{sec:conc}. 

%Mention??
%\cite{kanso2005}
%\cite{boyer2015}
%\cite{boyer2017}

%%%%%%%%%%%%%%%%%%%%%%%%%%%%%%%%%%%%%%%%%%%%%%%%%%%%%%%%%%%%%%%%%%%%%%%%%%%%%%%%
%% METHODOLOGY  
%%%%%%%%%%%%%%%%%%%%%%%%%%%%%%%%%%%%%%%%%%%%%%%%%%%%%%%%%%%%%%%%%%%%%%%%%%%%%%%%

\section{Methodology \label{sec:met}}

%%%%%%%%%%%%%%%%%%%%%%%%%%%%%%%%%%%%%%%%%%%%%%%%%%%%%%%%%%%%
% Problem Description
%%%%%%%%%%%%%%%%%%%%%%%%%%%%%%%%%%%%%%%%%%%%%%%%%%%%%%%%%%%%
\subsection{Problem description\label{ssec:met_problem}}

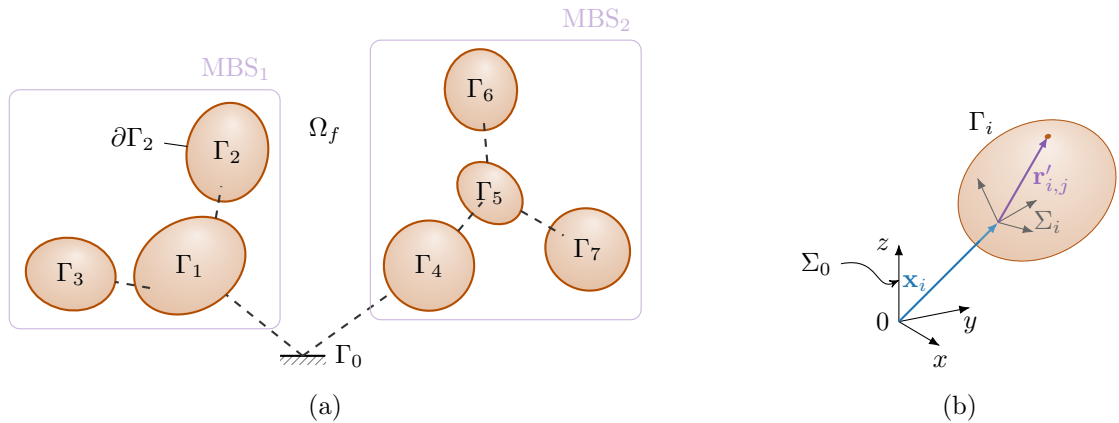
\begin{figure}
   \centering
   \begin{subfigure}[b]{.6\tw}
      \centering
      \pgfdeclareradialshading[mycolor]{sphere}{\pgfpoint{0.1cm}{0.5cm}}
{color(0cm)=(mycolor!10);
color(.3cm)=(mycolor!20);
color(1.2cm)=(mycolor!40)}

\colorlet{mycolor}{red!70!green}

\begin{tikzpicture}[scale=.6,
   joints/.style={black!80,thick,dashed},
   body/.style={shading=sphere,thick,draw=mycolor}]%,
   %background rectangle/.style={rounded corners=1cm,fill=blue!5!white},
   %show background rectangle]
   \begin{footnotesize}
%   \fill[rounded corners,blue!5!white] (-5.5,-2) rectangle (4.5,6);

   %% MULTIBODY SYSTEM 1
   \coordinate (C0) at (30:.5);
   \coordinate (C1) at (30:-1.0);
   \shadedraw[body,rotate=30,name path=B0] 
             (0,0) ellipse (1.3 and 1)node (B0) {};
   \shadedraw[body,shift={(C0)},rotate=80,name path=B1] 
             (2.3,0) ellipse (1.1 and .9)node (B1) {};
   \shadedraw[body,shift={(C1)},rotate=170,name path=B2] 
             (1.8,0) ellipse (1.0 and .8)node (B2) {};

   \path[name path=J01] (B1) -- (C0);
   \path[name intersections={of=J01 and B0,by=j01}]; 
   \path[name intersections={of=J01 and B1,by=j10}]; 
   \draw [joints,shorten >=-.2cm] (j01) -- (j10);

   \path[name path=J02] (B2) -- (C1);
   \path[name intersections={of=J02 and B0,by=j02}]; 
   \path[name intersections={of=J02 and B2,by=j20}]; 
   \draw [joints,shorten <=-.2cm] (j02) -- (j20);

   %% MULTIBODY SYSTEM 2
   \coordinate (D00) at (5.3,+0);
   \coordinate (D0) at (D00) + (00:.5);
   \coordinate (D1) at (D00) + (30:-1.0);
   \shadedraw[body,rotate=30,name path=E0] 
             (D00) circle(1)node (E0) {};
   %
%   \shadedraw[body,shift={(D1)},rotate=170,name path=E1] 
%             (2.3,0) ellipse (1.0 and .8)node (E1) {};
   %
   \shadedraw[body,shift={(D0)},rotate=50,name path=E2] 
             (2.1,0) ellipse (.6 and .8)node (E2) {};
   \shadedraw[body,shift={(E2)},rotate=95,name path=e21] 
             (2.3,0) ellipse (.9 and .8)node (E21) {};
   \shadedraw[body,shift={(E2)},rotate=-30,name path=E22] 
             (2.5,0) ellipse (.95 and .9)node (E22) {};

%   \path[name path=J03] (E1) -- (E0);
%   \path[name intersections={of=J03 and E0,by=g01}]; 
%   \path[name intersections={of=J03 and E1,by=g10}]; 
%   \draw [joints,shorten >=-.2cm] (g01) -- (g10);

   \path[name path=J04] (E2) -- (E0);
   \path[name intersections={of=J04 and E0,by=g02}]; 
   \path[name intersections={of=J04 and E2,by=g20}]; 
   \draw [joints,shorten >=-.2cm] (g02) -- (g20);

   \path[name path=J05] (E2) -- (E21);
   \path[name intersections={of=J05 and E2, by=ge21}]; 
   \path[name intersections={of=J05 and e21,by=g21}]; 
   \draw [joints,shorten >=-.2cm] (ge21) -- (g21);

   \path[name path=J06] (E2) -- (E22);
   \path[name intersections={of=J06 and E2, by=ge22}]; 
   \path[name intersections={of=J06 and E22,by=g22}]; 
   \draw [joints,shorten >=-.2cm] (ge22) -- (g22);

   % BASE
   \fill[pattern=north east lines, pattern color=black!50] (2,-2) rectangle (3,-2.2);
   \path[black,thick,draw] (2,-2) -- +(1,0.) coordinate [midway] (cc) node[anchor=west] {$\Bdy_0$};

   \path[name path=ccB] (cc) -- (B0);
   \path[name intersections={of=ccB and B0,by=cB}]; 
   \draw[joints] (cc) -- (cB);

   \path[name path=ccE] (cc) -- (E0);
   \path[name intersections={of=ccE and E0,by=cE}]; 
   \draw[joints] (cc) -- (cE);

   % TEXT
   \node at (3,3) {$\Omega_f$};
   \path (B1)+(-1.4,.3) node[anchor=east] (pO) {$\partial \Bdy_2$};  

   \path[name path=pOB1] (pO) -- (B1);
   \path[name intersections={of=pOB1 and B1,by=ppO}]; 
   \draw (pO) -- (ppO);

   \path (B0) node  {$\Bdy_1$};  
   \path (B1) node  {$\Bdy_2$};  
   \path (B2) node  {$\Bdy_3$};  
   \path (E0) node  {$\Bdy_4$};  
   \path (E2) node  {$\Bdy_5$};  
   \path (E21) node {$\Bdy_6$};  
   \path (E22) node {$\Bdy_7$};  

   \draw[rounded corners,thin,C4!50] (-4,-1.4) rectangle (2,3.9) 
         node [above left] {MBS$_1$};

   \draw[rounded corners,thin,C4!50] (+4,-1.2) rectangle +(6,6.2) 
         node [above left] {MBS$_2$};
   \end{footnotesize}
\end{tikzpicture}
      \caption{\label{fig:scheme}}
   \end{subfigure}\hfill
   \begin{subfigure}[b]{.4\tw}
      \centering
      \pgfdeclareradialshading[mycolor]{sphere}{\pgfpoint{0.1cm}{0.5cm}}
{color(0cm)=(mycolor!10);
color(.3cm)=(mycolor!20);
color(1.2cm)=(mycolor!40)}

\colorlet{mycolor}{red!70!green!90!white}

\tdplotsetmaincoords{-60}{+20}

\begin{tikzpicture}[scale=1.1,tdplot_screen_coords,
   >={Latex[length=.15cm]},
   ineax/.style={tdplot_main_coords,black},
   bdyax/.style={tdplot_screen_coords,black!60},
   body/.style={tdplot_screen_coords,shading=sphere,draw=mycolor}]

   \begin{footnotesize}

   \coordinate (x0) at (0.,0.,0.);        % origin
   \coordinate (xi) at (1.2,1.2);         % body control point
   \coordinate (xp) at ($(xi)+(60:1.2)$); % surface point

   \coordinate (xl) at (3.2,.1);          % 2nd body control point

   % INERTIAL REFERENCE FRAME
   \draw[ineax,->] (x0) --+ (1,0,0) node[below] {$x$};
   \draw[ineax,->] (x0) --+ (0,1,0) node[below] {$y$};
   \draw[ineax,->] (x0) node[left] {$0$} --+ (0,0,1) node[left] {$z$};

   \draw[thin,>={stealth'[length=3pt]},<-,shift=(x0)] (0,.5) .. 
         controls (-.3,.3) and (-.2,+.8).. (-.7,.7) node[left] {$\Sigma_0$};

   % BODY 
   \filldraw[name path=Bi,body,rotate=30] ($(xi)+(.6,.1)$) ellipse (1 and .8);
 
   % MOVING REFERENCE FRAME
   \draw[bdyax,->] (xi) --+ (30:.55); 
   \draw[bdyax,->] (xi) --+ (115:.6); 
   \draw[bdyax,->] (xi) --+ (-15:.45);
 
   \node[bdyax] at ($(xi)+(.6,0)$) {$\Sigma_i$};
   \node[black] at ($(xi)+(-.2,1.2)$) {$\Gamma_i$};

   % VECTORS
   \path[name path=p0i] (x0) -- (xi);
   \path[name intersections={of=p0i and Bi,by=e0i}]; 
   \draw[thick,C0!80,->] (e0i) -- (xi);
   \draw[thick,C0,shorten >=-.2em] (x0) -- (e0i) node[midway,left] {$\vvec{x}_i$};

   \draw[thick,C4!90!black,->] (xi) -- (xp) node[midway,right,C4] {$\vvec{r}^\prime_{i,j}$};;
   
   \fill[rotate=20,mycolor] (xp) ellipse (.04 and .03);

   %\DrawControl{(-.3,.3)}{blue}\DrawControl{(.2,.8)}{blue};

   \end{footnotesize}
\end{tikzpicture}
      \caption{\label{fig:singleb}}
   \end{subfigure}\hfill
   \caption{(a) Schematic representation of the interaction of two multi-body systems (MBS$_1$ and MBS$_2$) of rigid bodies, $\Bdy_i$, 
with a surrounding fluid domain, $\Fld$. 
The connections among the bodies are represented as dashed lines.
$\Bdy_0$ is a fixed inertial reference base.
(b) Definition of the reference systems, $\Sigma$, and vectors used to define the position and orientation of
a given body $\Gamma_i$.
}
\end{figure}

The physical problem considered in the present work is the interaction of a multi-body system (MBS) of rigid bodies with a surrounding fluid.
In this MBS, the rigid bodies are connected among them by joints (i.e., kinematic constraints) and are subject 
to the hydrodynamic forces and torques exerted by the surrounding fluid.
Note that a collection of MBSs can also be handled by the proposed method.
This is illustrated with a sketch in Figure~\ref{fig:scheme}, where $\Bdy_i$ stands for the $i$th rigid body of the MBS.
Note that $\Bdy_0$ in the sketch does not represent a rigid body, but a fixed inertial base. 
Consequently, the joints connecting $\Bdy_1$ and $\Bdy_4$ to $\Bdy_0$ do not necessarily restrict any degree of
freedom (i.e., free body motion).

The fluid is modelled as incompressible and Newtonian, whose governing equations are
\begin{subequations}\label{eq:NS}
\begin{align}
&\nabla \cdot \uv = 0, \\
&\frac{\partial \uv}{\partial t} + (\uv \cdot \nabla)\uv = - \frac{1}{\rhof}\nabla p + \nu \nabla^2 \uv + \fv,
\end{align}
\end{subequations}
where $\uv$ is the fluid velocity field, $\rhof$ is the fluid density, $p$ is the pressure, $\nu$ is the kinematic viscosity, and $\fv$ is the Immersed Boundary Method (IBM) forcing term.
The latter is calculated to fulfill the \emph{non-slip boundary condition} on the surface of the bodies,
\begin{equation}\label{eq:BC}
\uv(\vvec{x}) = \vvec{U}_{\partial \Bdy_i}(\vvec{x}) \quad \forall \vvec{x} \in \partial\Bdy_i, \quad \forall i\in B,
\end{equation}
where $\vvec{U}_{\partial \Bdy_i}$ is the
velocity on the 
%bounding 
surface of the rigid body $\Gamma_i$, and $B = \{1,...,\NB\}$, is the set of rigid bodies, being $\NB$ the total number of them.

%On the other hand, to compute the dynamics of the MBS, the equations of motion of the systems must be written.
Concerning the equations that govern the dynamics of the MBS, recall that six scalar equations, the so-called Newton-Euler equations, 
are needed to represent the dynamics of a single rigid body in three dimensions (3D).
Thus, in principle $6\times \NB$ equations fully describe the dynamics of the MBS.
However, the joints connecting bodies usually constrain their relative motion.
Consequently, it is possible to reduce the number of equations
to the number of the degrees of freedom ($\Ndof$) of the MBS.
Although several methodologies can be adopted to find these equations \citep{greenwood2006}, 
the final result is a system of ordinary differential equations which can be written in the form \cite{featherstone2014}:
\begin{equation}\label{eq:MB}
\Hmat(\qv) \qddv + \Cmat(\qv,\qdv) = \bm{\xi} + \bm{\xi}_{h},
\end{equation}
where $\qv$ is the vector of the generalized coordinates (of size $\Ndof\times1$),
$\Hmat$ is the joint space or generalized inertia matrix, 
$\Cmat$ is the generalized bias force (accounting for gravity, Coriolis and centrifugal forces),
$\bm{\xi}$ is the vector of generalized forces (e.g., springs and/or dampers in the joints, \emph{etc.}),
and $\bm{\xi}_{h}$ is the vector of generalized forces due to the surrounding fluid (i.e., hydrodynamic forces).
Note that, although only the dependence on $\qv$ and $\qdv$ of $\Hmat$ and $\Cmat$ is made explicit in \eqcite{eq:MB}, both also depend on the inertia properties of the bodies.
Note also that, $\qv$ represents the configuration of the MBS at a given time instant in the $\Ndof$-dimensional \emph{configuration} space \citep{greenwood2006,boyer2015}.

In the following subsections, the solvers employed to solve \eqcite{eq:NS} and \eqcite{eq:MB} are described (in \S~\ref{ssec:met_FS} and
\S~\ref{ssec:met_MB}, respectively),
followed by the description of the algorithm coupling both solvers (\S~\ref{ssec:met_coupling}).

%%%%%%%%%%%%%%%%%%%%%%%%%%%%%%%%%%%%%%%%%%%%%%%%%%%%%%%%%%%%
% Flow solver
%%%%%%%%%%%%%%%%%%%%%%%%%%%%%%%%%%%%%%%%%%%%%%%%%%%%%%%%%%%
\subsection{Flow solver\label{ssec:met_FS}}

Equation \eqref{eq:NS} is solved using the  numerical method proposed by \citet{uhlmann2005}, where the forcing term is 
computed using a direct forcing formulation of the IBM. 
The method requires the use of two grids. 
First, the fluid domain, $\Fld$, is discretized into a fixed, Cartesian grid, the so-called  Eulerian grid. 
Second, the surface of each rigid-body, $\partial\Bdy_i$, is discretized into $n_{i}$ 
evenly distributed points.
Therefore, the set of surface points for a body $\Bdy_i$ is defined as $L(i) = \{1,...,n_{i}\}$, and 
the position of grid points on $\partial\Bdy_i$ is labelled as $\xB_{i,j},\, j\in L(i)$.
This is the so-called Lagrangian grid.

The equations (\ref{eq:NS}) are solved using a fractional step method.
The spatial derivatives are discretized with 2nd order finite differences on a staggered grid.
The temporal scheme is a 3-stage low-storage, semi-implicit Runge-Kutta, in which the convective terms are treated explicitly and the viscous terms are treated implicitly.
For completeness, the discretized equations at the $k$th Runge-Kutta substep are provided below: 
\begin{subequations}\label{eq:NSdis}
\begin{align}
&\tilde{\uv} = \uv^{k-1} + \Delta t \left(2\alpha_k \nu \nabla^2 \uv^{k-1} - 
              2\alpha_k \rho^{-1}\nabla p^{k-1} - \gamma_k[(\uv\cdot\nabla)\uv]^{k-1} - 
	      \zeta_k[(\uv\cdot\nabla)\uv]^{k-2}\right), \label{eq:NSdisa}\\
&\nabla^2 \uv^* - \frac{\uv^*}{\alpha_k \nu \Delta t} = 
 -\frac{1}{\nu\alpha_k}\left(\frac{\tilde{\uv}}{\Delta t} + \fv^k\right) + \nabla^2 \uv^{k-1}, \label{eq:NSdisb}\\
&\nabla^2 \phi^k = \frac{\nabla \cdot \uv^*}{2\alpha_k \Delta t}, \\
&\uv^k = \uv^* - 2\alpha_k \Delta t \nabla \phi^k, \\
&p^k = p^{k-1} + \rho\left(\phi^k - \alpha_k  \nu \Delta t \nabla^2 \phi^k\right),
\end{align}
\end{subequations}
where $\tilde{\uv}$ is an estimated velocity without the forcing term (i.e., disregarding the solid surfaces), 
%$\phi$ is the pseudo-pressure and the Runge-Kutta coefficients ($\alpha_k,\,\gamma_k$ and $\zeta_k$) are taken from \citet{rai1991}.
$\phi$ is the pseudo-pressure and the Runge-Kutta coefficients ($\alpha_1 = 4/15$, $\alpha_2 = 1/15$,
$\alpha_3 = 1/6$; $\gamma_1 =8/15$, $\gamma_2 =5/12$, $\gamma_3 =3/4$; and $\zeta_1 = 0$,
$\zeta_2 = -17/60$, $\zeta_3 = -5/12$) are taken from \citet{rai1991}.
 
The forcing term in \eqcite{eq:NSdisb}, $\fv^k$, is obtained from estimating the necessary force to fulfil the boundary condition 
given by \eqcite{eq:BC}:
\begin{equation}\label{eq:for}
\vvec{F}^k(\xB_{i,j}) = \frac{\vvec{U}_{\partial\Bdy_i}^{k-1}(\xB_{i,j}) - \tilde{\vvec{U}}(\xB_{i,j})}{\Delta t}, \quad\forall j \in L(i), i \in B.
\end{equation}
In this equation, $\tilde{\vvec{U}}$ corresponds to $\tilde{\uv}$ interpolated to the Lagrangian points.
Note that this implementation of the IBM requires interpolations from the Eulerian grid to the Lagrangian grid ($\tilde{\uv} \mapsto \tilde{\vvec{U}}$), 
as well as a spreading operator from the Lagrangian grid to the Eulerian grid ($\vvec{F}^k \mapsto \fv^k$). 
These two operators are defined using the regularized delta functions introduced by 
\citet{peskin2002} and defined by \citet{roma1999}, which satisfy the necessary conditions in terms of conservation of  momentum, force and torque in the interpolation and spreading operations. 
%\textcolor{red}{mention 3 or 4 points?}

Note that, the explicit direct forcing IBM used herein is particularly suited
for the stiff systems inherent to rigid body problems; contrary  
to classical IBM, which may suffer for instabilities when dealing
with these systems \citep{sotiropoulos2014}.
For further details on the immersed boundary method described above, the reader is referred to \citet{uhlmann2005}.
This algorithm has been implemented in a flow solver called TUCAN, which has been successfully used for the simulation of rigid-bodies with 
prescribed kinematics \citep{moriche2016,moriche2017,gonzalo2018,arranz2020,moriche2021a,moriche2021b}.
Likewise, the free motion of a single-rigid body immersed in a fluid has been also successfully simulated \citep{arranz2018a,arranz2018b},
using the coupling method presented in \citet{uhlmann2005}.

%%%%%%%%%%%%%%%%%%%%%%%%%%%%%%%%%%%%%%%%%%%%%%%%%%%%%%%%%%%%
% Multi-body solver
%%%%%%%%%%%%%%%%%%%%%%%%%%%%%%%%%%%%%%%%%%%%%%%%%%%%%%%%%%%
\subsection{Multi-body solver\label{ssec:met_MB}}

The temporal integration of \eqcite{eq:MB} provides the $\Ndof$ components of the generalized velocities, which in turn, are integrated to compute the generalized coordinates at a given time instant, $t$.
Nonetheless, the kinematics of several DoFs are often known as a prescribed function of time 
(e.g., the  motion of a wing with respect to the flyer's body).
Therefore, it is convenient to write $\qv$
as $\qv = \begin{bmatrix}\qvf^ \T & \qvi^\T\end{bmatrix}^\T$, where $\qvi$ contains the $\Ninv$ generalized 
coordinates whose temporal evolution is prescribed.
Likewise,
\begin{align}\label{eq:ordered}
\Hmat &= \begin{bmatrix} \HmatD{u} & \HmatD{up} \\ \HmatD{pu} & \HmatD{p}\end{bmatrix}, &
\Cmat &= \begin{bmatrix} \CmatD{u} \\ \CmatD{p}\end{bmatrix}, &
\bm{\xi}   &=  \begin{bmatrix} \bm{\xi}_{u}   \\ \bm{\xi}_{p}  \end{bmatrix},  &
\bm{\xi}_h &=  \begin{bmatrix} \bm{\xi}_{h,u} \\ \bm{\xi}_{h,p}\end{bmatrix} 
\end{align}
Therefore,  a reduced system for the unknown generalized accelerations is found,
\begin{equation}\label{eq:MBf}
%\Hmat_f \qddvf = \bm{\xi}_f - \left(\Cmat_f + \Hmat_{fi}\qddvi\right) + \bm{\xi}_{e,f}
\HmatD{u}(\qv) \qddvf = \bm{\xi}_u - \CmatD{u}^*(\qv,\qdv) + \bm{\xi}_{h,u},
\end{equation}
where $\CmatD{u}^* = \CmatD{u} + \HmatD{up}\qddvi$.
The reduced system of \eqcite{eq:MBf} has the same form as \eqcite{eq:MB} but consists of $\Ndof - \Ninv$ algebraic equations 
(remember, $\Ninv$ is the number of DoF whose motion is prescribed).
Note also that $\HmatD{u}$ and $\CmatD{u}^*$ depend on all generalized coordinates, 
free and prescribed. 

Equation~\eqref{eq:MBf} is discretized using the same temporal scheme used for the convective terms in \eqcite{eq:NS},
\begin{equation}\label{eq:qddis}
\qdvf^{k} = \qdvf^{k-1}  + \Delta t 
\left( \gamma_k [\HmatD{u}^{-1}(\bm{\xi}_u - \CmatD{u}^*) ]^{k-1} + 
       \zeta_k  [\HmatD{u}^{-1}(\bm{\xi}_u - \CmatD{u}^*) ]^{k-2} +
       [\HmatD{u}^{-1}]^{k-1}\bm{\xi}^{k}_{h,u} \right),
\end{equation}
where the inverse of the reduced joint space matrix ($\HmatD{u}$) is computed using the Cholesky factorization. 
On the other hand, $\bm{\xi}^k_{h,u}$ are the generalized forces mapped from the physical hydrodynamic forces computed from $\vvec{F}^k$ of \eqcite{eq:for} 
as explained in the following section.
 
The generalized coordinates are computed implicitly, as in the coupling method proposed in \citet{uhlmann2005}, using the same scheme as for the viscous terms in \eqcite{eq:NSdis},
\begin{equation}\label{eq:qdis}
\qvf^{k} = \qvf^{k-1}  + \Delta t \alpha_k(\qdvf^{k} + \qdvf^{k-1}).
\end{equation}

As mentioned in the introduction, several methods can be used to derive \eqcite{eq:MB} and compute the corresponding 
joint space inertia matrix, $\Hmat$, and the bias force vector, $\Cmat$.
In the present work, the open-source Rigid Body Dynamics Library (RBDL) developed by \citet{felis2017} has been employed.
%%
%RBDL is a reduced coordinate algorithm implemented as recursive method, which uses the \emph{spatial vector algebra} described in \citet{featherstone2014}.
%%
In particular, $\Cmat$ is computed using the Recursive Newton-Euler algorithm (RNEA), and $\Hmat$ is computed by means of the 
Composite Rigid-Body algorithm (CRBA) \cite{featherstone2014,felis2017}.
These matrices are then reordered into \eqcite{eq:ordered} to obtain $\HmatD{u}$ and $\CmatD{u}^*$.
Lastly, 
the generalized forces are computed from the IBM forcing term, as discussed in next section and in \ref{sec:app_mapF}.
%the expression for the generalized forces is discussed in next section and in \ref{sec:app_mapF}.

Finally, it is worth mentioning that several degrees of freedom are allowed between two connected bodies.
In these cases, the joint that links the bodies is usually denoted as \emph{multiple DoF joint}.
Depending on the involved degrees of freedom, the definition of these joints can become cumbersome.
Therefore, for a higher versatility of the algorithm, multiple DoF joints are simulated using single DoF joints 
(i.e., prismatic or revolute joints) with \emph{virtual} bodies whose mass and inertia is zero \cite{felis2017} 
(see \ref{sec:app_joint} for further details).
The only exceptions are spherical joints (3 rotations), which are simulated using a quaternion formulation to avoid singularities \cite{featherstone2014}.

%%%%%%%%%%%%%%%%%%%%%%%%%%%%%%%%%%%%%%%%%%%%%%%%%%%%%%%%%%%%
% Coupling
%%%%%%%%%%%%%%%%%%%%%%%%%%%%%%%%%%%%%%%%%%%%%%%%%%%%%%%%%%%
\subsection{Coupling\label{ssec:met_coupling}}

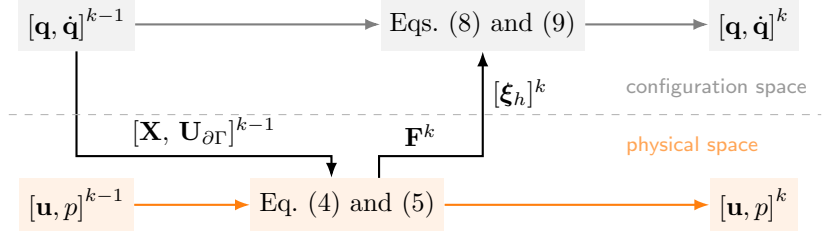
\begin{figure}
   \centering
   \pgfmathsetmacro{\hsp}{15}  % Horizontal spacing
\pgfmathsetmacro{\vsp}{4}   % Vertical spacing

\colorlet{qcol}{C7}
\colorlet{fcol}{C1}

\begin{tikzpicture}[scale=.6,
   arr/.style={thick,->},
   box/.style={minimum height=2em}]%,

   \begin{footnotesize}

   \coordinate (mid) at (0,-\vsp/2);
   \coordinate (phy) at (0,-.73*\vsp);

   \path (0,0) coordinate (cq0) -- (0,-\vsp) coordinate (cf0) 
      -- (\hsp,-\vsp) coordinate (cf1) -- (\hsp,0) coordinate(cq1);

   \node[fill=qcol!10,box] (q0) at (cq0) {$\left[\qv,\qdv\right]^{k-1}$};
   \node[fill=qcol!10,box] (q1) at (cq1) {$\left[\qv,\qdv\right]^{k}$};
    
   \path (cq0) -- (cq1) node[pos=0.6,name=Hq,fill=qcol!10,box] 
              {Eqs.~\eqref{eq:qddis} and \eqref{eq:qdis}};
              %{$\Hmat\qddv = \bm{\xi} - \Cmat + \bm{\xi}_e$};

   \path (cf0) -- (cf1) node[pos=0.4,name=NS,fill=fcol!10,box] 
              {Eq.~\eqref{eq:NSdis} and \eqref{eq:for}};

   \node[fill=fcol!10,box] (f0) at (cf0) {$\left[\uv,p\right]^{k-1}$};
   \node[fill=fcol!10,box] (f1) at (cf1) {$\left[\uv,p\right]^{k}$};

   \draw[arr] (q0.south) |- (phy)
          -| node[pos=.25,above] {$[\xB,\,\vvec{U}_{\partial\Bdy}]^{k-1}$} ([xshift=-1em]NS.north); 

%   \draw[arr] (q0.south) |- (q0 |- mid) 
%          -| node[pos=.25,above] {$\xB,\,\vvec{U}_{\partial\Bdy}$} ([xshift=-1em]NS.north); 

   \draw[arr] ([xshift=+2em]NS.north) |-
         (Hq |- phy) node [pos=.7,above] {$\vvec{F}^k$} -| (Hq.south) 
         node [pos=.8,right] {$[\bm{\xi}_h]^k$};

   \draw[arr,qcol] (q0) -- (Hq);
   \draw[arr,qcol] (Hq) -- (q1);
   \draw[arr,fcol] (f0) -- (NS);
   \draw[arr,fcol] (NS) -- (f1);

   \draw[dashed,black!30] (-.1*\hsp,-\vsp/2) -- (1.1*\hsp,-\vsp/2);

   \end{footnotesize}

   \begin{scriptsize}
      \node[qcol!80,anchor=west] at (.8*\hsp,-.35*\vsp) {\textsf{configuration space}};
      \node[fcol!70,anchor=west] at (.8*\hsp,-.67*\vsp) {\textsf{physical space}};
   \end{scriptsize}
\end{tikzpicture}
   \caption{Schematic diagram of the coupling between the fluid solver and the dynamic algorithm during a Runge-Kutta substep. 
%
%Assumed to be known the fluid state ($[\vvec{u},p]$) and the multi-body systems' state ($[\qv,\qdv]$), 
%the latter is used to compute the velocity and position of the bodies' surfaces to compute the new state of the fluid.
%
%In solving \eqcite{eq:NSdis}, the hydrodynamic forces are computed and expressed in terms of generalized forces;
 these are used to compute the new state of the multi-body system.
\label{fig:coupling}}
\end{figure}

The coupling of the algorithms corresponding to the fluid phase and to the MBS is depicted in Fig.~\ref{fig:coupling}.
%
%Let the state of fluid phase and MBS be known at Runge-Kutta substep $k-1$. The steps are as follows:
With the state of fluid phase and MBS known at Runge-Kutta substep $k-1$, 
the coupling algorithm is as follows:
\begin{enumerate}
\item The generalized coordinates and velocities ($\qv^{k-1}$ and $\qdv^{k-1}$) are used to compute the position 
and velocity of \emph{the Lagrangian points}, namely, $\xB_{i,j}$ and $\vvec{U}^{k-1}_{\partial \Bdy_i}(\xB_{i,j})$.%, $i \in B$.
\item The latter is used in \eqcite{eq:for} to compute $\vvec{F}^k$, which is then transferred to the Eulerian grid,
$\vvec{F}^k \mapsto \vvec{f}^k$.
\item With $\vvec{f}^k$, \eqcite{eq:NSdis} can be solved 
to obtain the state of the fluid phase at substep $k$, namely, $\uv^k$ and $p^k$ inside 
the fluid domain, $\Fld$.
\item The hydrodynamic forces and moments acting on the bodies ($\Fbdy$, $\Mbdy$) are computed from $\vvec{F}^k$, as detailed below. 
\item ($\Fbdy$, $\Mbdy$) are mapped as generalized external forces, $\bm{\xi}_h$. 
Then, \eqcite{eq:qddis} is solved, yielding $\qdv^{k}$.
\item Finally, $\qv^{k}$ is computed from \eqcite{eq:qdis}, fully determining the state of the 
MBS at substep $k$.
\end{enumerate}
Note that, 
in the fluid solver, $\xB$ and $\vvec{U}_{\partial\Bdy}$ are treated explicitly (i.e., at $k -1$), 
while in the multi-body solver $\bm{\xi}_h$ represents the hydrodynamic force integrated between $k-1$ and $k$.
This leads to a \emph{weak coupling} between the sub-systems, 
where the flow field at the solid interface may not be fully compatible with the solid's interface velocity 
at the end of the time step
\citep{uhlmann2005}.

To build the mapping $(\qv,\qdv) \mapsto (\xB_{i,j},\vvec{U}_{\partial\Bdy_i}(\xB_{i,j}))$, 
it is necessary to know the position of a control point of $\Bdy_i$, 
$\vvec{x}_i$; 
%\OF{$\vvec{x}_i^0$;}
%
the orientation of $\Bdy_i$ %(given by an coordinate basis $\Sigma_i$) 
with respect to the inertial coordinate basis, $\Sigma_0$ (given by the rotation matrix $\mathsf{E}_{i}$);
and the angular and linear velocity of the control point, $\bm{\omega}_i$ and $\vvec{v}_i$, respectively.
Thus, 
\begin{subequations}\label{eq:physXU}
\begin{align}
\xB_{i,j} &= \vvec{x}_i^0 + \mathsf{E}_i \vvec{r}^\prime_{i,j}, \\
%\vvec{U}_{\partial\Bdy_i} &= \mathsf{E}_i\vvec{v}_i^i + \mathsf{E}_i\bm{\omega}_i^i \times \vvec{r}^\prime_{i,j},
\vvec{U}_{\partial\Bdy_i} &= \vvec{v}_i^0 + \bm{\omega}_i^0 \times \mathsf{E}_i\vvec{r}^\prime_{i,j},
\end{align}
\end{subequations}
where the superscripts indicate the coordinate basis in which the vector is expressed, 
and $\vvec{r}_{i,j}^\prime$ is the relative position of the point $j \in L(i)$ on the surface of the $i$th body, 
with respect to the body's control point expressed in $\Sigma_i$ (hence, it is a fixed quantity for a rigid body).
Note that, $\vvec{x}_i^0$ and $\mathsf{E}_i$ can be calculated from the rotation matrices and position vectors of the joints that link $\Bdy_i$ to the ground.
Likewise, $\bm{\omega}_i$ and $\vvec{v}_i$ can be expressed as functions $f(\qv,\qdv)$.
In the present case, these variables can directly be extracted from the multi-body library, RBDL.
For the interested reader, \ref{sec:app_mapV} provides the expressions to compute $\vvec{x}_i^0$, $\bm{\omega}_i$ and $\vvec{v}_i$.

%The \emph{motion subspace} of the joint that connects to $\Bdy_{i}$ to its predecessor, $\Bdy_\lambda(i)$, defines the rotation matrix from $\Sigma_{\lambda(i)}$ to $\Sigma_i$; as well as the position of the control point with respect to $\Sigma_{\lambda(i)}$.
%%
%%The \emph{motion subspace} of the joints that connects to $\Bdy_{i}$ to its predecessor, $\Bdy_\lambda(i)$, defines the rotation matrix from $\Sigma_{\lambda(i)}$ to $\Sigma_i$, namely, $\mathsf{E}_{\lambda(i)\to i}$; as well as the position of the control point with respect to $\Sigma_{\lambda(i)}$, namely $\vvec{r}^{\lambda(i)}_{i}$.
%
%The position control point, $x_i$, the rotation matrices, $\mathsf{E}_i$, and the velocities can be computed from the \emph{motion subspaces} of the joints.

The hydrodynamic forces and moments acting upon the body $\Bdy_i$, namely, $\Fbdy_i$ and $\Mbdy_i$, 
can be shown to be \citep{uhlmann2005}:
\begin{subequations}\label{eq:FMi}
\begin{align}
\Fbdy_i &= \underbrace{-\rhof \sum_{j \in L(i)} \vvec{F}(\xB_{i,j})\Delta V_j}_{\Gbdy_{i}}
+ \frac{\rhof}{\rho_i} m_i \ddot{\vvec{x}}_{G,i}, \label{eq:Fi}\\
%
%\Mbdy_{i} &=  -\rhof \sum_{j \in L(i)} \left(\xB_j - \vvec{x}_{G,i}\right)\times
%\vvec{F}(\xB_j)\Delta V_j + 
%\frac{\rhof}{\rho_i} \frac{\text{d} \vvec{H}_{G,i}}{\text{d}t}.\\
%
\Mbdy_{i} &=  \underbrace{-\rhof \sum_{j \in L(i)} \left(\xB_{i,j} - \vvec{x}_{i}\right)\times
\vvec{F}(\xB_j)\Delta V_j}_{\Nbdy_{i}} + 
\rhof \int_{\partial\Bdy_i} \left(\vvec{x} - \vvec{x}_{i}\right) \times \uv \,\text{d}\vvec{x} ,
\label{eq:Mi}
\end{align}
\end{subequations}
where $\rho_i$, $m_i$ and $\vvec{x}_{G,i}$ are the density, mass and position of the gravity centre of $\Bdy_i$, respectively.
%, and $\vvec{x}_{i}$ is the point about which $\Mbdy_i$ is computed.
%
The integral term of \eqcite{eq:Mi} represents the rate of change of angular momentum of the fluid inside $\Bdy_i$ 
whose value has to be computed by numerical integration, as in other works \citep{tschisgale2017}. 
However, in the present work this term is approximated by supposing rigid-body motion of the fluid inside $\Bdy_i$,
%for efficiency reasons.
an assumption that is justified for small or relatively thin bodies (such as filaments, wings or fins).
This entails that the contribution of the last term of \ref{eq:Fi} and \ref{eq:Mi}
%\eqcite{eq:FMi} 
can be embedded in $\Hmat$ and $\Cmat$ if they are built using an effective density for each body equal to $(\rho_i - \rhof)$, 
in a similar fashion as for single rigid bodies.
This imposes a lower limit of the density ratios that can be simulated using the present approach of approximately $\rho_i/\rhof \geq 1.2$, based on \cite{uhlmann2005}.
The remaining terms of \eqcite{eq:FMi}, namely $\Gbdy_i$ and $\Nbdy_i$, constitute then $\bm{\xi}_h$  of \eqcite{eq:MB}, after being mapped to the space of generalized forces (see \ref{sec:app_mapF} for further details).

%%%%%%%%%%%%%%%%%%%%%%%%%%%%%%%%%%%%%%%%%%%%%%%%%%%%%%%%%%%%%%%%%%%%%%%%%%%%%%%%
%% VALIDATION  
%%%%%%%%%%%%%%%%%%%%%%%%%%%%%%%%%%%%%%%%%%%%%%%%%%%%%%%%%%%%%%%%%%%%%%%%%%%%%%%%
\section{Validation\label{sec:val}}

As detailed in \S~\ref{sec:met}, the multi-body algorithm developed herein has been coupled with a 
pre-existing flow solver which has already been used to solve the coupled fluid-structure interaction 
of single rigid bodies \cite{arranz2018a,arranz2018b}.
Since the multi-body algorithm can be also used to simulate the problem of a single rigid body with several DoFs (i.e., by defining virtual, mass-less bodies, linked by the single DoFs joints described in appendix \ref{sec:app_joint}), we can compare the results of the multi-body algorithm with those obtained with the pre-existing algorithm.
Since the flow solver is the same, the only differences should arise from the construction and solution of the dynamic equations.
Several test cases have been performed following this methodology (both in 2D and 3D), including the free fall of circular cylinders (2D) and spheroids (3D), and the auto-rotation of a winged \emph{samara} seed \cite{arranz2018a}, yielding an excellent agreement between the multi-body and single-rigid-body algorithms. 
These tests are not presented herein for the sake of brevity, and instead, we 
validate our multi-body solver against 2D multi-body problems previously reported 
in the literature \citep{toomey2008,arora2018}. 
Additionally, we exemplify the ability of the present methodology to simulate 3D flexible bodies by  comparing a multi-body model of a flexible flag with a finite-element model \citep{tian2014,lee2015}.

%present comparison with existing multi-body problems available in the literature.

\subsection{Flapping of a flexible airfoil\label{ssec:toomey}}

\begin{figure} 
   \centering
   \ig[width=.3\tw]{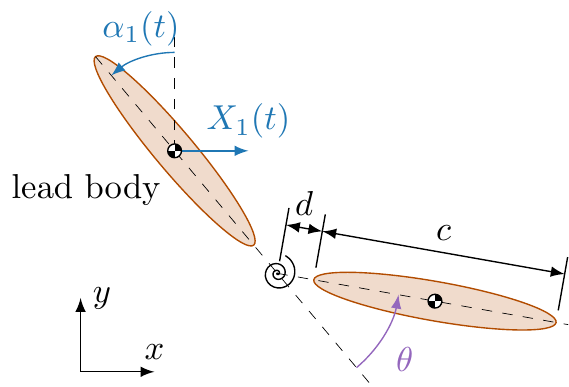}
   \caption{Sketch of the validation case adapted from \citet{toomey2008}.
\label{fig:toomeyscheme}}
\end{figure}

The first validation case is extracted from \citet{toomey2008}.
A sketch of the problem is shown in Fig.~\ref{fig:toomeyscheme}.
It consists of two 2D rigid bodies connected by a torsional spring and a damper, immersed in a quiescent fluid.
Both bodies are ellipses of aspect ratio $5:1$ of major axis $c$.
The distance between each body and the torsional spring is $d = 0.05c$.
The motion of the lead body is fully prescribed and given by the linear displacement
of its centre of mass, $X_1(t)$, and the orientation angle, $\alpha_1(t)$.
The motion of the follower body 
is given by the deflection angle, $\theta(t)$, between the follower and the lead body (see Fig.~\ref{fig:toomeyscheme}), which results
from the dynamic interaction with the lead body and the surrounding fluid.
Hence, the degrees of freedom of the MBS are $X_1, \alpha_1$ and $\theta$, 
while the only unknown
degree of freedom 
%of the problem 
is $\theta$.
Consequently, the vectors of generalized coordinates are $\qvi = [X_1,\,\alpha_1]$ and $\qvf = \theta$.
The prescribed motion of the lead body follows the time laws
\begin{align*}
   X_1(t) &= \frac{A_0}{2}\frac{G_t(ft) C(ft)}{\max{(G_t)}}, &
   \alpha_1(t) &= - \beta \frac{G_r(ft)}{\max{(G_r)}},
\end{align*}
where $f$ is the frequency  of oscillation 
($T=1/f$ is the period of oscillation),
and the translational and angular amplitudes are set to $A_0/c = 1.4$ and $\beta = \pi/4$, respectively.
Furthermore,
\begin{align*}
G_t(t) &= \int_t \tanh{(\sigma \cos{2\pi t')}} dt^\prime, &
G_r(t) &= \tanh{(\sigma \cos{2\pi t)}}, &
C(t) = \frac{\tanh{(8t-2) + \tanh{2}}}{1 + \tanh{2}}.
\end{align*}
 
\begin{figure}[h]
   \centering
   \begin{subfigure}{.48\tw}
      \ig[width=\tw]{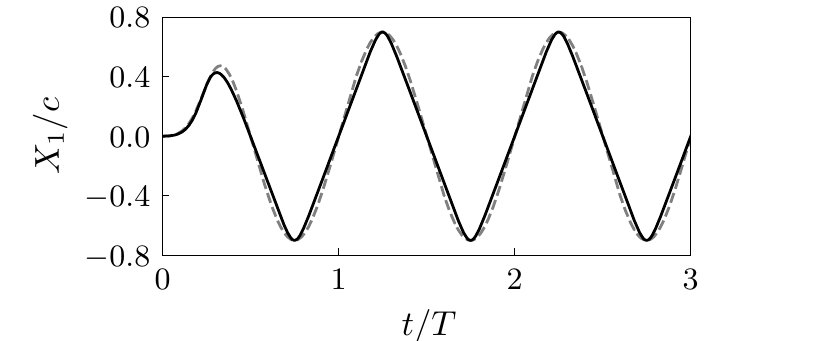}
      \caption{\label{fig:toomey_X1}}
   \end{subfigure}\hfill
   \begin{subfigure}{.48\tw}
      \ig[width=\tw]{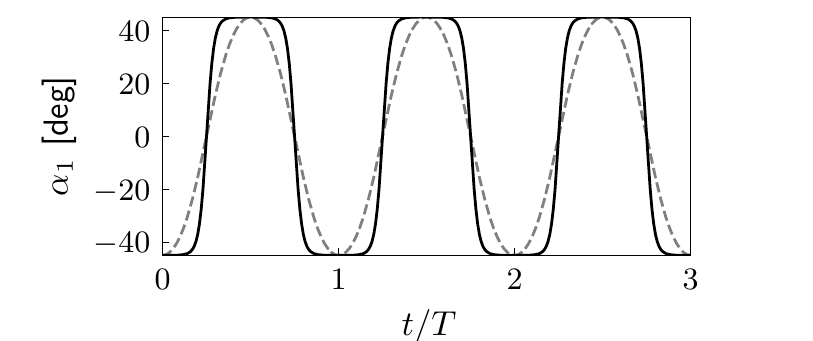}
      \caption{\label{fig:toomey_A1}}
   \end{subfigure}
   \caption{(a) Horizontal position and (b) pitch angle of the lead body as a function of time.
\lyy{--}{gray} Case 1; \lyy{-}{black} Case 4. \label{fig:toomey_kin}} 
\end{figure}

Two cases from \citet{toomey2008} are considered: Case 1 ($\sigma = 0.628$) and Case 4 ($\sigma = 3.770$). 
Fig. \ref{fig:toomey_kin} depicts the kinematics of the lead body for both cases.
The rest of parameters that fully define the problem are the rotation Reynolds number, 
$\mathit{Re}_r = \dot{\alpha}_{M} c^2/\nu = 100$ 
(where $\dot{\alpha}_{\textit{M}}$ is the maximum angular velocity); the body-to-fluid density ratio, $\rho_s/\rhof = 5$; 
the dimensionless spring stiffness, $K^* = K/(\rhof f^2 c^4) = 456$; 
and the damping coefficient, $R^* = R/(\rhof f c^4) = 3.95$.

The results obtained with the present algorithm are compared with the results reported by 
%\citet{toomey2008} using a viscous vortex particle method and with the results reported by \citet{wang2015} using a strongly coupled dynamic algorithm with the IB-projection method.
\citet{toomey2008} using a viscous vortex particle method, and with the results reported by \citet{wang2015} using the IB-projection method. Note that both cases use strongly coupled body dynamics, while in the present algorithm the body dynamics are weakly coupled with the fluid solver. 

In the present simulations, 
the computational domain is a square of side $32c$ (like in \citet{wang2015}) with periodic boundary conditions.
At $t = 0$, the centre of mass of the lead body is located at $8c$ from the top wall and centred in the horizontal direction.
The fluid domain is discretized with a uniform mesh of grid spacing $\Delta x = \Delta y = c/64 $. This is a slightly coarser resolution than the one used in previous studies \cite{toomey2008,wang2015} ($\Delta x \approx c/100$).
%l––
The time step is set to $\Delta t = 5\cdot10^{-4}T$, leading to a Courant-Friedrichs-Lewy 
condition number, $\mathit{CFL} = \frac{U_{\textit{max}} \Delta t}{\Delta x} \leq 0.4$ 
(where $U_{\textit{max}}$ is the instantaneous maximum flow velocity in the fluid domain).
Finally, the bodies are discretized using a Lagrangian mesh with evenly spaced points with distance $\Delta x$.

\begin{figure}[h]
   \centering
   \begin{subfigure}{.5\tw}
      \ig[width=\tw]{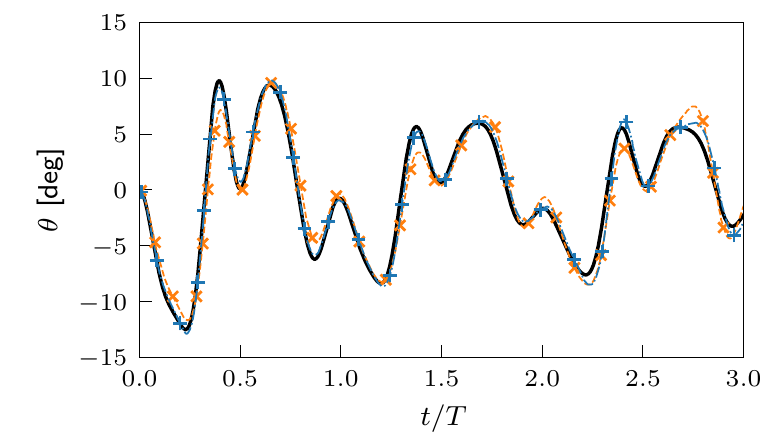}
      \caption{Case 1\label{fig:toomey_c1_th}}
   \end{subfigure}\hfill
   \begin{subfigure}{.5\tw}
      \ig[width=\tw]{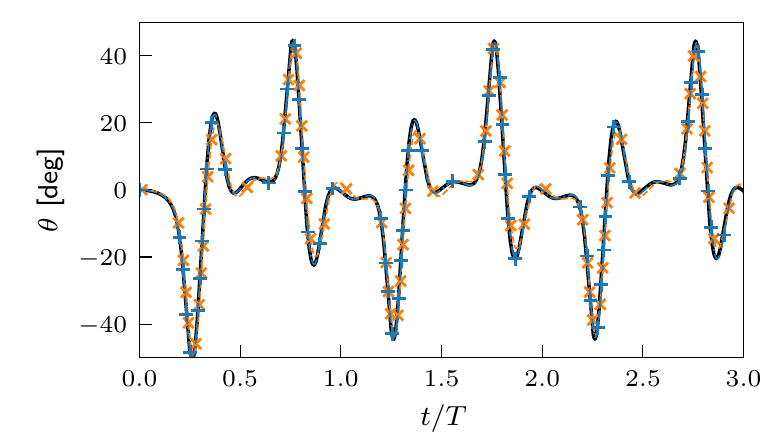}
      \caption{Case 4\label{fig:toomey_c4_th}}
   \end{subfigure}\vspace{1em}
   \begin{subfigure}{.5\tw}
      \ig[width=\tw]{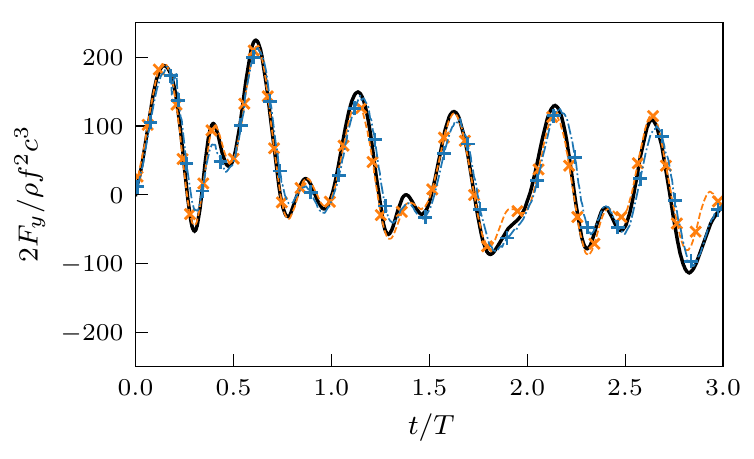}
      \caption{Case 1\label{fig:toomey_c1_fy}}
   \end{subfigure}\hfill
   \begin{subfigure}{.5\tw}
      \ig[width=\tw]{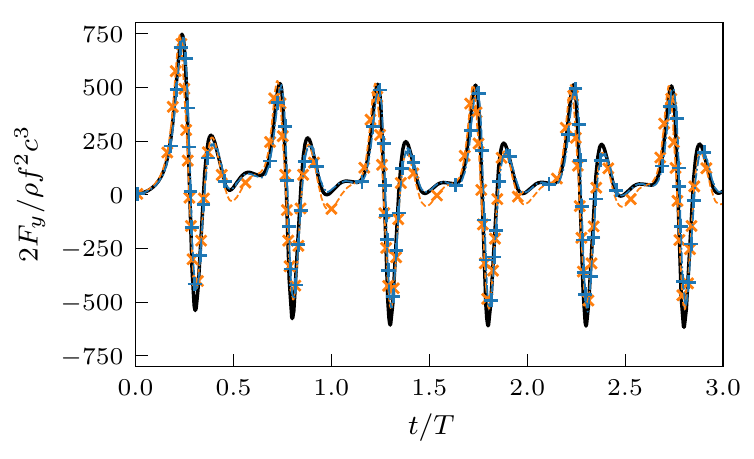}
      \caption{Case 4\label{fig:toomey_c4_fy}}
   \end{subfigure}   
\caption{(a-b) Deflection angle of the follower body, and (c-d) total vertical force acting upon the multi-body system
as a function of time.
Note the difference in scale between cases 1 and 4.
\lyy{-}{black} present results, \myls{+}{C0}{.-} \citet{toomey2008}, and \myls{x}{C1}{--} \citet{wang2015}.\label{fig:toomey}}
\end{figure}

Fig.~\ref{fig:toomey} depicts the comparison of the deflection angle, $\theta$,
and the vertical force, $F_y$, between the current results and the existing literature.
A good agreement of the evolution of the deflection angle, and the non-dimensional 
vertical force is observed for both cases despite the different numerical approaches and computational details.

\subsection{Self-propelling flexible plate\label{ssec:Varora}}

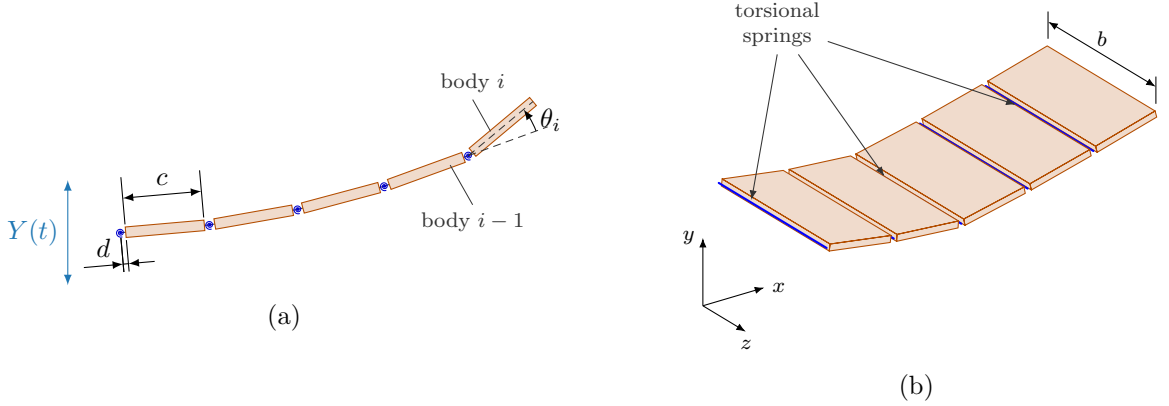
\begin{figure}
   \begin{subfigure}{.6\tw}
      \centering
      \pgfmathsetmacro{\ddis}{.15}
\pgfmathsetmacro{\theo}{-50}
\pgfmathsetmacro{\blen}{.3}
\pgfmathsetmacro{\thic}{.02}

\colorlet{mycolor}{red!70!green}
\begin{tikzpicture}[scale=3.5,>={latex[length=.2cm]},
   body/.style ={color=mycolor,fill=mycolor!20},
   CG/.pic= {
      \draw[fill=white] (0,0) circle (.1);
      \clip (0,0) circle (.1);
      \fill[black] (0,0) rectangle (.5,.5);
      \fill[black] (0,0) rectangle (-.5,-.5);},
   spring/.pic={
      \draw [domain=0:24,variable=\t,smooth,samples=75]
             plot ({\t r}: {0.0001*\t*\t});}]

   \begin{footnotesize}

   \coordinate (G) at (0,0);

   \coordinate (C) at (0,0);
   \coordinate (Ca) at (C);
   \foreach \j in {1,...,4} 
   { 

      \pgfmathparse{\blen+2*\thic};
      \coordinate (C) at ($(\j*5:\pgfmathresult)+(C)$);

      \draw[thin,black!80] (Ca) -- (C);
      %\fill[fill=blue!80!black] (Ca) circle (.008);
      \pic[blue!80!black] at (Ca) {spring};

      \draw[rotate around={\j*5:(Ca)},body] 
          (Ca) ++ (\thic+.5*\blen,0) ++ (-.5*\blen,-\thic) rectangle ++(\blen,2*\thic);

      \coordinate (Ca) at (C);
      \coordinate (C\j) at (C);
   }

   \foreach \j in {5} 
   { 
      \pgfmathparse{\thic};
      \coordinate (C) at ($(\j*8:\pgfmathresult)+(C)$);

      \draw[thin,black!80] (Ca) -- (C);
      %\fill[fill=blue!80!black] (Ca) circle (.008);
      \pic[blue!80!black] at (Ca) {spring};

      \draw[rotate around={\j*8:(Ca)},body] 
          (Ca) ++ (\thic+.5*\blen,0) ++ (-.5*\blen,-\thic) rectangle ++(\blen,2*\thic);

      \draw[rotate around={\j*8:(Ca)},very thin,densely dashed,black!80] (Ca) --+ (\thic+\blen,0);
      \draw[rotate around={4*5:(Ca)} ,very thin,densely dashed,black!80] (Ca) --+ (\thic+\blen,0);
      \draw[thin,->] (Ca)+(20:\blen/1.1) arc (20:\j*8:\blen/1.1) node [midway,right] {$\theta_i$};   

      \coordinate (Ca) at (C);
      \coordinate (C\j) at (C);
   }

   % First body:
   \begin{scope}[rotate=5]
      \draw (0,-.02) --  (0,-.15);   
      \draw (\thic,-.03) -- (\thic,-.15);   
      \draw (0,-.12) --+ (\thic,0);   
      \draw[->] (-.15,-.12)  -- (0,-.12) node[above, midway] {$d$};   
      \draw[<-] (\thic,-.12)  --+ (.1,0);

      \draw[<->] (\thic,.13) --+ (\blen,0)node[above, midway] {$c$};
      \draw (0,-.02) --  (0,-.15);   
      \draw (\thic,+.03) --+ (0,.18);   
      \draw (\thic+\blen,+.03) --+ (0,.18);   

   \end{scope}
 
      \draw[C0,<->] (-.2,-.2) --+ (0,.4) node [left,midway] {$Y(t)$};

   \end{footnotesize}
   \begin{scriptsize}
      %\node (Kspr) at (.8,.7) {$K$};
      %\draw[->,shorten >=.1cm] (Kspr) -- (C1);
      %\path (Kspr) -- ($(C2)!0.5!(C3)$)node[midway,rotate=20]{...};
      %\draw[->,shorten >=.1cm] (Kspr) -- (C4);
      \draw[shorten <=.2em,very thin,black!80] (C5)+(40:.1) --+ (up:.2)node [above] (bi) {body $i$};
      \draw[shorten <=.2em,very thin,black!80] (C5)+(20:-.1) --+ (down:.2)node [below] (bi) {body $i-1$};  
      %\path[] (C5) --+ (40:.1) --+ (up:.2)node [above] {body $i$} (bi);

   \end{scriptsize}
\end{tikzpicture}
      \caption{\label{fig:arorascheme}}
   \end{subfigure}
   \begin{subfigure}{.38\tw}
      \centering
      %\tdplotsetmaincoords{70}{28
\tdplotsetmaincoords{-55}{25}

\pgfmathsetmacro{\ddis}{.15}
\pgfmathsetmacro{\theo}{-50}
\pgfmathsetmacro{\blen}{2}
\pgfmathsetmacro{\thi}{.2}
\pgfmathsetmacro{\mthi}{.1}

\colorlet{mycolor}{red!70!green}
\begin{tikzpicture}[scale=.5,tdplot_main_coords,>={Latex[length=.13cm]},
   seg/.style = {rounded corners=.01,thin,color=mycolor,fill=mycolor!20,tdplot_rotated_coords,scale=.5},
   seg/.pic={
   %\draw[wing] (.01,-.5,0) --+ (0,1,0) --+ (0,1,.5) --+ (0,0,.5) --  cycle
   %            (-.01,-.5,0) --++ (0,1,0) --++ (.02,0,0) --++ (0,-1,0) -- cycle
   %            (-.01,-.5,0) --++ (.02,0,0) --++ (0,0,.5) --++ (-.02,0,0) -- cycle;
   \draw[seg] ( 0,-\blen-\mthi,\mthi)  --+ (0,\blen,0) --+ (5,\blen,0) --+ (5,0,0) --  cycle
              ( 5,-\blen-\mthi,\mthi) --++ (0,\blen,0) --++ (0,0,-\thi) --++ (0,-\blen,0) -- cycle 
              ( 0,-\blen-\mthi,\mthi) --++ (0,0,-\thi) --++ (5,0,0) --++ (0,0,\thi) -- cycle;
  }]

   \begin{scriptsize}

   %% FIRST BODY
   \tdplotsetrotatedcoords{90}{-15}{-90}
   \pic at (0,0,0) {seg};
   \draw[tdplot_rotated_coords,line cap=round,thick,blue] (0,-\blen-\thi,0) --+ (5,0,0) coordinate [pos=.3] (sp4);
   
   \tdplottransformrotmain{0}{-\thi/2}{0}
   \coordinate (co1) at (\tdplotresx,\tdplotresy,\tdplotresz);

   \tdplottransformrotmain{5}{-\thi/2}{0}
   \coordinate[tdplot_screen_coords] (co2) at (\tdplotresx,\tdplotresy,\tdplotresz);

   \draw[tdplot_main_coords,<->] ($(co1)+(0,0,.8)$) -- ($(co2)+(0,0,.8)$)
         node [midway,above] {$b$};
   \draw[tdplot_main_coords] ($(co1)+(0,0,.3)$) --+ (0,0,.8) 
                             ($(co2)+(0,0,.3)$) --+ (0,0,.8);
   %\draw[tdplot_main_coords,] ($(co1)+(0,0,.8)$) -- ($(co2)+(0,0,.8)$);
 
   %\draw[tdplot_rotated_coords,<->,>={Latex[length=.15cm]}] (0,0,.8) --+ (5,0,0);

   %% SECOND BODY
   % Get origin for RF
   \pgfmathparse{-\thi-\blen}
   \tdplottransformrotmain{0}{\pgfmathresult}{0}
   % Set origin and orientation of RF
   \coordinate (C0) at (\tdplotresx,\tdplotresy, \tdplotresz);
   \tdplotsetrotatedcoords{90}{-14}{-90}
   \tdplotsetrotatedcoordsorigin{(C0)}
   \pic at (0,0,0) {seg};
   \draw[tdplot_rotated_coords,line cap=round,thick,blue] (0,-\blen-\thi,0) --+ (5,0,0) coordinate [pos=.3] (sp3);

   %% THIRD BODY
   % Get origin for RF
   \pgfmathparse{-\thi-\blen}
   \tdplottransformrotmain{0}{\pgfmathresult}{0}
   % Set origin and orientation of RF
   \coordinate (C0) at ($(C0)+(\tdplotresx,\tdplotresy, \tdplotresz)$);
   \tdplotsetrotatedcoords{90}{-12}{-90}
   \tdplotsetrotatedcoordsorigin{(C0)}
   \pic at (0,0,0) {seg};
   \draw[tdplot_rotated_coords,line cap=round,thick,blue] (0,-\blen-\thi,0) --+ (5,0,0) coordinate [pos=.3] (sp2);

   %% FOURTH BODY 
   % Get origin for RF
   \pgfmathparse{-\thi-\blen}
   \tdplottransformrotmain{0}{\pgfmathresult}{0}
   % Set origin and orientation of RF
   \coordinate (C0) at ($(C0)+(\tdplotresx,\tdplotresy, \tdplotresz)$);
   \tdplotsetrotatedcoords{90}{+4}{-90}
   \tdplotsetrotatedcoordsorigin{(C0)}
   \pic at (0,0,0) {seg};
   \draw[tdplot_rotated_coords,line cap=round,thick,blue] (0,-\blen-\thi,0) --+ (5,0,0) coordinate [pos=.3] (sp1);

   %% FOURTH BODY 
   % Get origin for RF
   \pgfmathparse{-\thi-\blen}
   \tdplottransformrotmain{0}{\pgfmathresult}{0}
   % Set origin and orientation of RF
   \coordinate (C0) at ($(C0)+(\tdplotresx,\tdplotresy, \tdplotresz)$);
   \tdplotsetrotatedcoords{90}{8}{-90}
   \tdplotsetrotatedcoordsorigin{(C0)}
   \pic at (0,0,0) {seg};
   \draw[tdplot_rotated_coords,line cap=round,thick,blue] (0,-\blen-\thi,0) --+ (5,0,0) coordinate [pos=.3] (sp);

   \draw[tdplot_screen_coords,shorten <=.1em,<-,black!80] (sp) --+ (80:4) node [above,align=center]  (splab)
         {torsional \\ springs};
   \draw[tdplot_screen_coords,shorten <=.1em,<-,black!80] (sp2) --  (splab);
   \draw[tdplot_screen_coords,shorten <=.1em,<-,black!80] (sp4) --  (splab);

   \coordinate (C1) at (1,-12,-4);
   \draw[black,->] (C1) --+ (0,0,2) node [left] {$y$};
   \draw[black,->] (C1) --+ (0,2,0) node [right] {$x$};
   \draw[black,->] (C1) --+ (2,0,0) node [below] {$z$};
  
   \end{scriptsize}

\end{tikzpicture}
      \caption{\label{fig:arora3d}}
   \end{subfigure}
   \caption{(a) Lumped-torsional flexibility model of the flexible plate based on \citet{arora2018} consisting of 5 rectangular rigid bodies joined by torsional springs.
(b) Extension of the model of \citet{arora2018} to three-dimensions by considering rectangular plates
of width $b$.
}
\end{figure}

The second validation case is taken from \citet{arora2018} and consists of
a 2D self-propelling flexible plate.
The plate is modelled using a lumped-torsional flexibility model as shown in Fig.~\ref{fig:arorascheme}.
In particular, the plate of chord $C$ and thickness $e/C = 0.02$ is divided into five rectangular rigid bodies, of uniform density, $\rho_s$, separated by a distance $2d = e$, joint by torsional springs of torsional stiffness, $K$.
The plate is free to move in the horizontal direction, whereas the vertical position of the leading edge is prescribed as:
\begin{equation}\label{eq:Yarora}
Y(t) = A \cos(2\pi f t)
\end{equation} 
The relative deflection angle of body $i$ with respect its predecessor, $i-1$, is defined as $\theta_i$ (see Fig.~\ref{fig:arorascheme}).
Consequently, 
the generalized vectors of the multi-body system are $\qvi = Y$ and $\qvf = [X,\,\theta_1,...,\,\theta_5]$,
where $X$ is the horizontal coordinate of the leading edge of the plate. 

The parameters that govern the present problem are the non-dimensional amplitude, $A/C = 0.6$; 
the Reynolds number, $\mathit{Re} = AfC/\nu = 100$; the body-to-fluid density ratio, $\rho_s/\rhof = 10$; 
and the non-dimensional torsional stiffness, $K^* = K/\rhof f^2 C^4 = 52.242$.
The chosen parameters correspond to the case $\psi = {A\rhof}/(e\rho_s) = 3$, 
$\omega^* = \omega_n/(2\pi f) = 3.5$ (where $\omega_n$ is the first natural frequency of the multi-body system), 
shown in \citet[Fig.~15 and 16]{arora2018}.

The simulations are performed in a rectangular domain of size $20C\times12C$ uniformly discretized with a grid size $\Delta x = \Delta y = 0.004C$.
Note that, due to the plunging motion of the plate's leading edge (LE), it moves horizontally at an instantaneous speed $\dot{X}(t)$.
Therefore, within each cycle the plate travels an horizontal distance $U_p T$, 
where $U_p = f\int_{T-1}^T \dot{X} \mathrm{d}t$ is the mean propulsive speed.
In order to prevent the plate from leaving the computational domain, 
%for the plate not to leave the computational domain, 
the plate is immersed in a uniform stream flow 
of intensity $U_\infty$ such that $U_\infty \approx U_p$ and the mean horizontal displacement of the plate with respect to the computational domain is as small as possible. 
Consequently, Dirichlet boundary conditions are imposed at the inflow and lateral walls on the horizontal ($u = U_\infty$) and vertical ($v = 0$) fluid velocities.
The value of $U_\infty$ equals the 
estimated mean propulsive velocity extracted from \citep[Fig.~14]{arora2018}). 
An advective boundary condition is imposed at the outflow boundary.
The bodies are discretized using a Lagrangian grid with equidistant points separated $\Delta x = 0.004C$ and the time step is $\Delta t = 5 \cdot 10^{-5}T$ (where $T = f^{-1}$).
On the other hand, simulations in \citet{arora2018} are performed using a Lattice-Boltzmann method with a larger computational domain ($50C\times20C$) and a similar spatial resolution ($\Delta x = 0.005C$).

\begin{figure}[h]
   \centering
   \hspace{-2em}
   {\begin{tikzpicture}
      \node[mylab={a}] (fx) at (0,0)      {\ig[width=.35\tw]{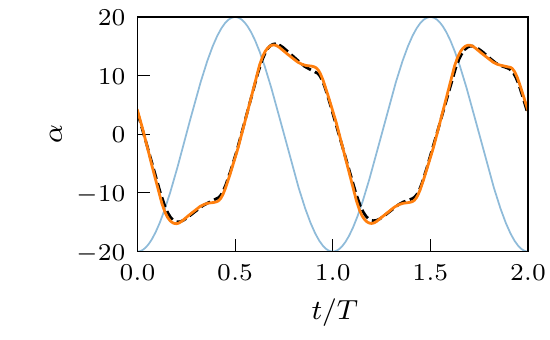}};
      \node[mylab={b}] (fy) at (.34\tw,0) {\ig[width=.35\tw]{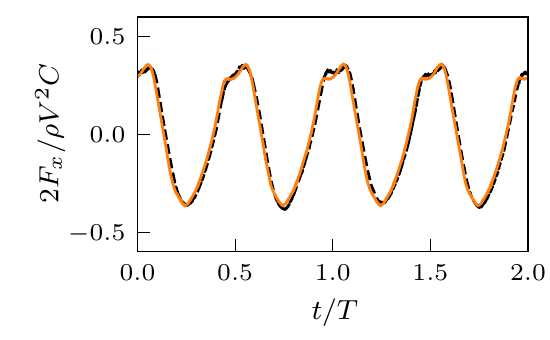}};
      \node[mylab={c}] (fz) at (.67\tw,0) {\ig[width=.35\tw]{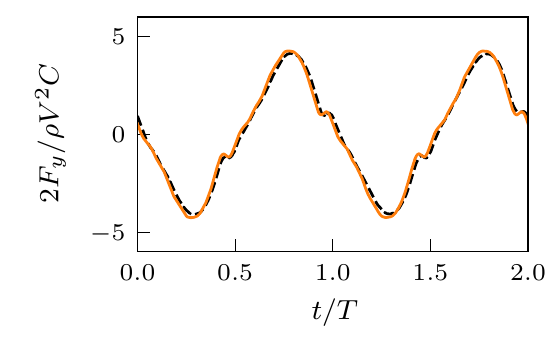}};
   \end{tikzpicture}
   \phantomsubcaption\label{fig:arora_al}
   \phantomsubcaption\label{fig:arora_cx}
   \phantomsubcaption\label{fig:arora_cy}
   }%
   \vspace{-1em}
   \caption{Comparison of the (a) tip deflection angle, (b) horizontal force and (c) vertical force: \lyy{-}{C1} present results, and \lyy{--}{black} \citet{arora2018}.
In Fig.~\ref{fig:arora_al}, \ly{-}{C0!50} corresponds to the $Y$-position of the leading edge (without scale).\label{fig:arora}}
\end{figure}

Fig.~\ref{fig:arora}a depicts the time evolution during two cycles of the tip deflection angle (defined as the angle between the horizontal and the line that joins the leading edge and the trailing edge).
Fig.~\ref{fig:arora}b-c shows the time evolution of
the horizontal and vertical forces exerted on the plate 
normalized by $\frac{1}{2}\rho V^2C$, where $V = 2\pi f A$ is the maximum vertical velocity of the leading edge.
It should be pointed out that results from \citet{arora2018} correspond to cycles $48-49$, whereas the present results are for cycles $12-13$, since no changes were observed with respect to previous cycles.
Nevertheless, a good agreement is observed between the present simulations and those from \citet{arora2018}, 
with relative differences in the peak-to-peak amplitudes of the forces of less than $3\%$, 
and an absolute difference of the maximum tip-deflection angle lower than $0.19^\circ$.

\subsection{Three-dimensional flapping flag in a free stream\label{ssec:flag}}
The third case considered is the 3D flow around a flag, flapping freely in a uniform stream.
This case has been studied by several authors \citep{tian2014,lee2015}
using finite-element structural solvers.
%
%Contrary to the previous examples, this case is three-dimensional, and the results are
%not compared against a multi-body solver, but to two higher fidelity, finite-element
%structural solvers.
%
%Thus, t
The objective of presenting this comparison is not to provide a direct validation of the
multi-body algorithm (already proven in the previous subsections), but to exemplify
the capabilities of a multi-body approximation to simulate the dynamics of flexible bodies.

The problem setup corresponds to the one reported by  \citet{lee2015}. 
It consists of a flexible, square flag of side length $C$ and thickness $e/C = 0.01$,
immersed in a uniform free-stream of velocity $U_\infty$.
The flag-to-fluid density ratio is $\rho_s/\rho = 100$, the non-dimensional 
elastic modulus of the flag is $E/\rho U_\infty^2 = 1008$, and 
$Re = U_\infty C/\nu = 200$.
The computational domain is a rectangular prism of size $8C\times8C\times2C$ in the 
streamwise ($x$), vertical ($y$) and spanwise directions ($z$), respectively.
Dirichlet boundary conditions ($u_x = U_\infty$, $u_y = u_z = 0$) are imposed at 
the top, bottom and inflow boundaries; 
periodic boundary conditions are imposed at the lateral boundaries; 
and an advective boundary condition is imposed at the outflow boundary.
The leading edge of the flag is fixed and parallel to the $z$-axis, with its centre 
located at a distance $C$ downstream of the inflow boundary and centered in the $y$ and
$z$ directions.
A grid with uniform spacing $\Delta x = \Delta y = \Delta z = C/64$ is used to
discretize a refined region which contains the flag, whereas a constant stretching 
factor of $1.5\%$ is applied to the grid in the streamwise and vertical directions 
outside the refined region.
The time step is $\Delta t = 0.002 C/U_\infty$ ensuring $\textit{CFL} < 0.28$.

The structural model of the flag only considers flexibility along the chordwise direction. 
It is an extension of the lumped-torsional model of \citet{arora2018} to three dimensions,
as depicted in Fig.~\ref{fig:arora3d}.
Since $e \ll C$, the flag is assumed to be an infinitely thin surface, as in \citet{tian2014}.
Thus, the rigid bodies in Fig.~\ref{fig:arora3d} become rectangular surfaces, discretized with a Cartesian distribution of Lagrangian points with a uniform spacing $\Delta x$.
The total number of bodies conforming the flag is set to 25, with $d/C = 0$.  
The torsional stiffness of the joints, $K$,
is computed to match the natural frequency of the flag, yielding
$K/\rho U_\infty^2 C^2 = 2.129 \cdot 10^{-3}$.
% https://www.overleaf.com/project/5db9c69fe9eec80001219f67

The self-sustained flapping motion of the flag is triggered by the initial condition, as in \citet{lee2015}.
Hence, at the beginning of the simulation the flag is flat, with an angle of attack $\alpha=0.1\pi$ rad with respect to the free-stream. 
%with respect to the
%free-stream velocity.
%
As a consequence of this initial asymmetry with respect to the midplane, the flag starts a self-sustained 
flapping motion about its leading edge with a constant frequency. 
A total of number of 5 flapping cycles have been simulated with the multi-body algorithm, as in \citet{lee2015}.

\begin{figure}[h]
   \centering
   \begin{subfigure}{.33\tw}
      \centering
      \ig[width=\tw]{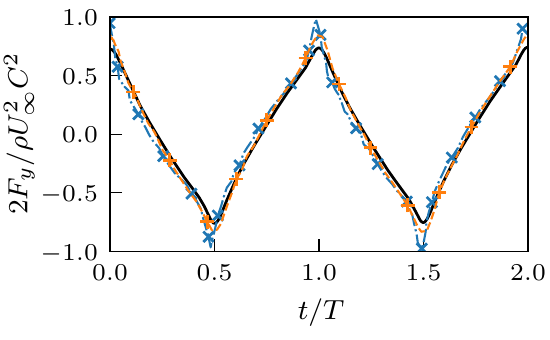}
      \caption{\label{figi:flag_cy}}
   \end{subfigure}\hfill
   \begin{subfigure}{.33\tw}
      \centering
      \ig[width=\tw]{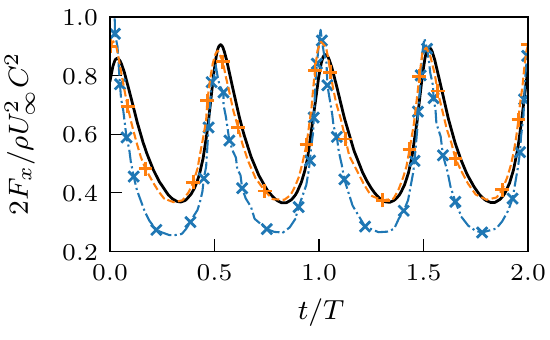}
      \caption{\label{fig:flag_cx}}
   \end{subfigure}
   \begin{subfigure}{.33\tw}
      \centering
      \ig[width=\tw]{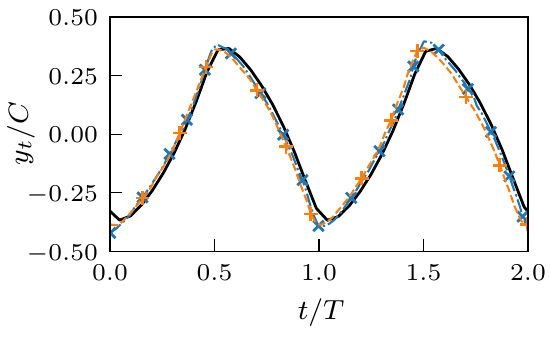}
      \caption{\label{fig:flag_def}}
   \end{subfigure}
   \caption{Comparison of the (a) Vertical force, (b) horizontal force and 
       (c) vertical position of the tip at midspan.
       \lyy{-}{black} present results, \myls{+}{C0}{.-} \citet{tian2014}, 
       and \myls{x}{C1}{--} \citet{lee2015}.\label{fig:flag}}
\end{figure}

Figure~\ref{fig:flag} displays the temporal evolution of the vertical and horizontal
force coefficients, and the vertical displacement of the tip at midspan.
The plot shows the last two cycles of the simulations. 
The results from \citet{tian2014} and \citet{lee2015} are also shown for comparison.
Note that, \citet{tian2014} and \citet{lee2015} employ finite-element methods to model
the flag, thus accounting for spanwise flexibility and twisting 
of the flag.
However, for the case considered, the deformations along the span are small 
(see Fig. 14b of \citet{lee2015}).
Hence, even if the present multi-body model of the flag does not allow for spanwise flexibility, Fig.~\ref{fig:flag} shows that both the forces and the vertical displacement at the tip computed with the 
proposed algorithm are in good agreement with the previous references, in particular
with those reported by \citet{lee2015}.
Discrepancies with the results provided by \citet{tian2014} might be due to the differences in the computational setup.
%
%This agreement can be in part attributed to the small deformations of the flag
%in the spanwise direction (as shown in Fig.~15 of \citet{lee2015}) 
%but shown the capabilities of the proposed algorithm to simulate flexible bodies.
%
Note that the time in Fig.~\ref{fig:flag} is normalized with the flapping period ($T$), which 
is a result of the simulation.
Table~\ref{tab:flag_St} shows the Strouhal number ($St = fC/U_\infty$, where $f = 1/T$ is the flapping frequency) 
for the present case and those from \citet{tian2014} and\cite{lee2015}.
The differences in the $St$ of multi-body and finite-element models of the flag are smaller than $2\%$, 
consistent with the agreement shown in the amplitude of force coefficient and tip displacements in Fig. ~\ref{fig:flag}.

%Table~\ref{tab:flag_St} gathers the Strouhal number ($St = fC/U_\infty$) and the 
%tip deflection amplitude ($y_t$) at mid-span for the present case and those from 
%\citet{tian2014,lee2015}.
%%
%Table~\ref{tab:flag_St} shows that difference in $St$ is less than $2\%$ when compared
%to results obtained from finite-element methods.

\begin{table}
   \centering
   \begin{tabular}{L{3.5cm}C{1cm}}
      & $St$ \\
   \hline
   \citet{tian2014}  & $0.263$ \\
   \citet{lee2015}   & $0.265$ \\ 
   present results   & $0.261$ \\ 
   \end{tabular}
   \caption{Comparison of the Strouhal number ($St$).\label{tab:flag_St}}
\end{table}

%\begin{table}
%   \centering
%   \begin{tabular}{L{3.5cm}C{1cm}C{1cm}}
%      & $St$ & $y_t/C$  \\
%   \hline
%   \citet{tian2014}  & $0.263$ & $0.812$ \\
%   \citet{lee2015}   & $0.265$ & $0.75$ \\ 
%   present results   & $0.261$ & $0.74$ \\ 
%   \end{tabular}
%   \caption{Comparison of the Strouhal number ($St$) and tip deflection amplitude
%   at midspan ($y_t$).\label{tab:flag_St}}
%\end{table}

%Rojo -> Lee, Choi (2015). St = 0.265
%Azul -> MB. St = 0.2615
%Verde -> Tian et al (2014), modelo estructural 1. St = 0.263
%Negro -> Tian et al (2014), modelo estructural 2. St = 0.266

%%%%%%%%%%%%%%%%%%%%%%%%%%%%%%%%%%%%%%%%%%%%%%%%%%%%%%%%%%%%%%%%%%%%%%%%%%%%%%%%
%%%%%%%%%%%%%%%%%%%%%%%%%%%%%%%%%%%%%%%%%%%%%%%%%%%%%%%%%%%%%%%%%%%%%%%%%%%%%%%%
%
%  RESULTS
%
%%%%%%%%%%%%%%%%%%%%%%%%%%%%%%%%%%%%%%%%%%%%%%%%%%%%%%%%%%%%%%%%%%%%%%%%%%%%%%%%
%%%%%%%%%%%%%%%%%%%%%%%%%%%%%%%%%%%%%%%%%%%%%%%%%%%%%%%%%%%%%%%%%%%%%%%%%%%%%%%%
\section{Results\label{sec:res}}

In the next sections, two examples of the capabilities of the proposed methodology are presented.
In \S\ref{ssec:Rselfprop} the study from \citet{arora2018} presented in 
\S\ref{ssec:Varora} for validation is extended to a 3D configuration following the 
approach in \S\ref{ssec:flag}.
In \S\ref{ssec:Rspider} the proposed algorithm is employed to model a deformable filament attached to a sphere, as an idealized model of the \emph{spider ballooning} problem.

\subsection{Self propelling finite aspect ratio plate\label{ssec:Rselfprop}}

We now extend the study of \citet{arora2018} by considering a flexible plate of finite span,
%
%In particular, the performance of a two-dimensional (2D) flexible plate is compared to that of a 
%flexible plate of finite span 
$b = 0.5C$, undergoing the same plunging motion given by eq.~\eqref{eq:Yarora}.
The flexible plate with finite aspect ratio ($\AR = 0.5$) is modelled using the lumped-torsional 
model of \citet{arora2018} (see Fig.~\ref{fig:arorascheme}) extended to three dimensions,
as depicted in Fig.~\ref{fig:arora3d}.

In order to reduce the computational cost of the 3D configuration, a lower $Re$ 
than in \S~\ref{ssec:Varora} is considered, allowing for a coarser spatial grid.
Hence, both 2D and 3D flexible plates are simulated, to compare both configurations under the same conditions. 
The Reynolds number and the plunging amplitude in eq.~\eqref{eq:Yarora} are set to $Re = 20$ and $A/C = 0.6$, respectively.
On the other hand, the plunging frequency is selected as that of maximum propulsive speed for the 2D plate, namely $\omega^* = \omega_n/(2\pi f) \approx 5$ \cite{arora2018}. 
This leads to a torsional stiffness parameter $K^* = 106.617$ for the 2D plate and a torsional stiffness parameter
$K^*_\textit{3D} = K_\textit{3D}/(\rhof f^2 C^5) = K^* \AR = 53.309$ for the torsional springs of the 3D plate.
%
%Note that, the finite aspect-ratio flapper have the same non-dimensional frequency if $K^*$ is set as the normalized torsional stiffness per unit span.

\subsubsection{Computational set-up}
Since the Reynolds number of these cases is five times smaller than that of the
validation case discussed in \S~\ref{ssec:Varora}, a grid refinement study is 
performed for the 2D configuration
to select the grid spacing, $\Delta x$, and the size of the 
computational domain.

\begin{figure}[h]
   \centering
   \begin{subfigure}{.5\tw}
      \centering
      \ig[width=\tw]{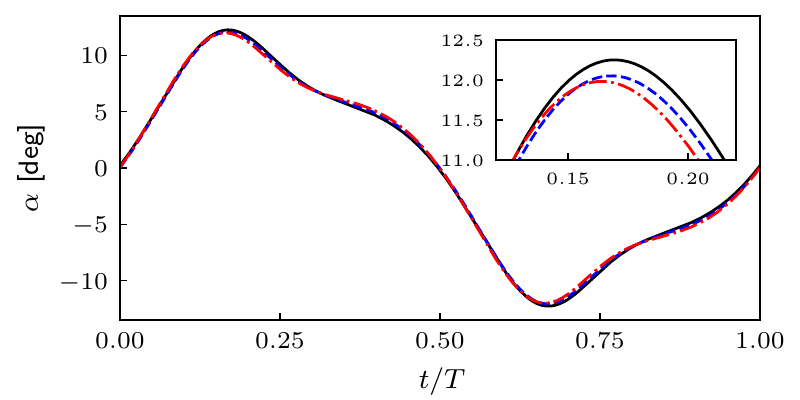}
      \caption{\label{fig:res_N}}
   \end{subfigure}\hfill
   \begin{subfigure}{.5\tw}
      \centering
      \ig[width=\tw]{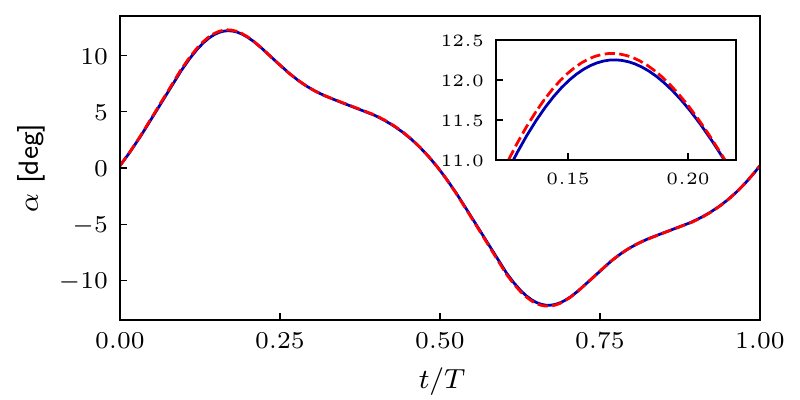}
      \caption{\label{fig:res_D}}
   \end{subfigure}
   \caption{(a) Grid sensitivity analysis on the tip deflection angle, $\alpha$: \lyy{-}{black} $\Delta x = 0.02C$, \lyy{--}{blue} $\Delta x = 0.0125C$, and \lyy{.-}{red} $\Delta x = 0.00625C$.
(b) Effect of the fluid domain on $\alpha$: \lyy{-}{blue} $16C\times8C$, and \lyy{--}{red} $40\times16C$.}
\end{figure}

Figure~\ref{fig:res_N} displays the tip deflection angle during a cycle for 3 different values of $\Delta x = \Delta y$ 
in a computational domain of size $16C\times8C$.
As it can be appreciated, the trend of the tip angle is well captured even for the lowest grid resolution, $\Delta x = 0.02 C$. 
In particular, the relative error in the maximum tip angle is of $2\%$ and $0.5\%$ for $\Delta x = 0.02C$ and $0.0125C$, respectively.
%
%{\color{blue} MGV: Puedes dar valores de porcentajes de variacion en el maximo, por ejemplo? o bien hacer un zoom y ponerlo en un inset de la figura?}
%
Likewise, the effect of the external boundaries is evaluated by considering two sizes of 
the computational domain, $16C\times8C$ and $40C\times16C$, both discretized 
with a uniform grid spacing $\Delta x = \Delta y = 0.02C$.
The evolution of the tip angle during a cycle is displayed in Fig.~\ref{fig:res_D} for both computational domains.
The variation of the maximum tip angle with the fluid domain is less than $1\%$, implying that 
%
%Therefore, it is clearly observed that 
the location of the far field boundaries is not affecting the computed solution.

In sight of the previous results, 
the computational domain is chosen to be $16C \times 8C$ for the 2D simulation 
and $16C \times 8C \times 8C$ for the 3D simulation.
The computational domain of the 2D simulations is discretized with a uniform grid of resolution $\Delta x = \Delta y = 0.0125C$.
For the 3D simulation, a uniform grid of resolution $\Delta x = \Delta y = \Delta z = 0.02C$ is used to discretize a refined region which contains the plate, whereas a constant stretching factor of $1\%$ is applied to the grid in all directions outside the refined region.
The size of the refined region is $4C\times2C\times2C$, being centred along the $y$ and $z$ directions and starting at $3C$ from the inflow boundary along the streamwise direction.

The boundary conditions of the 2D case are the same used for the validation case 
reported in  section~\ref{ssec:Varora}.
For the 3D simulation, free-slip boundary conditions are imposed at all lateral boundaries, uniform streamwise flow of intensity $U_\infty$ at the inflow boundary, and an advective boundary condition at the outflow boundary.
Note that, while $U_p$ is known from \citet{arora2018} for the 2D case and we can set $U_\infty = U_p$; it is not known \emph{a priori} for its 3D counterpart.
In order to estimate 
$U_p$, we first performed simulations fixing the horizontal position of the flapper, varying $U_\infty$ until the mean horizontal force over a cycle was approximately zero. 
%$U_\infty$ such that the plate remains at a constant mean horizontal position
%
%within the computational domain, 
%
%simulations fixing the horizontal position of the 3D plate are performed in which
%the $U_\infty$ is varied until the mean horizontal force over a cycle was approximately zero.
%
Then, the simulation is restarted 
with this value of $U_\infty$, 
allowing the horizontal displacement of the flapper.
Hence, 
%Then 
the propulsive speed can be computed as:
\begin{equation}\label{eq:Upcalc}
U_p = U_\infty - \langle \dot{X} \rangle = U_\infty - \frac{1}{T}\int_{T-1}^{T} \dot{X} \mathrm{d}t, 
\end{equation}
where $\dot{X}$ is the instantaneous velocity of the leading edge of the plate with respect to the computational domain.

In terms of the IBM, all the surfaces are discretized into evenly distributed points separated by a distance $\Delta x$.
The simulations are run until the forces on the plate are periodic 
and the value of $U_p$, computed with eq.~\eqref{eq:Upcalc}, 
does not vary with respect to the previous cycle.

\subsubsection{Discussion of the results}

One of the most noticeable differences between both cases is the change of the mean propulsive speed, $U_p$. 

\begin{figure}[h]
   \hspace{-1em}
   \noindent%
   {\begin{tikzpicture}
      \node[mylab1={a}{1.2em}{+.3em}] (fx) at (0,0)      {\ig[width=.35\tw]{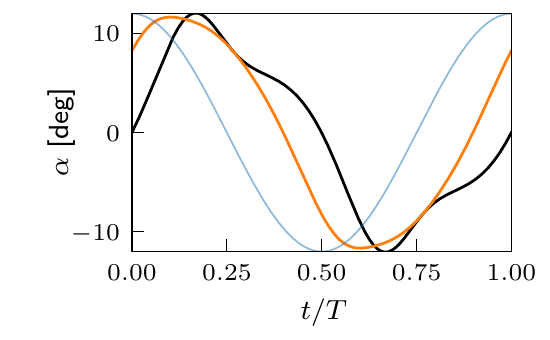}};
      \node[mylab1={b}{1.2em}{+.3em}] (fy) at (.34\tw,0) {\ig[width=.35\tw]{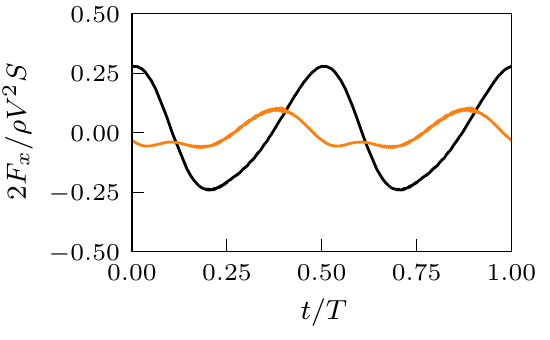}};
      \node[mylab1={c}{1.2em}{+.3em}] (fz) at (.67\tw,0) {\ig[width=.35\tw]{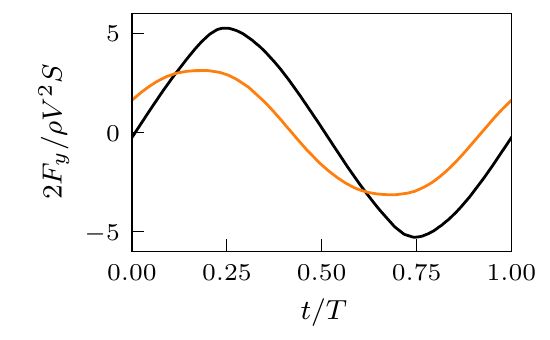}};
   \end{tikzpicture}
   \phantomsubcaption\label{fig:res_sp_al}
   \phantomsubcaption\label{fig:res_sp_cx}
   \phantomsubcaption\label{fig:res_sp_cy}
   }%
   \vspace{-1em}
   \caption{Comparison of the (a) tip deflection angle, (b) horizontal force and (c) vertical force: \lyy{-}{black}, 2D configuration  and \lyy{-}{C1} $\AR = 0.5$ plate.
    In the normalization of the forces, $S=C$ for the 2D case and $S=bC$ for the 3D cases. 
    In Fig.~\ref{fig:res_sp_al}, \ly{-}{C0!50} corresponds to the position of the leading edge (without scale).\label{fig:res_sp}}
\end{figure}

Table~\ref{tab:res_sp} shows that $U^*_p = U_p/V$ is three times lower for the $\AR = 0.5$ flapper than for the 2D configuration.
%which significantly decreases from the 2D configuration to the finite aspect ratio case.
%
This result is consistent with 
the reduction in propulsion speed when $b/C$ decreases reported by  
%the findings of 
\citet{yeh2016} for plunging flexible plates at smaller plunging 
amplitude $A/C = 0.1$ and at slightly higher $\mathit{Re}$ than the present one.
%
%\citet{yeh2016} reported that, for a given $\omega^*$, the propulsive speed decreased with decreasing $b/C$ (note that the 2D case would correspond to $\AR \to \infty$).
%
%
%Thus, it is remarkable that 
Remarkably, Fig.~\ref{fig:res_sp_al} shows that the maximum tip deflection angle is similar in the 2D and 3D flappers, 
even if the phase-shift between the vertical position of the leading edge and the tip deflection angle is different: 
%differs between both cases: 
for the 2D case this phase-shift is close to $\pi/2$ 
($\alpha \approx 0$ when $Y$ is maximum), 
but it is smaller for the 3D case.

\begin{table}
   \centering
%   \begin{tabular}{L{.9cm} C{.8cm} C{.8cm} C{.8cm}}
   \begin{tabular}{L{.9cm}C{.9cm}C{.9cm}C{.9cm}}
      & $U^*_p$ & $\langle P^*_i \rangle$ & $\varepsilon$ \\
      \hline
   2D & $1.50$              & $2.52$               & $0.60$       \\
   3D & $0.48$              & $1.44$               & $0.33$      
   \end{tabular}
   \caption{Non-dimensionalized values of the propulsive speed, $U_p^*$, average input power, $\langle P_i^* \rangle$, 
   and effectiveness, $\varepsilon$, for both cases.\label{tab:res_sp}}
\end{table}

The aerodynamic forces also change significantly from the 2D to the 3D scenario.
This is appreciated in Fig.~\ref{fig:res_sp_cx} and \ref{fig:res_sp_cy}, which depict the aerodynamic forces 
normalized with the maximum vertical velocity and the reference surface, $S$.
This reference surface is $C$ for the 2D case (since $F_x$ are forces per unit span) and $bC$ for the 3D case.
In both cases, the tip deflection angle and the vertical force are in phase, suggesting that 
both, tip deflection and lift force, are
direct consequences of the pressure difference between the upper and lower pressure acting upon the plate.
A similar rationale holds for the horizontal force (Fig.~\ref{fig:res_sp_cx}); 
in both cases the peak thrust (negative $F_x$) occurs at the maximum tip 
deflection, meanwhile the drag (positive $F_x$) is maximum for $\alpha \approx 0$.
Nonetheless, it can be observed that the drag and thrust peak have a similar amplitude for the 2D case,
whereas the thrust peak for the 3D case is less pronounced.
The smaller amplitude of $F_x$ in the 3D case is directly  linked to a more steady instantaneous horizontal velocity (not shown).

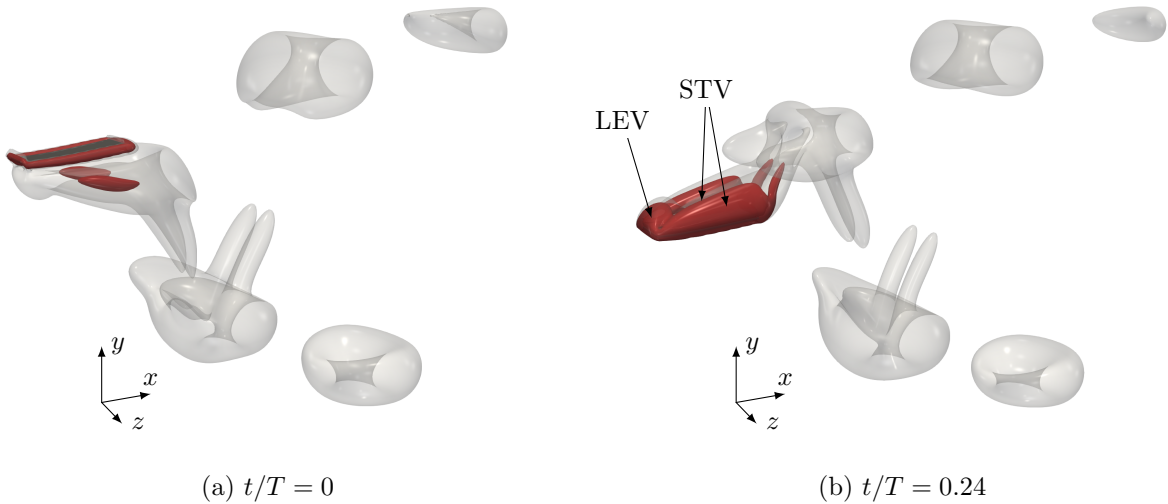
\begin{figure}
   \centering%
   {\begin{tikzpicture}[line cap=round,scale=1,>={latex[length=.2cm]}]

    \begin{footnotesize}

    \node[anchor=south west,outer sep=0pt] (pic1) at (0,0) {\ig[width=.48\tw]{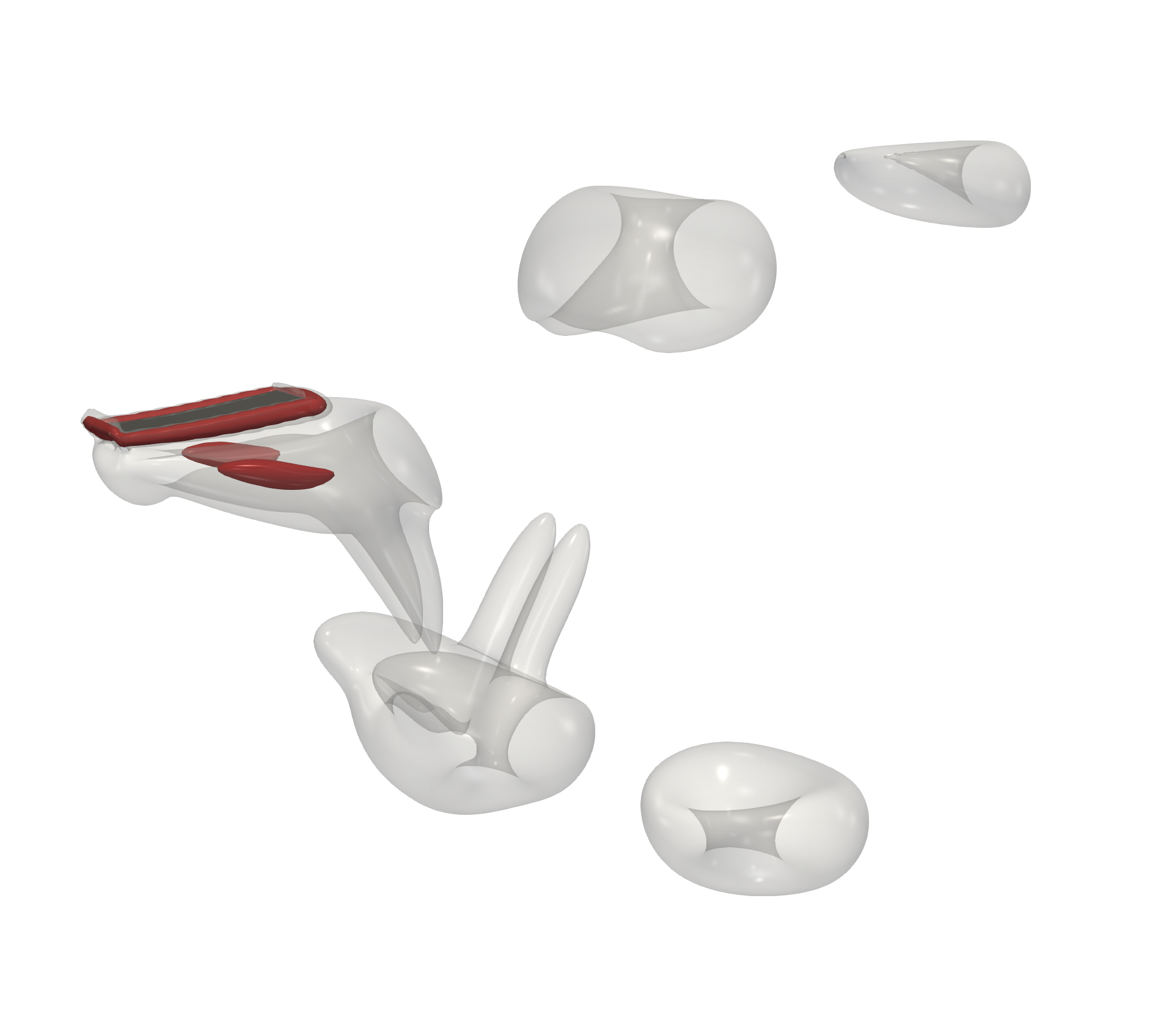}};
    \begin{scope}[x={(pic1.south east)},y={(pic1.north west)}]
       %\draw[ultra thin,gray!50,step=.01] (.2,.1) grid (.4,.3); 
       \fill[white] (.2,.1) rectangle (.31,.25);
       \coordinate (C0) at (.235,.153);
       \draw[->] (C0) --+ (11:.08) node[above]{$x$};
       \draw[->] (C0) --+ (90:.1)  node[right]{$y$};
       \draw[->] (C0) --+ (311:.05)node[right] {$z$};
    \end{scope}
      
    \node[anchor=south west,outer sep=0pt] (pic2) at (.5\tw,0) {\ig[width=.48\tw]{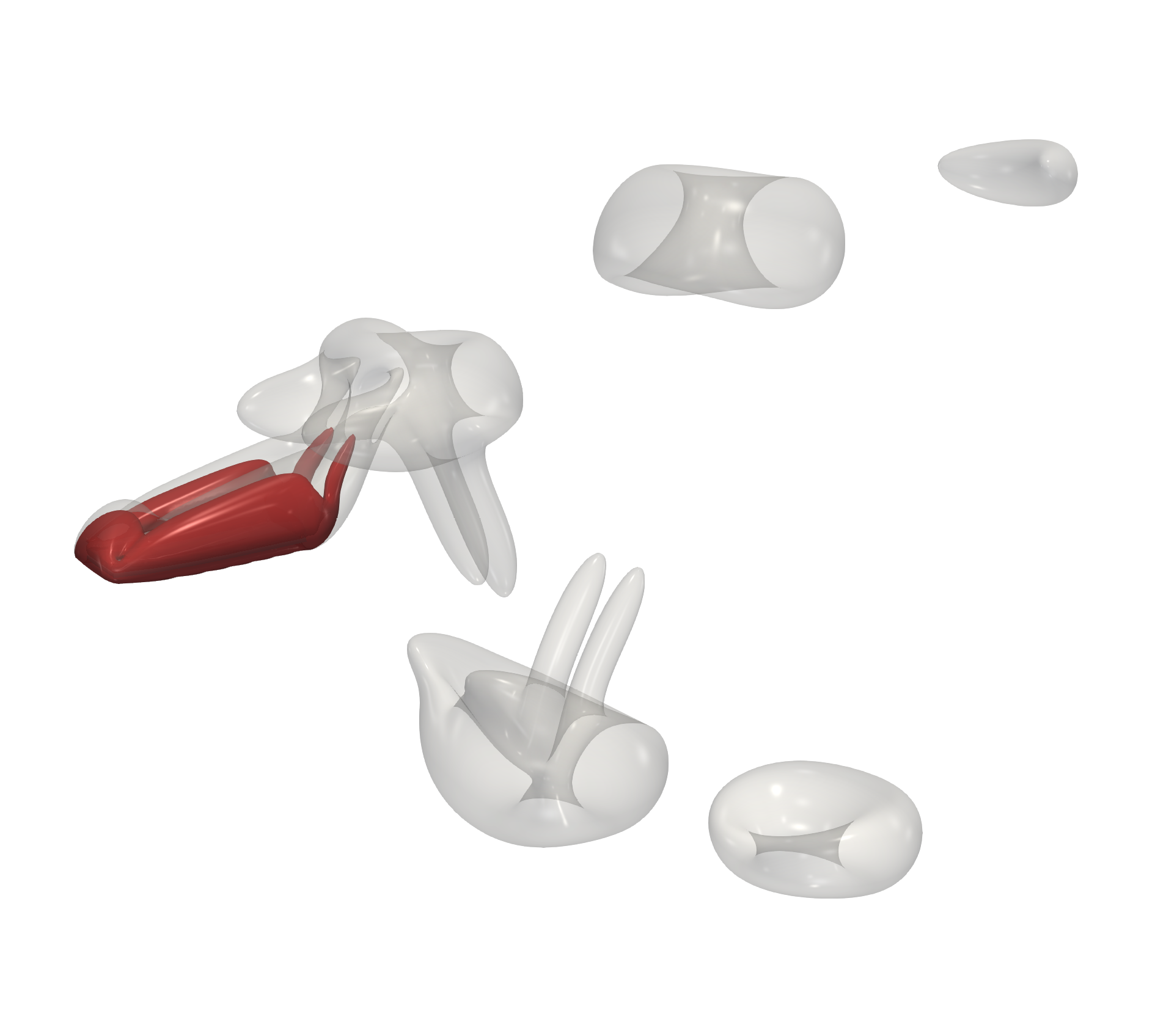}};
    \begin{scope}[shift=(pic2.south west),x={(pic2.south east)},y={(pic2.north west)}]
       %\draw[ultra thin,gray!50,step=.1] (.0,.0) grid (1,1); 
       \fill[white] (.2,.1) rectangle (.31,.25);
       \coordinate (C0) at (.235,.153);
       \draw[->] (C0) --+ (11:.08) node[above]{$x$};
       \draw[->] (C0) --+ (90:.1)  node[right]{$y$};
       \draw[->] (C0) --+ (311:.05)node[right] {$z$};

       \draw[<-] (.1,.48) --+ (105:.18) node[fill=white] {LEV};
       \draw[<->] (.18,.51) --+ (88:.2) node[fill=white] {STV} -- (.22,.5);
    \end{scope}

    \node at (pic1.south) {(a) $t/T = 0$};
    \node at (pic2.south) {(b) $t/T = 0.24$};

    \end{footnotesize}

\end{tikzpicture}
   \phantomsubcaption\label{fig:spQ_t00}
   \phantomsubcaption\label{fig:spQ_t24}
   }
   \caption{Visualization of the flow structures around the $\AR=0.5$ flapper at two time instants. Vortical structures are displayed as iso-contours of the $Q$ criterion \cite{hunt1988}: light grey structures correspond to $Q = 0.004 V^2/C^2$ and red structures to $Q = 7V^2/C^2$.\label{fig:spQ}}
\end{figure}

Figure~\ref{fig:spQ} displays the vortical structures around the 3D flapper at the beginning of the downstroke (Fig.~\ref{fig:spQ_t00}) and roughly at mid-downstroke (Fig.~\ref{fig:spQ_t24}).
The observed structures are qualitatively similar to those reported in the literature of 
similar flexible flappers but at post-resonance plunging frequencies \citep{yeh2014,yeh2016}.
In particular, a leading edge vortex (LEV) and a pair of side tip vortices (STV) are developed at each stroke of the flapper.
These vortices are shed at the end of each stroke and become a vortex ring which is convected downstream.

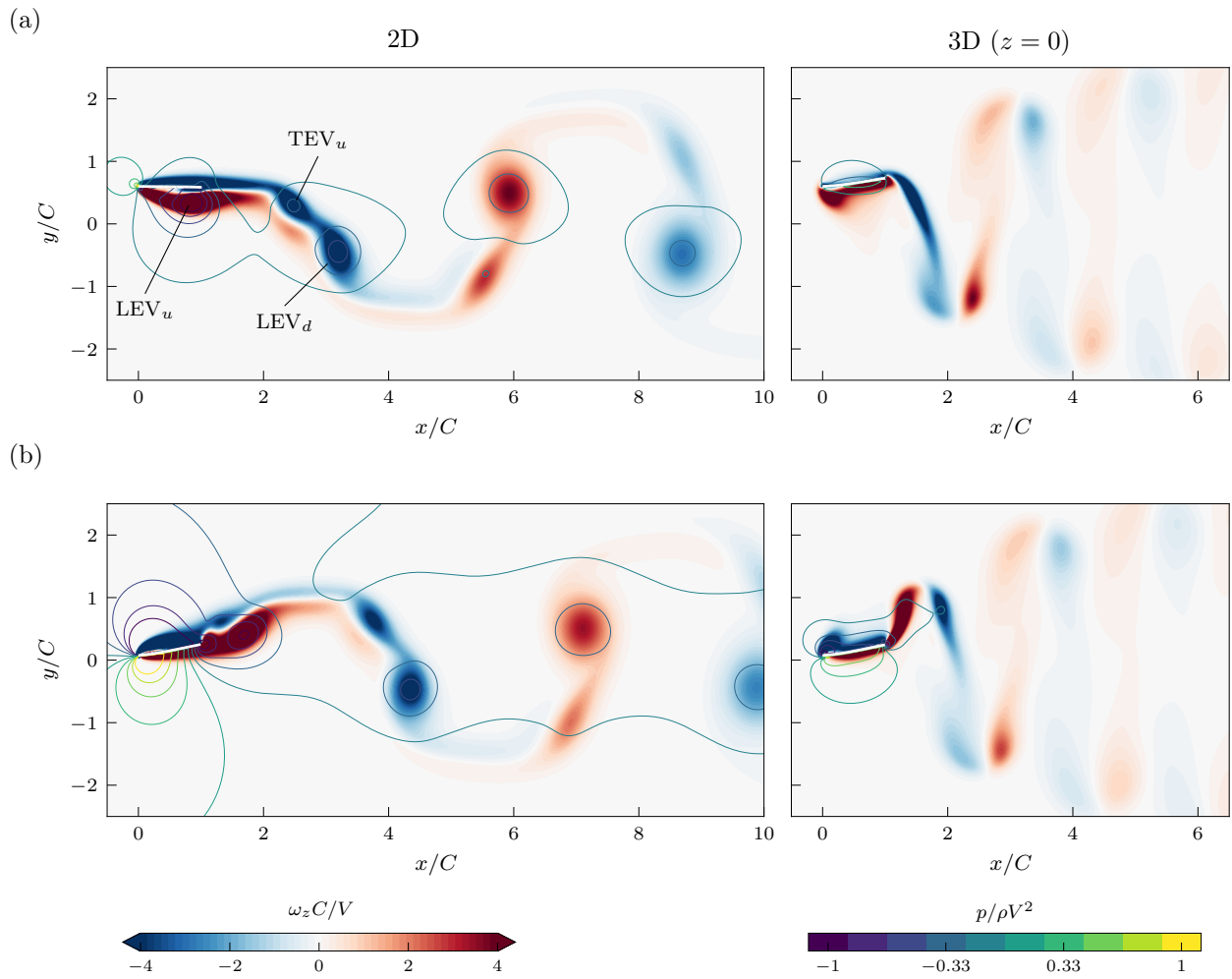
\begin{figure}
   \centering
   {\phantomsubcaption\label{fig:wzp_t00}%
   \phantomsubcaption\label{fig:wzp_t24}}%
   \begin{tikzpicture}%[framed]

   \begin{scriptsize}

   \node[anchor=south west] (pic1) at (0,0)  {\ig[width=\tw]{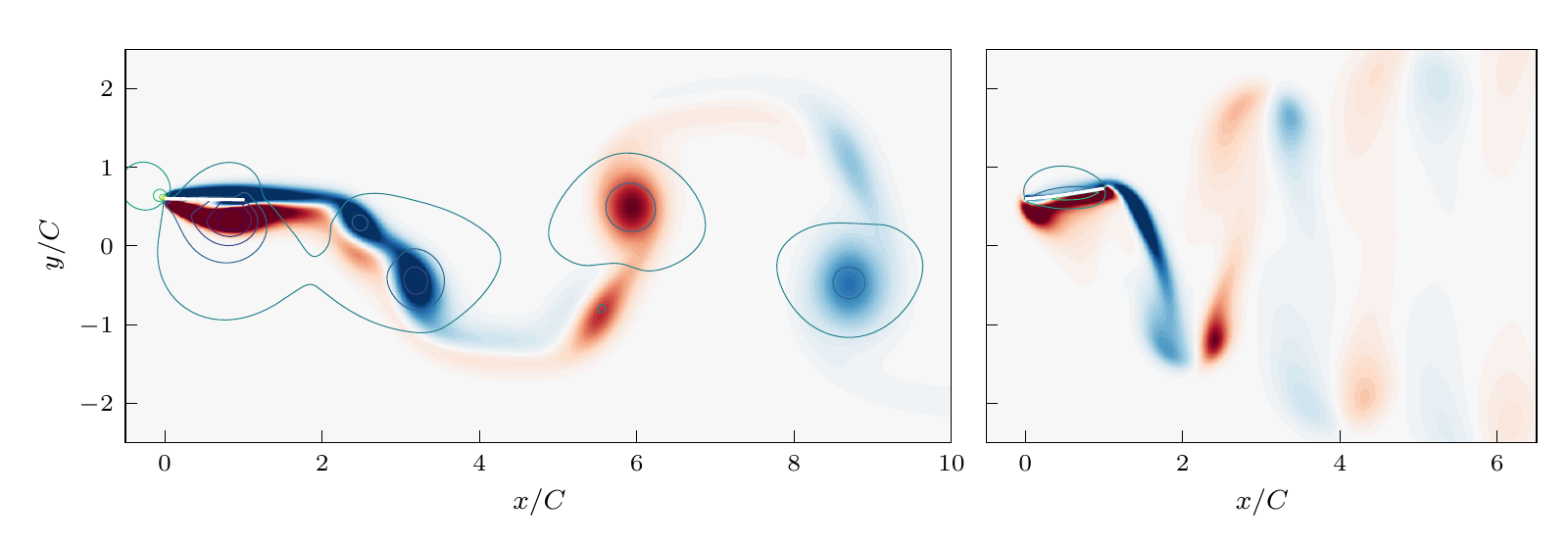}};
   \node[anchor=south west] (pic2) at (0,-.35\tw) {\ig[width=\tw]{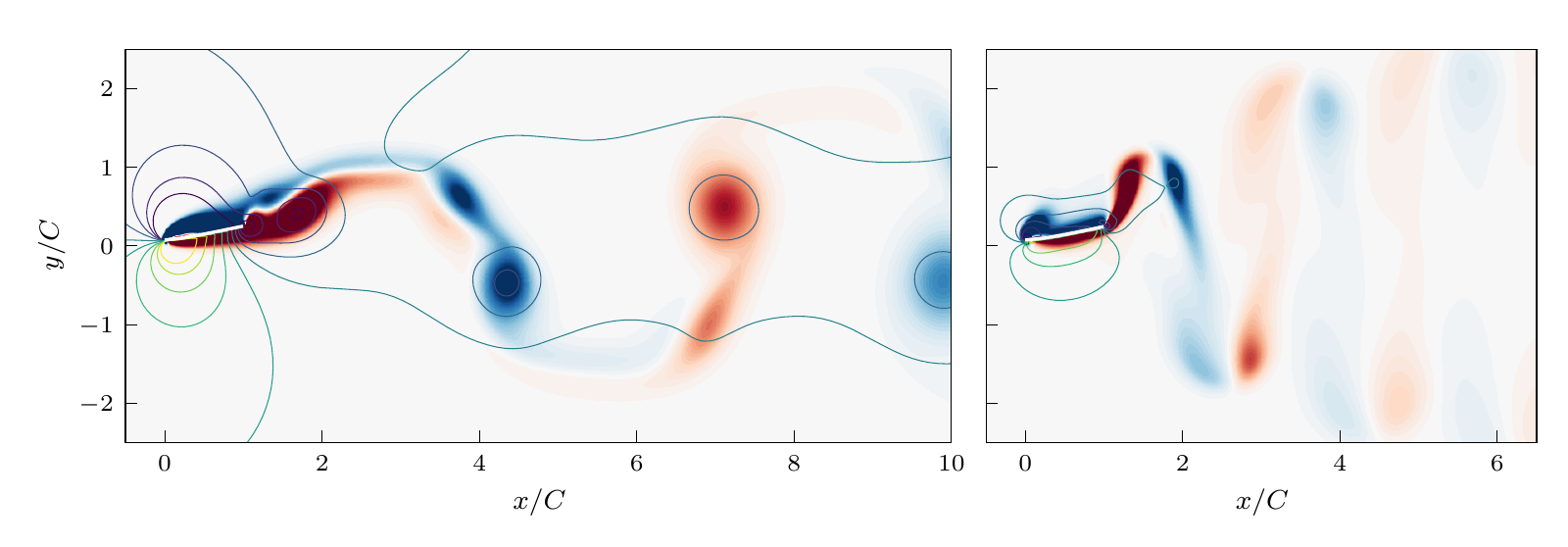}};

   \begin{scope}[x=(pic1.south east),y=(pic1.north west)]
      %\draw [help lines,step=.1] (0,0) grid (1,1);
      \draw [thin,-] (.235,.6) --+ (80:.1) node[above] {$\mathrm{TEV}_u$}; 
      \draw [thin,-] (.26,.46) --+ (-110:.1) node[below] {$\mathrm{LEV}_d$}; 
      \draw [thin,-] (.15,.59) --+ (-100:.2) node[below] {$\mathrm{LEV}_u$}; 
   \end{scope}
 
   \end{scriptsize}

   \begin{footnotesize}
   \path (pic1.north west) -- (pic1.north east) node[below,pos=.32] {2D} node[below,pos=.8] {3D ($z = 0$)}  ;
   %      (pic2.north west) -- (pic2.north east) node[below,pos=.32] {2D} node[below,pos=.74] {3D ($z = 0$)}; 
   %\node at (pic2.north) {(b) $t/T = 0.24$};   

   %\node[below] at (pic1.south) {(a)};% $t/T = 0$};
   %\node[below] at (pic2.south) {(b)};% $t/T = 0.24$};   
   \node[anchor=west] at (pic1.north west) {(a)};% $t/T = 0$};
   \node[anchor=west] at (pic2.north west) {(b)};% $t/T = 0.24$};   
   \end{footnotesize}
   %
   %\node[anchor=north west] at (.3\tw,-.4) {\ig[width=.3\tw]{figs/wzcb.pdf}};
   \node[yshift=-1em] at ($(pic2.south)+(left:.25\tw)$) {\ig[width=.35\tw]{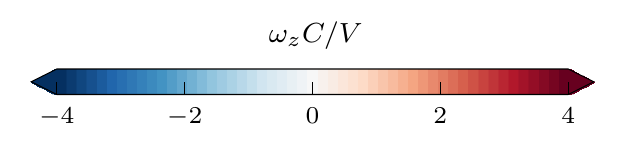}};
   \node[yshift=-1em] at ($(pic2.south)+(right:.3\tw)$) {\ig[width=.35\tw]{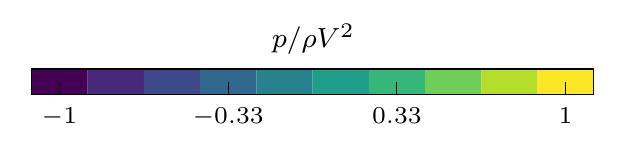}};
\end{tikzpicture}%
   \caption{Spanwise vorticity, $\omega_z$, and pressure contours for the 2D case and at mid-span of the 3D case at (a) beginning of the downstroke ($t/T = 0$) and (b) mid-downstroke ($t/T = 0.24$).\label{fig:wzp}
   }
\end{figure}

In order to compare the 
flow structure in
both configurations, Fig.~\ref{fig:wzp} depicts the spanwise vorticity $\omega_z$ and the pressure for the 2D case and for the mid-span ($z=0$) plane of the 3D case.
It can be appreciated that the flow structure is clearly different in 2D and 3D, particularly in the wake of the flappers.  
%
%In particular, the 2D wake topology is different from the 3D wake described in the previous paragraph.
%
Instead of the train of vortex dipoles observed in the 3D case (which correspond to the intersection of the vortex 
rings at $z = 0$), the 2D wake consists of pairs of vortices with the same vorticity sign: 
an LEV$_d$ formed during the downstroke and a TEV$_u$ formed during the previous upstroke. 
%orientation are shed during each stroke; 
%namely: an LEV and a trailing edge vortex (TEV).
%
The higher $U_p$ can be appreciated as a larger separation between vortices shed during each stroke
in the 2D scenario, as compared to the 3D wake.
Note also that in the 2D case, the LEV developed during upstroke (LEV$_u$) is shed at mid-downstroke, right before the shedding of the TEV that develops during the downstroke (TEV$_d$), as depicted in Fig.~\ref{fig:wzp_t24}.
%

%{\color{red} OF: El siguiente parrafo no me acaba de convencer. No se si deberiamos de explicarlo mejor en terminos de desfase entre desplazamiento y deflexion ... en cualquier caso, creo que lo podemos dejar para la revision, si es que los referees nos dicen algo sobre esto.} 

It is important to note that the LEV developed by the 2D plate is more intense, with a lower associated pressure, than to its 3D counterpart (see Fig.~\ref{fig:wzp_t00}).
Indeed, the lower pressure region associated to the 2D LEV could explain the smaller wing tip deflection 
at the beginning of a stroke: Fig.~\ref{fig:wzp_t00} reveals that at $t/T =0$, the 2D LEV$_u$ is close 
to the trailing edge, making the plate to remain nearly horizontal.
On the contrary, 
the LEV$_u$ of the 3D flapper is still very close to the leading edge at the begining of the downstroke ($t=0$), resulting into a weaker suction on the lower surface of the flapper and correspondingly to a higher tip deflection. 
%the absence of LEV$_u$ at the trailing of the 3D flapper, leads to a higher pressure 
%(i.e, weaker suction)
%at the lower surface, yielding a higher tip deflection.

Finally, it is interesting to analyse the variation in the performance of the flappers.
To that end, the \emph{effectiveness}, or \emph{swimming economy} of each plate is computed as \cite{yeh2014,yeh2016}:
\begin{equation}
\varepsilon = \frac{U_p^*}{\langle P_i^* \rangle},
\end{equation}
where $\langle P_i^* \rangle$ is the average non-dimensional input power, 
namely, $P_i^* = 2F_y \dot{Y} / (\rhof V^3 S)$, over the last cycle \cite{arora2018}.
The values of $\langle P_i^* \rangle$ and $\varepsilon$ for both cases are gathered in Table~\ref{tab:res_sp}.
Although the required input power for the finite aspect ratio plate is lower than for its 2D counterpart, 
the reduction is not large enough to compensate for the lower propulsive speed.
As a consequence, the \emph{swimming economy} of the finite plate is significantly smaller than that of the 2D plate.
Previous studies have linked the detriment of the swimming performance with decreasing $\AR$ with the STV \cite{raspa2014,yeh2016}.
The absence of STV in the 2D case ($\AR \to \infty$), together with its greater performance, 
are in agreement with this hypothesis.

%%%%%%%%%%%%%%%%%%%%%%%%%%%%%%%%%%%%%%%%%%%%%%%%%%%%%%%%%%%%%%%%%%%%%%%%%%%%%%%%%%%%%%%%
% SPIDER BALLOONING
%%%%%%%%%%%%%%%%%%%%%%%%%%%%%%%%%%%%%%%%%%%%%%%%%%%%%%%%%%%%%%%%%%%%%%%%%%%%%%%%%%%%%%%%
\subsection{Spider ballooning\label{ssec:Rspider}}

The second example of application of the methodology proposed here is inspired by the ability of 
some spiders to disperse aerially by releasing one or several silk filaments.
These filaments act as \emph{drag lines} when they encounter a wind current, allowing spiders to achieve 
long dispersal distances. 
This mechanism is usually known as \emph{spider ballooning}  \cite{humphrey1987,zhao2017,cho2018}.
Several studies addressing this phenomena can be found in the literature, either using experimental methods  \cite{suter1991,cho2018,courtney2020} or numerical methods  \cite{humphrey1987,reynolds2006,zhao2017}, with the latter restricted to 2D configurations.
These studies use actual samples or simplified models to characterize different parameters and performance metrics of spider ballooning (like the effective length of the filaments, or dispersal lengths and terminal descent velocities). 
%
%These studies are mostly focused on characterizing the dispersal properties 
%(in terms of dispersal length, terminal descent velocity, effective length of the filaments, \emph{etc.})
%of spider ballooning, both with actual samples or simplified models.
 
Motivated by this problem, we present here a fundamental study 
%is devised
%where 
of the flow around a 
deformable filament of length $L$ attached to a sphere of diameter $D$. 
%with a 
%which has a filament of length $L$ attached to it, 
%and is immersed in a uniform flow, 
%is studied.
%
In particular, the objective of the study is to determine what is the effect of the filament on the flow around the sphere, as well as the fluid forces acting upon it.
From the point of view of the filament, 
this problem can be classified as an 
\emph{extraneously induced excitation} (EIE) fluid-induced vibration problem \cite{paidoussis1998,yu2019}.

\begin{figure}
   \centering 
   \begin{subfigure}[b]{.6\tw}
      \tdplotsetmaincoords{-40}{20}

\pgfmathsetmacro{\myrad}{.45}
\pgfmathsetmacro{\myrod}{.6}

\pgfdeclareradialshading[mycolor]{sphere}{\pgfpoint{0.1cm}{0.5cm}}
{color(0cm)=(mycolor!10);
color(.3cm)=(mycolor!20);
color(1.2cm)=(mycolor!40)}

\colorlet{mycolor}{red!70!green}

\begin{tikzpicture}[rotate=0,scale=2.5,tdplot_main_coords,>={latex[length=.1cm]},
   joints/.style={very thin,C0!90!black,inner color=C0,outer color=C0!80!black},
   links/.style={draw=black!70,semithick,line cap=round},
   body/.style={shading=sphere,draw=mycolor}]%,

   \begin{scriptsize}

   \draw[tdplot_screen_coords,body] (0,0) circle (\myrad);
   \draw[tdplot_main_coords,very thin,mycolor!70] (0,0,0) circle (\myrad);

   \draw[->] (0,0,0) -+ (.3,0,0) node [below] {$x$};
   \draw[->] (0,0,0) -+ (0,.3,0) node [above] {$y$};
   \draw[->] (0,0,0) -+ (0,0,.3) node [left ] {$z$};

   \draw[black!60,dashed] (.3,0,0) -- (\myrad,0,0);
   \tdplotdrawarc[tdplot_main_coords,black]
   {(0,0,0)}{\myrad}{0}{+5}{}{};

   \tdplotsetthetaplanecoords{0}
   \tdplotdrawarc[tdplot_rotated_coords,black]
   {(0,0,0)}{\myrad}{+95}{+85}{}{};

   \coordinate[tdplot_rotated_coords] (P0) at (\myrad,0,0);

   \fill[tdplot_screen_coords,joints] (P0) circle (.02);

   \draw[black!60,very thin,dashdotted] (.3,0,0) -- (\myrad,0,0);

   \tdplotdrawarc[tdplot_main_coords,black]
   {(0,0,0)}{\myrad}{-5}{-2}{}{};

   \draw[tdplot_rotated_coords,links,shorten <= .05em] (P0)
                                 --++ (80:\myrod) coordinate (P1) --++ (75:\myrod) coordinate (P2)
                                 --++ (76:\myrod) coordinate (P3) --++ (80:\myrod) coordinate (P4)
                                 --++ (84:\myrod) coordinate (P5) --++ (92:\myrod) coordinate (P6);
                                           % --++ (76:\myrod) --++ (80:\myrod);
   
   \foreach \j in {1,...,5}{
      \fill[tdplot_screen_coords,joints] (P\j) circle (.02);}

   \draw[tdplot_screen_coords,<-,shorten <= .4em] (P4) --+ (120:.4) node[fill=white] {joint $i$};
   \draw[tdplot_screen_coords,<-,shorten <= .2em] ($(P4)!.5!(P5)$) --+ (60:.5) node[fill=white] {link $i$};

   \end{scriptsize}

\end{tikzpicture}
      \caption{\label{fig:spiderscheme}}
   \end{subfigure}~
   \begin{subfigure}[b]{.38\tw}
      \centering
      %\tdplotsetmaincoords{70}{28}
\tdplotsetmaincoords{-15}{+20}

\pgfmathsetmacro{\phio}{+35}
\pgfmathsetmacro{\theo}{+30}

\pgfmathsetmacro{\lenR}{1.2}
\pgfmathsetmacro{\radJ}{.05}

\begin{tikzpicture}[scale=2.5,tdplot_main_coords,>={Latex[length=.5em]},
   joints/.style={very thin,C0!90!black,inner color=C0,outer color=C0!80!black},
   links/.style={draw=black!70,line width=2pt,line cap=round},
   fillarc/.style={thin,#1,fill=#1!50},
   myarc/.style={thick,#1,-{[length=.4em,flex]>}}]
   
   \begin{scriptsize}
 
   \draw[tdplot_main_coords,links] (0,0,0) --+ (-\lenR*.3,0,0) coordinate[midway]  (aux1);  
   \draw[tdplot_screen_coords,joints] (0,0) circle (\radJ);

   % ORIGINAL REFERENCE FRAME 
   \draw[black,semithick,->] (\radJ,0,0)  --+ (\lenR*.8,0,0) node[below] {$x_i$};   
   \draw[black,semithick,->] (0,-\radJ,0) --+ (0,-\lenR*.8,0) node[below right] {$z_i$};   
   \draw[black,semithick,->] (0,0,\radJ)  --+ (0,0,\lenR*.6) node[left] {$y_i$};   

   % FIRST ROTATION ABOUT Y AXES
   \tdplotdrawarc[myarc={C0},tdplot_main_coords]
                  {(0,0,0)}{\lenR*.6}{0}{\phio}{right,black}{$\theta_i$};

   \tdplotsetrotatedcoords{\phio}{0}{0}
   \draw[black!50,dashed,thin,tdplot_rotated_coords] (\radJ,0,0) --+ (\lenR,0,0);   
   \draw[black!50,->,tdplot_rotated_coords] (0,-\radJ,0) --+ (0,-\lenR*.8,0) node[right] {$z_i^\prime$};   
   \tdplotdrawarc[myarc={black!50},dashed,thin,tdplot_main_coords]
                  {(0,0,0)}{\lenR*.6}{-90}{\phio-90}{}{}

   % SECOND ROTATION ABOUT Z AXES
   \tdplotsetrotatedthetaplanecoords{0}
   \tdplotdrawarc[myarc={C1},tdplot_rotated_coords]
                  {(0,0,0)}{\lenR*.6}{90}{90-\theo}{right,black}{$\varphi_i$}

   \tdplotsetrotatedcoords{\phio}{-\theo}{0}
   \draw[tdplot_rotated_coords,links] (\radJ,0,0) --+ (\lenR,0,0) coordinate [pos=.8] (aux2);
   \draw[tdplot_screen_coords,<-,shorten <= .2em] (aux2) --+ (-10:.5) node[fill=white] {link $i$};
   \draw[tdplot_screen_coords,<-,shorten <= .2em] (aux1) --+ (+110:.5) node[fill=white] {link $i-1$};

%
%   %\tdplotsetrotatedcoords{0}{\phio}{0}
%   %\tdplotfillarc[C2,fill=C2!50,tdplot_rotated_coords]
%   %               {(0,0,0)}{.2}{0}{180}{}{};%{above}{$\beta$};
%
%  
%   % DEFINE PLANE TO DRAW IN XZ
%   \tdplotsetthetaplanecoords{0}
%   %\draw[tdplot_rotated_coords,C0,fill=C0!60] (0,0,0) circle (.2);
%   \tdplotdrawarc[C0,tdplot_rotated_coords,->,>={Latex[length=.3em]}]
%                  {(0,0,0)}{.4}{0}{\phio}{}{};%{above}{$\theta$};
%   \tdplotdrawarc[C0,tdplot_rotated_coords,->,>={Latex[length=.3em]}]
%                  {(0,0,0)}{.4}{-90}{-90+\phio}{}{};
%   %\tdplotdrawarc[C3,tdplot_rotated_coords,-{>[length=.3em,flex=.70]}]
%   %               {(0,0,.4)}{.07}{-50}{230}{}{};
%
%   %\draw[tdplot_rotated_coords,ultra thin,->] (.1,0,.5) arc (0:250:.1);
%
%    
%   \tdplotsetrotatedcoords{0}{\phio}{0}
%   \tdplotsetrotatedthetaplanecoords{90}
%   %\tdplotfillarc[C2,fill=C2!50,tdplot_rotated_coords]
%   %               {(0,0,0)}{.2}{0}{-180}{}{};%{above}{$\beta$};
%
%
%   \draw[black,tdplot_rotated_coords,thick,round cap-round cap] (\theo:.05) -- (\theo:1);
%   \tdplotdrawarc[C2,tdplot_rotated_coords,->,>={Latex[length=.3em]}]
%                  {(0,0,0)}{.4}{0}{\theo}{}{};%{above}{$\beta$};
%   %\tdplotdrawarc[C5,tdplot_rotated_coords,{<[length=.3em,flex=.70]}-]
%   %               {(0,0,.4)}{.07}{-50}{230}{}{};
%
%   %\tdplotfillarc[tdplot_rotated_coords,blu,fill=blu!30,thick]
%   %{(0,0,1)}{.7}{-90}{\storeresult}{anchor=east}{$\theta_2$};
%
%   \fill[C1,tdplot_screen_coords] (0,0) circle (.03);
%   \draw[black,tdplot_main_coords,thick,round cap-round cap] (0,0,-.05) --+ (0,0,-.2);
   
   \end{scriptsize}
\end{tikzpicture}
      \caption{\label{fig:wire}}
   \end{subfigure}
   \caption{(a) Sketch of the multi-body system composed of a fixed sphere and filament consisting of $N = 6$ links. 
The reference frame depicted corresponds to the inertial reference frame.
(b) Schematic view of the parameters that define the joint between two linkages.
The Cartesian frame $x_iy_iz_i$ is fixed to the link $i-1$.
}
\end{figure}
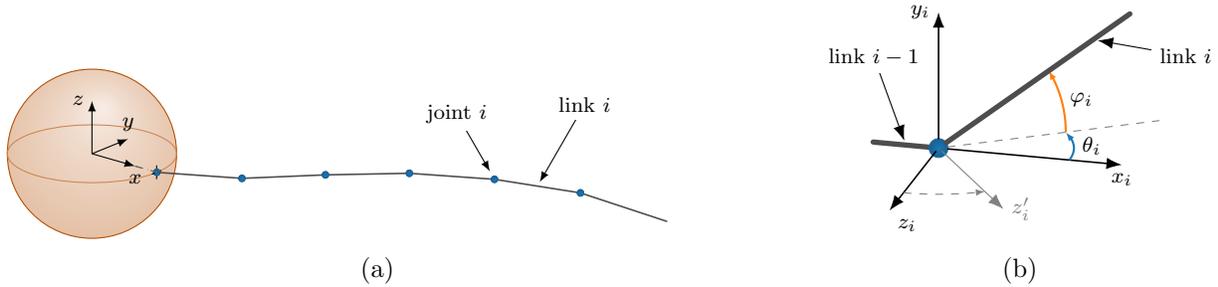

Two simulations are performed: a fixed sphere immersed in a uniform flow (case S), and the same problem but with a deformable filament attached to the surface of the sphere (case F).
%
%For the sake of clarity, the first case is labelled as S and the second case is labelled as F.
%
The filament is modelled as $N$ rigid links connected among them by joints which do not restrain the rotation (see Fig.~\ref{fig:spiderscheme}).
For the dynamical model, the links are modelled as cylindrical rods of constant density $\rho_s = 6\rho$, 
length $l = L/N$ and diameter, $d = \Delta x$ (where $\Delta x$ is the grid size).
For the fluid coupling, each link is discretized into a 
1D 
array of $l/\Delta x$ points evenly distributed. 
Consequently, the fluid does not exert any torque along the longitudinal axis of the filament.
This enables to define the position of a given link $i$ with respect to its predecessor by means of two angles, 
$\theta_i$ and $\varphi_i$ (see Fig.~\ref{fig:wire}), instead of 3, as should be required to define the orientation of a rigid body. 
Accordingly, the joint connecting a given link, $i$, with its predecessor, $i-1$, is a multi-DOF joint consisting of 
a revolute joint about the $y_i-$axis, followed by another revolute joint about the rotated $z_i-$axis, 
namely, $z_i^\prime$, as sketched in Fig.~\ref{fig:wire}.

The sphere is immersed in a uniform flow parallel to the $x$-axis (see Fig.~\ref{fig:spiderscheme}) of magnitude $U$.
The point at which the filament is attached is $x = -D/2$, that is, at the downstream end of the sphere.
For the present study, $L = 5D$, $\mathit{Re} = DU/\nu = 300$, and $N = 24$.
This Reynolds number corresponds to a flow regime for the case of the isolated sphere in which vortex shedding 
starts to occur \cite{bouchet2006}, 
and it has been selected to explore the interference between the filament and the vortex shedding process.

\subsubsection{Computational set-up}
For both simulations, the computational domain is a rectangular prism 
of dimensions $14D\times8D\times8D$ in the streamwise and lateral directions, respectively.
A refined region of size $7D\times2D\times2D$
is defined, with a uniform resolution $\Delta x = \Delta y = \Delta z = D/48$.
This region is located $3D$ downstream of the inflow, 
centered within the lateral directions of the computational domain.
Outside of this region, a stretching factor of $1\%$ is applied in all directions.
A uniform free stream of magnitude $U$ is imposed at the inflow boundary, free-slip boundary conditions are 
imposed at the lateral boundaries and an advective boundary conditions is implemented at the outflow boundary.

For the IBM, the sphere is discretized into $N_s$ evenly distributed points, 
such that $N_s \approx \pi D^2 / \Delta x^2 $, similarly to \citet{uhlmann2005}.
%
%On the other hand, the filament is discretized as an array of points equally spaced, as explained above.
On the other hand, each segment of the filament is discretized by equally spaced points, separated a distance $\Delta x$, without gaps between adjacent segments. 

The time step is fixed to $\Delta t = 0.0025 U/D$, ensuring $\textit{CFL} < 0.2$.
Finally, the simulation of the isolated sphere (case S) is started from scratch, whereas the sphere with the attached filament (case F) is started from a flow field of case S when vortex shedding was present.

\subsubsection{Discussion of the results}

\begin{figure}
   \centering
   \ig[trim={12cm 20cm 4cm 14cm},clip,width=.6\tw]{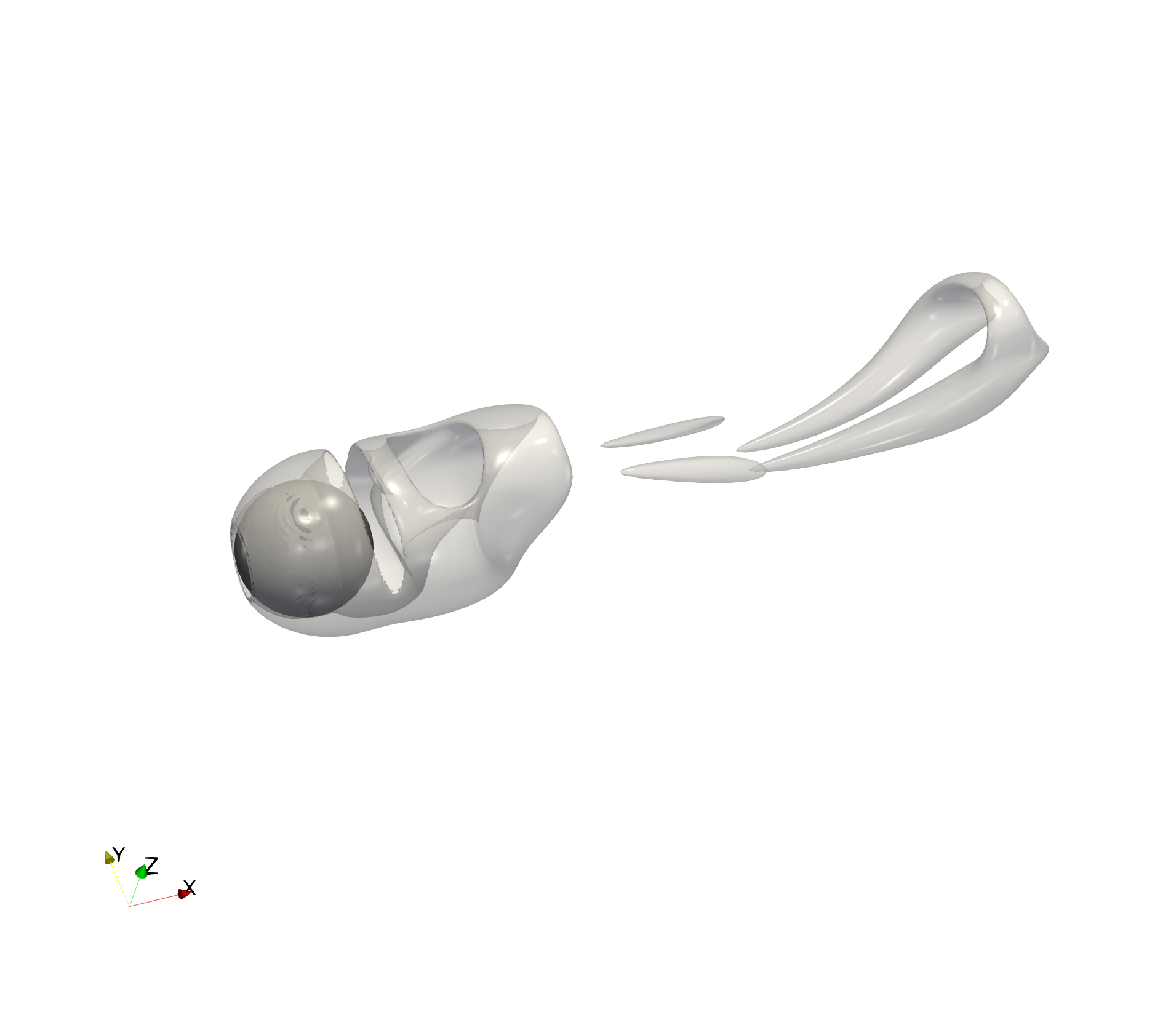}
   \caption{Instantaneous visualization of the flow structures around the isolated sphere at $\mathit{Re}=300$.
Vortical structures correspond to iso-contours of the $Q$-criterion, $Q = 0.2 U^2/D^2$. \label{fig:spi-sphQ}
%
%\mycor{The sketched reference frame is an inertial Cartesian frame whose $x$-axis is parallel to the free-stream, 
%and $e_\parallel$-axis is contained in the plane of symmetry of the wake.}
}
\end{figure}

\begin{figure}
    \hspace{-2em}
   {\begin{tikzpicture}
      \node[anchor=south west,mylab1={a}{.03\tw}{+.2em}] at (0,0)      {\ig[height=3.8cm]{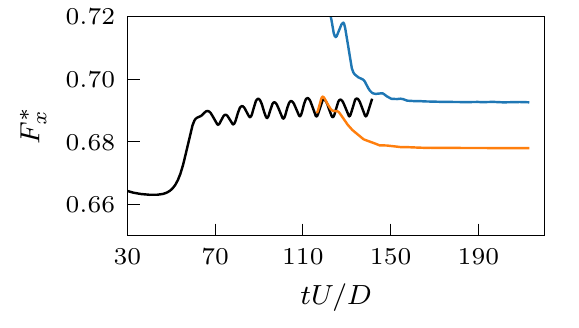}};
      \node[anchor=south west,mylab1={b}{.03\tw}{+.2em}] at (.39\tw,0) {\ig[height=3.8cm]{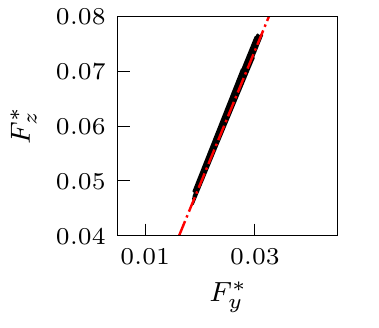}};
      \node[anchor=south west,mylab1={c}{.04\tw}{+.2em}] at (.62\tw,0) {\ig[height=3.8cm]{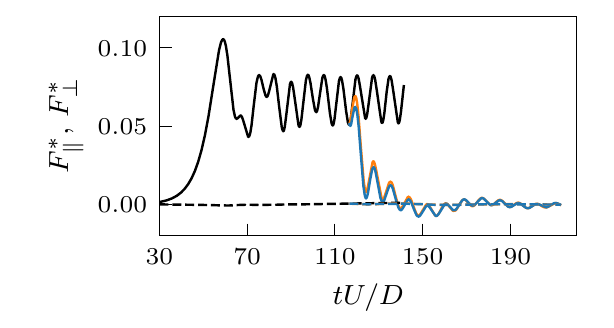}};
   \end{tikzpicture}
   \phantomsubcaption\label{fig:spi_fx}
   \phantomsubcaption\label{fig:spi_fyz}
   \phantomsubcaption\label{fig:spi_fs}
   }%
   \vspace{-1em}  
   \caption{(a) Temporal evolution of the non-dimensional streamwise force. 
\lyy{-}{black} case S; \lyy{-}{C1} case F (force over the sphere); 
\lyy{-}{C0} case F (force over the sphere and the filament).
(b) \lyy{-}{black} Non-dimensional transverse force on the sphere for case S for $t > 90D/U$ (after onset of vortex shedding); and its \lyy{.-}{red} least square regression.
(c) Temporal evolution of the non-dimensional transverse forces expressed into its parallel and 
normal components. Color legend is as \ref{fig:spi_fx}, line styles stands for \lyy{-}{black!50} $F_{\parallel}^*$, and 
\lyy{--}{black!50} $F_{\bot}^*$.}
\end{figure}

At the selected Reynolds number, the flow over the sphere exhibits periodic shedding of vortices after an initial transient.
Fig.~\ref{fig:spi-sphQ} displays an snapshot of the wake of the sphere after the onset of vortex shedding.
This leads to oscillatory hydrodynamic forces over the sphere, as observed from Fig.~\ref{fig:spi_fx}, 
which depicts the non-dimensional streamwise (i.e., drag) force, $F_x^* = F_x/(\rho U^3 \pi D^2/ 8)$.
It can also be observed that the mean value of $F_x^*$ increases after the onset of vortex shedding. 
The oscillation frequency is found to be $0.135 U/D$, in agreement with existing literature at the same Reynolds number \citep{tomboulides1993,johnson1999}.
Furthermore, the non-axisymmetric wake leads to the appearance of a transversal force 
contained in the plane of symmetry of the wake.
Note that the location of this plane of symmetry with respect to the inertial axis arises naturally, only forced by numerical biases \cite{johnson1999}.
In the present case, the angle between the $(x,y)-$plane and the symmetry plane is approximately $67.7^\circ$. 
This angle, 
is computed by a least square regression of the transversal forces,
as shown in in Fig.~\ref{fig:spi_fyz}.
Figure~\ref{fig:spi_fs} depicts the transverse force when expressed in parallel ($F_\parallel$) 
and perpendicular ($F_\bot$) components with respect to the symmetry plane of the wake.
The plot shows oscillations of $F^*_\parallel$, with the same frequency of oscillation as $F^*_x$, 
and a net non-zero $F^*_\parallel$ when averaged over several periods.

\begin{figure}%
\begin{subfigure}{.5\tw}
   \centering
   \ig[trim={16cm 20cm 16cm 14cm},clip,width=\tw]{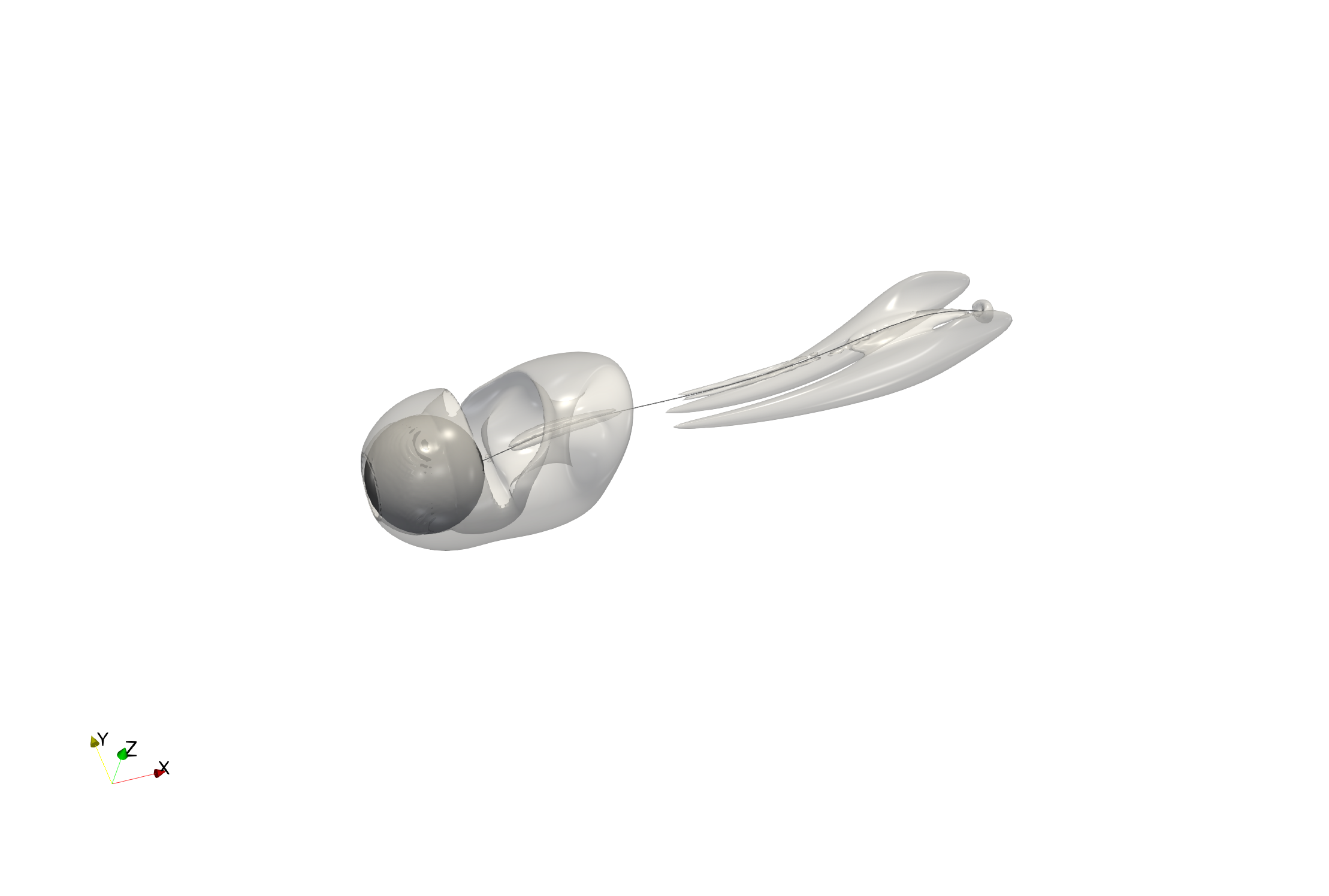}
   \caption{\label{fig:spi-mbQi}}
\end{subfigure}\hfill
\begin{subfigure}{.5\tw}  
   \centering
   \ig[trim={16cm 20cm 16cm 14cm},clip,width=\tw]{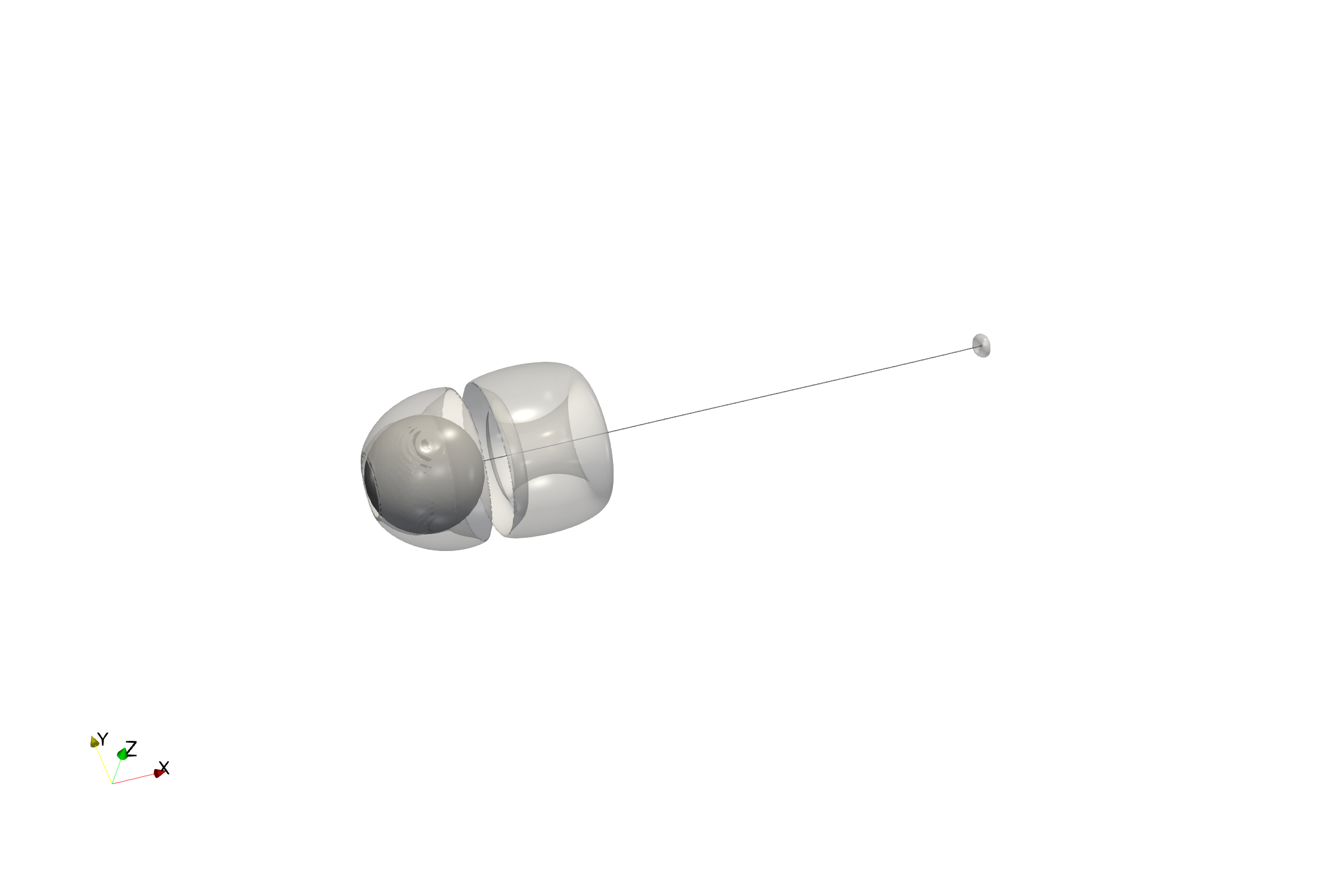}
   %\ig[trim={10cm 20cm 10cm 14cm},clip,width=\tw]{figs/mbsQ02-0100.png}
   \caption{\label{fig:spi-mbQf}}
\end{subfigure}
%%
% \begin{tikzpicture}
%   \centering
%   \node at (0,0) {\ig[trim={12cm 20cm 4cm 14cm},clip,width=.5\tw]{figs/mbsQ02-0005.png}};
%   \node at (.5\tw,0) {\ig[trim={12cm 20cm 4cm 14cm},clip,width=.5\tw]{figs/mbsQ02-0100.png}};
%%
%\end{tikzpicture}
   \caption{Instantaneous visualization of the flow structures around the sphere with the deformable filament at (a) initial time
instants, and (b) after reaching and stable state.
Vortical structures correspond to iso-contours of the $Q$-criterion, $Q = 0.2 U^2/D^2$, as in Fig.~\ref{fig:spi-sphQ}. 
\label{fig:spi-mbQ}}
\end{figure}

\begin{figure}
   \centering
   \ig[width=.8\tw]{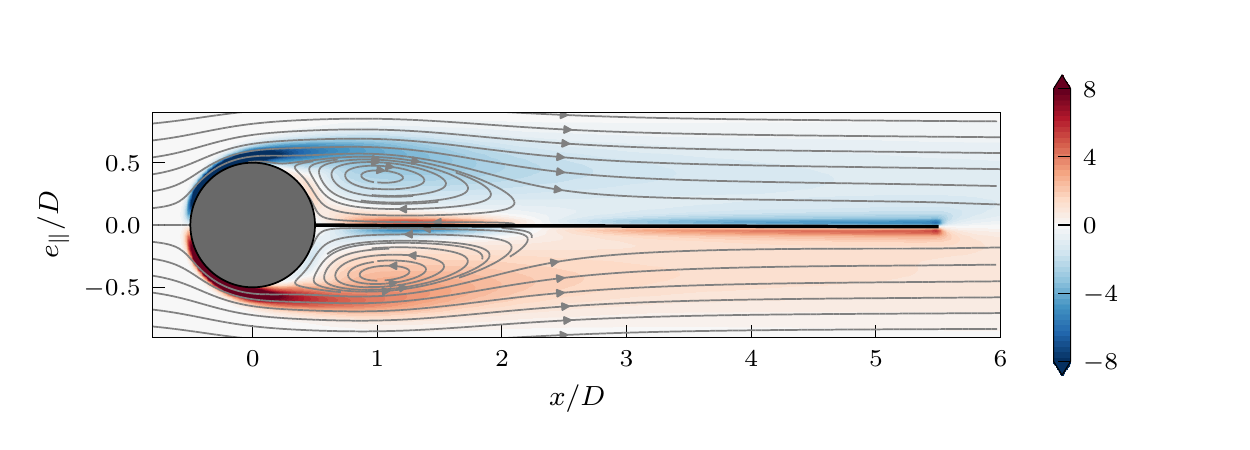}
   \caption{Instantaneous streamlines past the sphere and the filament in the $(x,e_\parallel)$-plane, together with
the vorticity perpendicular to the plane.\label{fig:streamlines}}
\end{figure}

When the deformable filament is attached to the posterior part of the sphere, the flow topology is greatly modified.
%
%it is observed that no vortex shedding occurs, 
%and existing shedding is inhibited, as shown in Fig.~\ref{fig:spi-mbQ}.
%
If the simulation is started from a flow field with vortical structures (Fig.~\ref{fig:spi-mbQi}) 
the filament starts oscillating, and after 2-3 shedding cycles, 
vortex shedding is suppressed, and a stable flow around the 
sphere-filament is developed (Fig.~\ref{fig:spi-mbQf}).
If the simulations are started from scratch, no vortex shedding occurs.
The topology of the flow in this new regime is characterized by the development of an axisymmetric recirculation 
region attached to the sphere, similarly to the case of the isolated sphere 
at $\mathit{Re} < 210$ \cite{johnson1999}.
This can be clearly appreciated in Fig.~\ref{fig:streamlines}, which depicts the instantaneous streamlines past the
sphere and the filament.
Note that a shear layer is developed along the filament, which changes direction in the recirculation bubble.

The vortex shedding inhibition has a noticeable effect on the forces acting over the sphere.
Firstly, the mean drag force acting over the sphere is reduced and the oscillations are hindered, 
as shown in Fig.~\ref{fig:spi_fx}, leading to a steady value of the drag force over the sphere.
However, the total drag force (i.e., sphere + filament) increases, due to 
the skin friction of the filament, which acts as a \emph{drag line}.
Secondly, 
the amplitude of the oscillations of the transverse force 
decreases with respect to the case of the isolated sphere, 
as shown in Fig.~\ref{fig:spi_fs}.
The frequency of oscillation of the transverse forces 
(which was previously linked to the shedding frequency of vortices over the sphere)
is also reduced, from $0.135U/D$ without the filament to $0.120U/D$ with the filament. 

%withwhich decreases to $0.120U/D$.
%
In addition, the mean value of the transverse force over an oscillation cycle 
has a zero mean value (Fig.~\ref{fig:spi_fs}), 
which suggest that the flow becomes symmetric, in a time-average sense, across the $(x,e_\parallel)-$plane 
(where $e_\parallel$ is the direction perpendicular to $x-$axis and contained in the wake's symmetry plane).
%
%Finally, it is noteworthy to recall that no perpendicular forces are developed, as for the isolated sphere.
%

\begin{figure}
   \centering
   \ig[width=.8\tw]{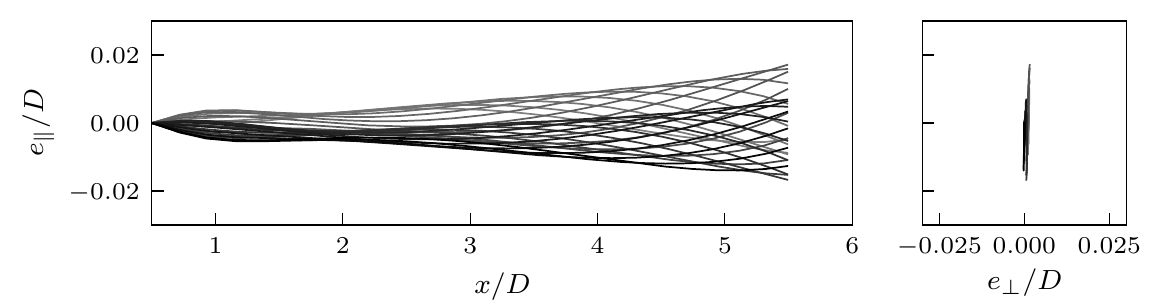}
   \caption{Oscillation pattern of the filament during the last two oscillation cycles of the transverse forces.
Line color indicates the time instant (from grey to black).
Note the difference in the scale of the axes.\label{fig:filamentdef}}
\end{figure}

Regarding the dynamics of the filament, Fig.~\ref{fig:filamentdef} depicts its deformation pattern during the 
last two oscillation cycles of the transverse force.
It can be observed that the filament is not steady but it oscillates with a low amplitude.
It is interesting to note that these oscillations are contained in a plane, which correspond to the symmetry plane
of the wake of the isolated sphere.

\section{Conclusions\label{sec:conc}}

A methodology to solve fluid-structure interaction problems with multi-body systems has been presented.
The proposed methodology follows a partitioned approach.
The flow is solved using a conventional fractional-step method, while the presence of
the bodies of the MBS in the fluid is imposed by means of the immersed boundary method proposed by \citet{uhlmann2005}.
On the other hand, the dynamic equations of the rigid bodies are computed in terms of the reduced coordinates 
of the MBS by the CRBA and RNEA recursive algorithms proposed by \citet{felis2017}.
The coupling between flow equations and the MBS equations is \emph{weak}, extending the approach of \citet{uhlmann2005}
for single rigid bodies to MBSs.

Since the flow solver has been already validated elsewhere, the validation focuses on the multi-body dynamics and the coupling.
Three cases from the literature have been selected to that end.
The first validation case corresponds to a system of two bodies joined by a torsional spring.
Very good agreement is obtained when comparing to a vortex particle method \cite{toomey2008} and to a vorticity-based IBM with strong coupling \cite{wang2015}.
The second case corresponds to a flexible, self-propelled plate modelled as several rigid bodies connected 
by torsional springs \cite{arora2018}.
Again, the agreement between the results of the present methodology and the Lattice-Boltzmann simulation of the reference is very good. 
%The obtained results show again very good agreement with those obtained from a Lattice-Boltzmann simulation \cite{arora2018}.
%
In the third case, the proposed methodology is used to simulate the dynamics 
of a three-dimensional flexible flag and compare against results using a finite-element formulation of the structure.
A remarkable good agreement is found, in terms of the kinematics and dynamics,
between the present methodology and the finite-element structural solvers.

Two additional bio-inspired examples are analized to illustrate the capabilities of the present methodology.
The first example is a three-dimensional extension of the case presented by \citet{arora2018}.
It is observed that 3D effects are detrimental in terms of propulsive speed and efficiency, although 
the deflection of the plate is not significantly modified.
The results are in accordance with those reported by \citet{yeh2014} for flexible self-propelled plates.
The second example is loosely inspired by the \emph{ballooning} mechanism of several spiders to disperse aerially.
The problem is modelled as a deformable filament attached to a fixed sphere and immersed in a free-stream.
The flexibility of the filament is modelled as a chain of rigid links connected with multi-DoF joints.
When compared to an isolated sphere at the same Reynolds number, it is shown that the
vortex shedding is suppressed by the filament,  
which oscillates with very low amplitudes in the wake of the sphere. 
%of oscillation of the filament. 
The reduction in the unsteadiness of the flow results in a decrease of the drag contribution from the sphere. 
However, a larger total drag is obtained, due to the extra friction introduced by the filament.

In summary, it has been shown that the proposed methodology allows the definition and analysis of a multitude of diverse configurations of MBS, thanks to the use of generalised recursive algorithms.
Moreover, the coupling between the flow equations and the MBS equations is very simple, yet robust enough to provide very good agreement with existing results from the literature.
Nonetheless, the \emph{weak coupling} imposes a lower limit on the density ratios of the bodies which can be 
simulated with the present methodology.
Although, recent works have proven to successfully tackle arbitrary density ratios for single rigid bodies using 
a non-iterative version of the \emph{weak coupling} approach presented here \cite{tschisgale2017}, 
the technical details are not trivial for arbitrary geometries so that this 
extension of the methodology is left for future work.

\section*{Acknowledgements\label{sec:ack}}

This work was supported by grant DPI2016-76151-C2-2-R (AEI/FEDER, UE). 
The computations were partially performed at the supercomputer Caesaraugusta from the {\it Red Espa\~nola
de Supercomputaci\'on} in activity IM-2020-1-0008.
We thank Dr. N. Arora, Dr. A. Gupta and  Dr. J. Eldredge for providing their data in electronic form and for fruitful discussions.

\appendix

\section{Joint modelling\label{sec:app_joint}}

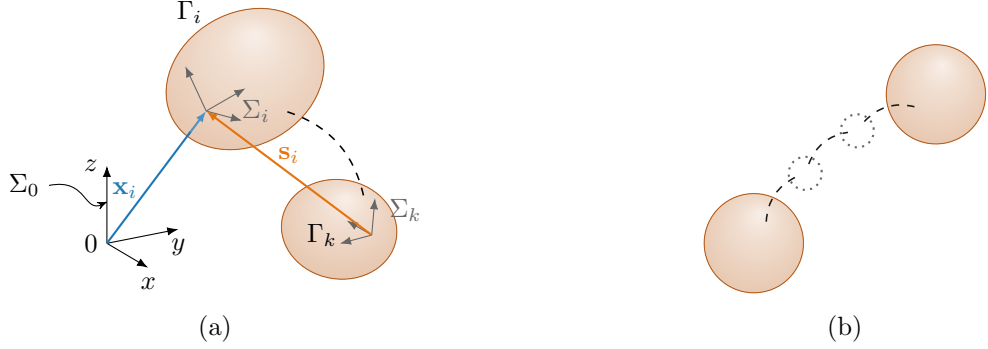
\begin{figure}
      \begin{subfigure}[b]{.48\tw}
         \centering
         \pgfdeclareradialshading[mycolor]{sphere}{\pgfpoint{0.1cm}{0.5cm}}
{color(0cm)=(mycolor!10);
color(.3cm)=(mycolor!20);
color(1.2cm)=(mycolor!40)}

\colorlet{mycolor}{red!70!green!90!white}

\tdplotsetmaincoords{-60}{+20}

\begin{tikzpicture}[scale=1.1,tdplot_screen_coords,
   >={Latex[length=.15cm]},
   ineax/.style={tdplot_main_coords,black},
   bdyax/.style={tdplot_screen_coords,black!60},
   body/.style={tdplot_screen_coords,shading=sphere,draw=mycolor}]

   \begin{footnotesize}

   \coordinate (x0) at (0.,0.,0.);        % origin
   \coordinate (xi) at (1.2,1.6);         % body control point
   \coordinate (xp) at ($(xi)+(60:1.2)$); % surface point

   \coordinate (xl) at (3.2,.1);          % 2nd body control point

   % INERTIAL REFERENCE FRAME
   \draw[ineax,->] (x0) --+ (1,0,0) node[below] {$x$};
   \draw[ineax,->] (x0) --+ (0,1,0) node[below] {$y$};
   \draw[ineax,->] (x0) node[left] {$0$} --+ (0,0,1) node[left] {$z$};

   \draw[thin,>={stealth'[length=3pt]},<-,shift=(x0)] (0,.5) .. 
         controls (-.3,.3) and (-.2,+.8).. (-.7,.7) node[left] {$\Sigma_0$};

   % BODY 
   \filldraw[name path=Bi,body,rotate=30] ($(xi)+(.6,.1)$) ellipse (1 and .8);
 
   % MOVING REFERENCE FRAME
   \draw[bdyax,->] (xi) --+ (30:.55); 
   \draw[bdyax,->] (xi) --+ (115:.6); 
   \draw[bdyax,->] (xi) --+ (-15:.45);
 
   \node[bdyax] at ($(xi)+(.6,0)$) {$\Sigma_i$};
   \node[black] at ($(xi)+(-.2,1.2)$) {$\Gamma_i$};

   % 2 BODY 
   \filldraw[name path=Bl,body,rotate=-10] ($(xl)-(.4,.0)$) ellipse (.7  and .6);
 
   % 2 MOVING REFERENCE FRAME
   \draw[bdyax,->] (xl) --+ (195:.4); 
   \draw[bdyax,->] (xl) --+ (150:.35); 
   \draw[bdyax,->] (xl) --+ (+85:.45);
 
   \node[bdyax,anchor=west] at ($(xl)+(.1,.3)$) {$\Sigma_{\Bpre}$};
   \node[black] at ($(xl)+(-.6,0)$) {$\Gamma_\Bpre$};

   % VECTORS
   \path[name path=p0i] (x0) -- (xi);
   \path[name intersections={of=p0i and Bi,by=e0i}]; 
   \draw[thick,C0!80,->] (e0i) -- (xi);
   \draw[thick,C0,shorten >=-.2em] (x0) -- (e0i) node[midway,left] {$\vvec{x}_i$};

   \draw [thick,C1!90!black,->] (xl) -- (xi) node[midway,above] {$\vvec{s}_{i}$};

   \draw[semithick,dashed] (xl)+(-.1,.48) to [bend right] ($(xi)+(.95,0)$);

   %\DrawControl{(-.3,.3)}{blue}\DrawControl{(.2,.8)}{blue};

   \end{footnotesize}
\end{tikzpicture}
         \caption{\label{fig:app2body}}
      \end{subfigure}~
      \begin{subfigure}[b]{.48\tw}
         \centering
         \pgfdeclareradialshading[mycolor]{sphere}{\pgfpoint{0.1cm}{0.5cm}}
{color(0cm)=(mycolor!10);
color(.3cm)=(mycolor!20);
color(1.2cm)=(mycolor!40)}

\colorlet{mycolor}{red!70!green!90!white}

\begin{tikzpicture}[scale=1.1,
   >={Latex[length=.15cm]},
   body/.style={shading=sphere,draw=mycolor},
   joints/.style={draw=black!90,semithick,dashed},
   vbody/.style={thick,draw=black!60,dotted}]

   \begin{footnotesize}

   \coordinate (x1) at (0,0);             % 1st body control point
   \coordinate (x4) at (2.2,1.8);          % 2nd body control point

   \path (x1) to [bend left=20]  node[pos=.35] (x2) {}
                                 node[pos=.65] (x3) {} (x4);

   \filldraw[name path=B1,body,rotate=+10] (x1) ellipse (.6);
   \filldraw[name path=B4,body,rotate=+10] (x4) ellipse (.6);

   \draw[name path=B2,vbody,rotate=+5]  (x2) ellipse (.2);
   \draw[name path=B3,vbody,rotate=+5]  (x3) ellipse (.2);

   \draw[joints] (x1)+(60:.3) to [bend left] (x2);
   \draw[joints] (x2)         to [bend left] (x3);
   \draw[joints] (x3) to [bend left] ($(x4) + (-150:.3)$);
   
   \end{footnotesize}
\end{tikzpicture}
         \caption{\label{fig:virtbody}}
      \end{subfigure}
      \caption{(a) Sketch of body $\Bdy_i$ and its predecessor, $\Bdy_k$, illustrating the elements that define
the joint between them.
(b) Representation of a joint with $3$ DoF, simulated by means of 2 virtual bodies (represented as dashed circles)
which have no mass and associated Lagrangian points.}
\end{figure}

A joint that connect two bodies can also be regarded 
as the constraint of the relative motion between two Cartesian reference frames, 
attached to each body \cite{featherstone2014}.
Figure \ref{fig:app2body} illustrates this concept: 
%, namely: 
body $\Bdy_i$ has an attached reference frame, $\Sigma_i$, and is linked to its predecessor body, $\Bdy_{\Bpre}$, which has its own attached reference frame, $\Sigma_{\Bpre}$.
Therefore, a joint can be defined by the rotation matrix, $\mathsf{E}_{\Bpre,i}$, 
from $\Sigma_i$ to $\Sigma_\Bpre$; 
and the vector $\vvec{s}_i$, which links the origin of both Cartesian frames and is implicitly 
expressed in $\Sigma_{\Bpre}$.
Note that $\mathsf{E}_{\Bpre,i}$ and $\vvec{s}_i$ only depend on the degrees of freedom allowed by the joint,
but their definition depend on the kind of joint.

Single DoF joints of two types are considered:
prismatic (i.e., translation) joints along any axis of $\Sigma_k$;
and revolute (i.e. rotation) joints about any axis of $\Sigma_k$.
For a prismatic joint which allows translation along the $x$-axis of $\Sigma_k$, $\mathsf{E}_{\Bpre,i}$ is the identity matrix 
of size $3$; and $\vvec{s}_i(q_i) = q_i \vvec{e}_x$, where 
$\vvec{e}_x$ is the unitary vector parallel to $x$-axis, and $q_i$ is the joint's degree of freedom and 
corresponds to the magnitude of the translation.
Then, the relative velocity of body $\Bdy_i$ with respect to $\Sigma_k$ is:
\begin{align*}
   \bm{\omega}_i^\prime &= \vvec{0}, & 
   \vvec{v}_i^\prime &= \dot{\vvec{s}}_i = \dot{q}_i \vvec{e}_x.
\end{align*}
On the other hand, for a joint which allows the rotation about the $x$-axis, $\mathsf{E}_{\Bpre,i}(q_i)$ is a
matrix belonging to the 3D rotation group, $SO(3)$, namely
\begin{equation*}
   \mathsf{E}_{\Bpre,i}(q_i) =
   \begin{bmatrix} 1 & 0 & 0  \\
   0 & \cos{q_i} & -\sin{q_i} \\    
   0 & \sin{q_i} &  \cos{q_i} \\    
   \end{bmatrix};
\end{equation*}
and $\vvec{s}_i$ is a constant vector.
Note that, in this case $q_i$ stands for the rotation angle.
In this case, the relative velocity of body $\Bdy_{i}$ with respect to $\Sigma_k$ takes the form:
\begin{align*}
   \bm{\omega}_i^\prime &= \dot{q}_i \vvec{e}_x, &
   \vvec{v}_i^\prime  &= \bm{\omega}_i^\prime \times \vvec{s}_i.
\end{align*}
Similar definitions stand for translations and rotations about $y$ and $z$ axes.

In order to model joints which allow multiple degrees of freedom 
between two bodies,
several virtual bodies can be linked sequentially using single DoF joints (prismatic or revolute), 
as illustrated in Fig.~\ref{fig:virtbody}.
For the dynamical model, these virtual bodies have no mass; and for the fluid coupling, they have no associated Lagrangian points (i.e., no volume).
Under this approach, the connection between the two physical bodies is equivalent to a multi DoF joint.
%allowing for a simpler configuration of the algorithm inputs.
%
Hence, the present methodology allows a simple implementation of any kind of kinematic joint with a
negligible increase of the computational cost.

\section{Mapping between generalized and physical coordinates\label{sec:app_map}}

\subsection{From generalized coordinates to physical coordinates\label{sec:app_mapV}}

Computation of $\vvec{X}_{i,j}$ and $\vvec{U}_{\partial\Bdy_{i}}$, according to \eqcite{eq:physXU}, requires 
the calculation of $\mathsf{E}_i$, $\vvec{x}_i^0$, $\vvec{v}_i^0$, and $\vvec{\omega}_i^0$.
These variables can be derived from the joint variables derived in \ref{sec:app_joint}.
To that end, we exploit the fact that, for a \emph{kinematic tree} like the one considered in Fig.~\ref{fig:scheme},
we can define a unique predecessor for each body, $\Bdy_{i}$, which can be hence denoted as $\Bdy_{\lambda(i)}$.
Likewise, a set $\mu(i)$ can be defined containing all the bodies which precede $\Bdy_{i}$.
As an example, for body $\Bdy_{7}$ in Fig.~\ref{fig:scheme}, $\mu(7) = \{4,\,5,\,7\}$.
Under these definitions, the rotation matrix of body $\Bdy_{i}$ is computed as:
\begin{equation}
\mathsf{E}_i =  \prod_{j \in \mu(i)} \mathsf{E}_{\lambda(j),j} %= \mathsf{E}_{0,j} ... \mathsf{E}_{\lambda(i),i}
\end{equation}
Likewise, 
\begin{equation} \label{eq:app-xi0}
\vvec{x}_i^0 =  \sum_{j \in \mu(i)} \mathsf{E}_{\lambda(j)} \vvec{s}_j (q_j),
\end{equation}
whereas, $\vvec{v}_i^0$ and $\bm{\omega}_i^0$ can be computed from \eqcite{eq:app-xi0} by substituting
$\vvec{s}_j$ by $\vvec{v}_j^\prime$ and $\bm{\omega}_j^\prime$, respectively.

\subsection{From physical coordinates to generalized coordinates\label{sec:app_mapF}}
\providecommand{\fspa}{\hat{\mathbf{f}}}
\providecommand{\gspa}{\hat{\mathbf{g}}}

In order to compute $\bm{\xi}_h$ to solve \eqcite{eq:MBf}, the hydrodynamic forces acting upon the bodies 
must be expressed in terms of generalized coordinates.
Note that, with the present coupling, the component of the hydrodynamic forces that have to be mapped are
$\Gbdy_i$ and $\Nbdy_i$ from \eqcite{eq:Fi} and \eqcite{eq:Mi}, respectively.
For the sake of efficiency, it is convenient to gather both forces and moments acting on body $\Bdy_i$ into
a single vector:
\begin{equation}
\fspa_{i}^h = \begin{bmatrix} \Nbdy_i \\ \Gbdy_i \end{bmatrix},
\end{equation}
where it is implicitly assumed that both $\Nbdy_i$ and $\Gbdy_i$ are expressed in $\Sigma_0$ and moments
are computed about the origin.

We also define the matrix transform of $\Bdy_i$ as:
\begin{equation}
\vvec{X}^*_i = \begin{bmatrix} \mathsf{E}_i^\top & -\mathsf{E}_i^\top {\vvec{x}_i^0 \times} \\
                               \mathsf{0}        & \mathsf{E}_i^\top \end{bmatrix},           
\end{equation}
where ${\vvec{x}_i^0\times}$ is a skew-symmetric matrix
belonging to the Lie algebra of the $SO(3)$ rotation group.

%\begin{align*}
%\vvec{X}^*_i &= \begin{bmatrix} \mathsf{E}_i^\top & -\mathsf{E}_i^\top {\vvec{x}_i^0 \times} \\
%                               \mathsf{0}        & \mathsf{E}_i^\top \end{bmatrix},           &
%\vvec{X}^*_{\lambda(i),i} &= \begin{bmatrix} \Eji^\top &  {\vvec{s}_i\times}\Eji^\top         \\
%                               \mathsf{0}        & \Eji^\top          \end{bmatrix},           
%\end{align*}
%%
%where ${\vvec{x}_i^0\times}$ and ${\vvec{s}_i\times}$ are $3\times3$ skew-symmetric matrices
%belonging to the Lie algebra of $SO(3)$ rotation group.

With the previous definitions, a simplified version of the RNEA can be implemented to compute the 
components of $\bm{\xi}_h$, namely:
\begin{algorithmic}
\For{$i \gets 1,N_B$}
   \State $\gspa_i = \vvec{X}^*_{i}\fspa_i^h$
\EndFor
\For{$i \gets N_B,1$}
   \State $\xi_{h,i} = \mathbf{S}_i^\top \gspa_i$
   \If{$\lambda(i) \neq 0$}
      \State $\gspa_{\lambda(i)} = \gspa_{\lambda(i)} + \vvec{X}^*_{\lambda(i)}\fspa_i^h$
   \EndIf
\EndFor
\end{algorithmic}
where $\lambda(i)$ is the index of the predecessor body of $\Bdy_i$.
The previous algorithm simply transfers the forces acting upon a given body across its supporting tree 
(i.e., the set of bodies that connect it to base, $\Sigma_0$).
Note that, $\vvec{S}_i$ is the motion subspace of the joint.
For prismatic and revolute joints, $\vvec{S}_i$ is a unitary column vector of size $6 \times 1$, whose only non-zero
component is the axis along which rotation/translation is allowed.
In particular, its first 3 components are associated to rotations about the $x$, $y$ or $z$ axes of the joints;
meanwhile its 3 last components are associated with translations along the aforementioned axes.

%%       \citet{<label>} ==> Jones et al. [21]
%%       \citep{<label>} ==> [21]
%%

\bibliographystyle{elsarticle-num-names} 
\bibliography{bibliography}

\begin{thebibliography}{65}
\expandafter\ifx\csname natexlab\endcsname\relax\def\natexlab#1{#1}\fi
\providecommand{\url}[1]{\texttt{#1}}
\providecommand{\href}[2]{#2}
\providecommand{\path}[1]{#1}
\providecommand{\DOIprefix}{doi:}
\providecommand{\ArXivprefix}{arXiv:}
\providecommand{\URLprefix}{URL: }
\providecommand{\Pubmedprefix}{pmid:}
\providecommand{\doi}[1]{\href{http://dx.doi.org/#1}{\path{#1}}}
\providecommand{\Pubmed}[1]{\href{pmid:#1}{\path{#1}}}
\providecommand{\bibinfo}[2]{#2}
\ifx\xfnm\relax \def\xfnm[#1]{\unskip,\space#1}\fi
%Type = Article
\bibitem[{de~Croon et~al.(2009)de~Croon, de~Clercq, Ruijsink, Remes, and
  de~Wagter}]{decroon2009}
\bibinfo{author}{G.~de~Croon}, \bibinfo{author}{K.~de~Clercq},
  \bibinfo{author}{R.~Ruijsink}, \bibinfo{author}{B.~Remes},
  \bibinfo{author}{C.~de~Wagter},
\newblock \bibinfo{title}{Design, aerodynamics, and vision-based control of the
  {DelFly}},
\newblock \bibinfo{journal}{Int. J. Micro Air Veh.} \bibinfo{volume}{1}
  (\bibinfo{year}{2009}) \bibinfo{pages}{71--97}.
  \DOIprefix\doi{10.1260/175682909789498288}.
%Type = Article
\bibitem[{Richter and Lipson(2011)}]{rithcher2011}
\bibinfo{author}{C.~Richter}, \bibinfo{author}{H.~Lipson},
\newblock \bibinfo{title}{Untethered hovering flapping flight of a 3{D}-printed
  mechanical insect},
\newblock \bibinfo{journal}{Artif. Life} \bibinfo{volume}{17}
  (\bibinfo{year}{2011}) \bibinfo{pages}{73--86}.
  \DOIprefix\doi{10.1162/artl\_a\_00020}.
%Type = Inproceedings
\bibitem[{Keennon et~al.(2012)Keennon, Klingebiel, and Won}]{keennon2012}
\bibinfo{author}{M.~Keennon}, \bibinfo{author}{K.~Klingebiel},
  \bibinfo{author}{H.~Won},
\newblock \bibinfo{title}{Development of the nano hummingbird: A tailless
  flapping wing micro air vehicle},
\newblock in: \bibinfo{booktitle}{50th AIAA Aerospace Sciences Meeting
  including the New Horizons Forum and Aerospace Exposition},
  \bibinfo{year}{2012}, pp. \bibinfo{pages}{1--12}.
  \DOIprefix\doi{10.2514/6.2012-588}.
%Type = Article
\bibitem[{Triantafyllou and Triantafyllou(1995)}]{triantafyllou1995}
\bibinfo{author}{M.~S. Triantafyllou}, \bibinfo{author}{G.~S. Triantafyllou},
\newblock \bibinfo{title}{An efficient swimming machine},
\newblock \bibinfo{journal}{Sci. Am.} \bibinfo{volume}{272}
  (\bibinfo{year}{1995}) \bibinfo{pages}{64--70}.
%Type = Inproceedings
\bibitem[{Hirata et~al.(2000)Hirata, Takimoto, and Tamura}]{hirata2000}
\bibinfo{author}{K.~Hirata}, \bibinfo{author}{T.~Takimoto},
  \bibinfo{author}{K.~Tamura},
\newblock \bibinfo{title}{Study on turning performance of a fish robot},
\newblock in: \bibinfo{booktitle}{Proc. 1st Int. Symp. Aqua Bio-Mechanisms},
  \bibinfo{year}{2000}, pp. \bibinfo{pages}{287--292}.
%Type = Article
\bibitem[{Yu et~al.(2004)Yu, Tan, Wang, and Chen}]{yu2004}
\bibinfo{author}{J.~Yu}, \bibinfo{author}{M.~Tan}, \bibinfo{author}{S.~Wang},
  \bibinfo{author}{E.~Chen},
\newblock \bibinfo{title}{Development of a biomimetic robotic fish and its
  control algorithm},
\newblock \bibinfo{journal}{IEEE Transactions on Systems, Man, and Cybernetics,
  Part B (Cybernetics)} \bibinfo{volume}{34} (\bibinfo{year}{2004})
  \bibinfo{pages}{1798--1810}. \DOIprefix\doi{10.1109/TSMCB.2004.831151}.
%Type = Article
\bibitem[{Deng et~al.(2013)Deng, Xu, Chen, Dai, Wu, and Tian}]{deng2013}
\bibinfo{author}{H.-B. Deng}, \bibinfo{author}{Y.-Q. Xu},
  \bibinfo{author}{D.-D. Chen}, \bibinfo{author}{H.~Dai},
  \bibinfo{author}{J.~Wu}, \bibinfo{author}{F.-B. Tian},
\newblock \bibinfo{title}{On numerical modeling of animal swimming and flight},
\newblock \bibinfo{journal}{Comput. Mech.} \bibinfo{volume}{52}
  (\bibinfo{year}{2013}) \bibinfo{pages}{1221--1242}.
  \DOIprefix\doi{10.1007/s00466-013-0875-2}.
%Type = Inproceedings
\bibitem[{Donea et~al.(2017)Donea, Huerta, Ponthot, and
  Rodr\'{i}guez-Ferran}]{donea2017}
\bibinfo{author}{J.~Donea}, \bibinfo{author}{A.~Huerta}, \bibinfo{author}{J.-P.
  Ponthot}, \bibinfo{author}{A.~Rodr\'{i}guez-Ferran},
\newblock \bibinfo{title}{{Arbitrary Lagrangian–Eulerian Methods}},
\newblock in: \bibinfo{booktitle}{Encyclopedia of Computational Mechanics
  Second Edition}, \bibinfo{publisher}{Wiley Online Library},
  \bibinfo{year}{2017}, pp. \bibinfo{pages}{1--23}.
  \DOIprefix\doi{10.1002/9781119176817.ecm2009}.
%Type = Article
\bibitem[{Tschisgale and Fr\"{o}hlich(2020)}]{tschisgale2020}
\bibinfo{author}{S.~Tschisgale}, \bibinfo{author}{J.~Fr\"{o}hlich},
\newblock \bibinfo{title}{An immersed boundary method for the fluid-structure
  interaction of slender flexible structures in viscous fluid},
\newblock \bibinfo{journal}{J. Comput. Phys.} \bibinfo{volume}{423}
  (\bibinfo{year}{2020}) \bibinfo{pages}{109801}.
  \DOIprefix\doi{10.1016/j.jcp.2020.109801}.
%Type = Article
\bibitem[{Mittal and Iaccarino(2005)}]{mittal2005}
\bibinfo{author}{R.~Mittal}, \bibinfo{author}{G.~Iaccarino},
\newblock \bibinfo{title}{Immersed boundary methods},
\newblock \bibinfo{journal}{Annu. Rev. Fluid Mech.} \bibinfo{volume}{37}
  (\bibinfo{year}{2005}) \bibinfo{pages}{239--261}.
  \DOIprefix\doi{10.1146/annurev.fluid.37.061903.175743}.
%Type = Article
\bibitem[{Griffith and Patankar(2020)}]{griffith2020}
\bibinfo{author}{B.~E. Griffith}, \bibinfo{author}{N.~A. Patankar},
\newblock \bibinfo{title}{Immersed methods for fluid–structure interaction},
\newblock \bibinfo{journal}{Annu. Rev. Fluid Mech.} \bibinfo{volume}{52}
  (\bibinfo{year}{2020}) \bibinfo{pages}{421--448}.
  \DOIprefix\doi{10.1146/annurev-fluid-010719-060228}.
%Type = Article
\bibitem[{Uhlmann(2005)}]{uhlmann2005}
\bibinfo{author}{M.~Uhlmann},
\newblock \bibinfo{title}{An immersed boundary method with direct forcing for
  the simulation of particulate flows},
\newblock \bibinfo{journal}{J. Comput. Phys.} \bibinfo{volume}{209}
  (\bibinfo{year}{2005}) \bibinfo{pages}{448--476}.
  \DOIprefix\doi{10.1016/j.jcp.2005.03.017}.
%Type = Article
\bibitem[{Pinelli et~al.(2010)Pinelli, Naqavi, Piomelli, and
  Favier}]{pinelli2010}
\bibinfo{author}{A.~Pinelli}, \bibinfo{author}{I.~Z. Naqavi},
  \bibinfo{author}{U.~Piomelli}, \bibinfo{author}{J.~Favier},
\newblock \bibinfo{title}{Immersed-boundary methods for general
  finite-difference and finite-volume {N}avier--{S}tokes solvers},
\newblock \bibinfo{journal}{J. Comput. Phys.} \bibinfo{volume}{229}
  (\bibinfo{year}{2010}) \bibinfo{pages}{9073--9091}.
  \DOIprefix\doi{10.1016/j.jcp.2010.08.021}.
%Type = Article
\bibitem[{Breugem(2012)}]{breugem2012}
\bibinfo{author}{W.-P. Breugem},
\newblock \bibinfo{title}{A second-order accurate immersed boundary method for
  fully resolved simulations of particle-laden flows},
\newblock \bibinfo{journal}{J. Comput. Phys.} \bibinfo{volume}{231}
  (\bibinfo{year}{2012}) \bibinfo{pages}{4469--4498}.
  \DOIprefix\doi{10.1016/j.jcp.2012.02.026}.
%Type = Article
\bibitem[{Kempe and Fr{\"o}hlich(2012)}]{kempe2012}
\bibinfo{author}{T.~Kempe}, \bibinfo{author}{J.~Fr{\"o}hlich},
\newblock \bibinfo{title}{An improved immersed boundary method with direct
  forcing for the simulation of particle laden flows},
\newblock \bibinfo{journal}{J. Comput. Phys.} \bibinfo{volume}{231}
  (\bibinfo{year}{2012}) \bibinfo{pages}{3663--3684}.
  \DOIprefix\doi{10.1016/j.jcp.2012.01.021}.
%Type = Article
\bibitem[{Bhalla et~al.(2013)Bhalla, Bale, Griffith, and Patankar}]{bhalla2013}
\bibinfo{author}{A.~P.~S. Bhalla}, \bibinfo{author}{R.~Bale},
  \bibinfo{author}{B.~E. Griffith}, \bibinfo{author}{N.~A. Patankar},
\newblock \bibinfo{title}{A unified mathematical framework and an adaptive
  numerical method for fluid–structure interaction with rigid, deforming, and
  elastic bodies},
\newblock \bibinfo{journal}{J. Comput. Phys.} \bibinfo{volume}{250}
  (\bibinfo{year}{2013}) \bibinfo{pages}{446--476}.
  \DOIprefix\doi{10.1016/j.jcp.2013.04.033}.
%Type = Article
\bibitem[{Wiens and Stockie(2015)}]{wiens2015}
\bibinfo{author}{J.~K. Wiens}, \bibinfo{author}{J.~M. Stockie},
\newblock \bibinfo{title}{An efficient parallel immersed boundary algorithm
  using a pseudo-compressible fluid solver},
\newblock \bibinfo{journal}{J. Comput. Phys.} \bibinfo{volume}{281}
  (\bibinfo{year}{2015}) \bibinfo{pages}{917--941}.
  \DOIprefix\doi{10.1016/j.jcp.2014.10.058}.
%Type = Article
\bibitem[{{de Tullio} and Pascazio(2016)}]{detullio2016}
\bibinfo{author}{M.~{de Tullio}}, \bibinfo{author}{G.~Pascazio},
\newblock \bibinfo{title}{A moving-least-squares immersed boundary method for
  simulating the fluid–structure interaction of elastic bodies with arbitrary
  thickness},
\newblock \bibinfo{journal}{J. Comput. Phys.} \bibinfo{volume}{325}
  (\bibinfo{year}{2016}) \bibinfo{pages}{201--225}.
  \DOIprefix\doi{10.1016/j.jcp.2016.08.020}.
%Type = Article
\bibitem[{Kim and Peskin(2009)}]{kim2009}
\bibinfo{author}{Y.~Kim}, \bibinfo{author}{C.~S. Peskin},
\newblock \bibinfo{title}{3-d parachute simulation by the immersed boundary
  method},
\newblock \bibinfo{journal}{Comput. Fluids} \bibinfo{volume}{38}
  (\bibinfo{year}{2009}) \bibinfo{pages}{1080--1090}.
  \DOIprefix\doi{10.1016/j.compfluid.2008.11.002}.
%Type = Article
\bibitem[{Zhang et~al.(2004)Zhang, Gerstenberger, Wang, and Liu}]{zhang2004}
\bibinfo{author}{L.~Zhang}, \bibinfo{author}{A.~Gerstenberger},
  \bibinfo{author}{X.~Wang}, \bibinfo{author}{W.~K. Liu},
\newblock \bibinfo{title}{Immersed finite element method},
\newblock \bibinfo{journal}{Comput. Methods Appl. Mech. Eng.}
  \bibinfo{volume}{193} (\bibinfo{year}{2004}) \bibinfo{pages}{2051--2067}.
  \DOIprefix\doi{10.1016/j.cma.2003.12.044}.
%Type = Article
\bibitem[{Tian et~al.(2014)Tian, Dai, Luo, Doyle, and Rousseau}]{tian2014}
\bibinfo{author}{F.-B. Tian}, \bibinfo{author}{H.~Dai},
  \bibinfo{author}{H.~Luo}, \bibinfo{author}{J.~F. Doyle},
  \bibinfo{author}{B.~Rousseau},
\newblock \bibinfo{title}{Fluid–structure interaction involving large
  deformations: 3d simulations and applications to biological systems},
\newblock \bibinfo{journal}{J. Comput. Phys.} \bibinfo{volume}{258}
  (\bibinfo{year}{2014}) \bibinfo{pages}{451--469}.
  \DOIprefix\doi{10.1016/j.jcp.2013.10.047}.
%Type = Article
\bibitem[{Liu(2009)}]{liu2009}
\bibinfo{author}{H.~Liu},
\newblock \bibinfo{title}{Integrated modeling of insect flight: From
  morphology, kinematics to aerodynamics},
\newblock \bibinfo{journal}{J. Comput. Phys.} \bibinfo{volume}{228}
  (\bibinfo{year}{2009}) \bibinfo{pages}{439 -- 459}.
  \DOIprefix\doi{10.1016/j.jcp.2008.09.020}.
%Type = Article
\bibitem[{Suzuki et~al.(2015)Suzuki, Minami, and Inamuro}]{suzuki2015}
\bibinfo{author}{K.~Suzuki}, \bibinfo{author}{K.~Minami},
  \bibinfo{author}{T.~Inamuro},
\newblock \bibinfo{title}{Lift and thrust generation by a butterfly-like
  flapping wing–body model: immersed boundary–lattice {B}oltzmann
  simulations},
\newblock \bibinfo{journal}{J. Fluid Mech.} \bibinfo{volume}{767}
  (\bibinfo{year}{2015}) \bibinfo{pages}{659 -- 695}.
  \DOIprefix\doi{10.1017/jfm.2015.57}.
%Type = Book
\bibitem[{Featherstone(2014)}]{featherstone2014}
\bibinfo{author}{R.~Featherstone}, \bibinfo{title}{Rigid body dynamics
  algorithms}, \bibinfo{publisher}{Springer}, \bibinfo{year}{2014}.
%Type = Article
\bibitem[{Zhang et~al.(2013)Zhang, Yu, Zhang, and Zhang}]{zhang2013}
\bibinfo{author}{S.~Zhang}, \bibinfo{author}{J.~Yu},
  \bibinfo{author}{A.~Zhang}, \bibinfo{author}{F.~Zhang},
\newblock \bibinfo{title}{Spiraling motion of underwater gliders: Modeling,
  analysis, and experimental results},
\newblock \bibinfo{journal}{Ocean Eng.} \bibinfo{volume}{60}
  (\bibinfo{year}{2013}) \bibinfo{pages}{1--13}.
  \DOIprefix\doi{10.1016/j.oceaneng.2012.12.023}.
%Type = Article
\bibitem[{Arora et~al.(2018)Arora, Kang, Shyy, and Gupta}]{arora2018}
\bibinfo{author}{N.~Arora}, \bibinfo{author}{C.-K. Kang},
  \bibinfo{author}{W.~Shyy}, \bibinfo{author}{A.~Gupta},
\newblock \bibinfo{title}{Analysis of passive flexion in propelling a plunging
  plate using a torsion spring model},
\newblock \bibinfo{journal}{J. Fluid Mech.} \bibinfo{volume}{857}
  (\bibinfo{year}{2018}) \bibinfo{pages}{562--604}.
  \DOIprefix\doi{10.1017/jfm.2018.736}.
%Type = Article
\bibitem[{Suzuki et~al.(2019)Suzuki, Okada, and Yoshino}]{suzuki2019}
\bibinfo{author}{K.~Suzuki}, \bibinfo{author}{I.~Okada},
  \bibinfo{author}{M.~Yoshino},
\newblock \bibinfo{title}{Effect of wing mass on the free flight of a
  butterfly-like model using immersed boundary–lattice {B}oltzmann
  simulations},
\newblock \bibinfo{journal}{J. Fluid Mech.} \bibinfo{volume}{877}
  (\bibinfo{year}{2019}) \bibinfo{pages}{614--647}.
  \DOIprefix\doi{10.1017/jfm.2019.597}.
%Type = Article
\bibitem[{Yao and Yeo(2019)}]{yao2019}
\bibinfo{author}{J.~Yao}, \bibinfo{author}{K.~S. Yeo},
\newblock \bibinfo{title}{Free hovering of hummingbird hawkmoth and effects of
  wing mass and wing elevation},
\newblock \bibinfo{journal}{Comput. Fluids} \bibinfo{volume}{186}
  (\bibinfo{year}{2019}) \bibinfo{pages}{99 -- 127}.
  \DOIprefix\doi{10.1016/j.compfluid.2019.04.007}.
%Type = Article
\bibitem[{Wang and Eldredge(2015)}]{wang2015}
\bibinfo{author}{C.~Wang}, \bibinfo{author}{J.~D. Eldredge},
\newblock \bibinfo{title}{Strongly coupled dynamics of fluids and rigid-body
  systems with the immersed boundary projection method},
\newblock \bibinfo{journal}{J. Comput. Phys.} \bibinfo{volume}{295}
  (\bibinfo{year}{2015}) \bibinfo{pages}{87--113}.
  \DOIprefix\doi{10.1016/j.jcp.2015.04.005}.
%Type = Article
\bibitem[{Li et~al.(2018)Li, Xiao, Liu, Hu, Li, Li, Liu, Hu, and Wen}]{li2018}
\bibinfo{author}{R.~Li}, \bibinfo{author}{Q.~Xiao}, \bibinfo{author}{Y.~Liu},
  \bibinfo{author}{J.~Hu}, \bibinfo{author}{L.~Li}, \bibinfo{author}{G.~Li},
  \bibinfo{author}{H.~Liu}, \bibinfo{author}{K.~Hu}, \bibinfo{author}{L.~Wen},
\newblock \bibinfo{title}{A multi-body dynamics based numerical modelling tool
  for solving aquatic biomimetic problems},
\newblock \bibinfo{journal}{Bioinspir. Biomim.} \bibinfo{volume}{13}
  (\bibinfo{year}{2018}) \bibinfo{pages}{056001}.
  \DOIprefix\doi{10.1088/1748-3190/aacd60}.
%Type = Article
\bibitem[{Bernier et~al.(2019)Bernier, Gazzola, Ronsse, and
  Chatelain}]{bernier2019}
\bibinfo{author}{C.~Bernier}, \bibinfo{author}{M.~Gazzola},
  \bibinfo{author}{R.~Ronsse}, \bibinfo{author}{P.~Chatelain},
\newblock \bibinfo{title}{Simulations of propelling and energy harvesting
  articulated bodies via vortex particle-mesh methods},
\newblock \bibinfo{journal}{J. Comput. Phys.} \bibinfo{volume}{392}
  (\bibinfo{year}{2019}) \bibinfo{pages}{34 -- 55}.
  \DOIprefix\doi{10.1016/j.jcp.2019.04.036}.
%Type = Article
\bibitem[{Gazzola et~al.(2011)Gazzola, Chatelain, Wim, and
  Koumoutsakos}]{gazzola2011}
\bibinfo{author}{M.~Gazzola}, \bibinfo{author}{P.~Chatelain},
  \bibinfo{author}{M.~v.~R. Wim}, \bibinfo{author}{P.~Koumoutsakos},
\newblock \bibinfo{title}{Simulations of single and multiple swimmers with
  non-divergence free deforming geometries},
\newblock \bibinfo{journal}{J. Comput. Phys.} \bibinfo{volume}{230}
  (\bibinfo{year}{2011}) \bibinfo{pages}{7093 -- 7114}.
  \DOIprefix\doi{10.1016/j.jcp.2011.04.025}.
%Type = Article
\bibitem[{Felis(2017)}]{felis2017}
\bibinfo{author}{M.~Felis},
\newblock \bibinfo{title}{{RBDL}: an efficient rigid-body dynamics library
  using recursive algorithms},
\newblock \bibinfo{journal}{Auton. Robot.} \bibinfo{volume}{41}
  (\bibinfo{year}{2017}) \bibinfo{pages}{495--511}.
  \DOIprefix\doi{10.1007/s10514-016-9574-0}.
%Type = Book
\bibitem[{Greenwood(2006)}]{greenwood2006}
\bibinfo{author}{D.~Greenwood}, \bibinfo{title}{Advanced dynamics},
  \bibinfo{publisher}{Cambridge University Press}, \bibinfo{year}{2006}.
%Type = Article
\bibitem[{Boyer and Porez(2015)}]{boyer2015}
\bibinfo{author}{F.~Boyer}, \bibinfo{author}{M.~Porez},
\newblock \bibinfo{title}{{M}ultibody system dynamics for bio-inspired
  locomotion: {F}rom geometric structures to computational aspects},
\newblock \bibinfo{journal}{Bioinspir. Biomim.} \bibinfo{volume}{10}
  (\bibinfo{year}{2015}) \bibinfo{pages}{1--21}.
  \DOIprefix\doi{10.1088/1748-3190/10/2/025007}.
%Type = Article
\bibitem[{Rai and Moin(1991)}]{rai1991}
\bibinfo{author}{M.~Rai}, \bibinfo{author}{P.~Moin},
\newblock \bibinfo{title}{Direct simulations of turbulent flow using
  finite-difference schemes},
\newblock \bibinfo{journal}{J. Comput. Phys.} \bibinfo{volume}{96}
  (\bibinfo{year}{1991}) \bibinfo{pages}{15 -- 53}.
  \DOIprefix\doi{10.1016/0021-9991(91)90264-L}.
%Type = Article
\bibitem[{Peskin(2002)}]{peskin2002}
\bibinfo{author}{C.~Peskin},
\newblock \bibinfo{title}{The immersed boundary method},
\newblock \bibinfo{journal}{Acta Numer.} \bibinfo{volume}{11}
  (\bibinfo{year}{2002}) \bibinfo{pages}{479–51}.
  \DOIprefix\doi{10.1017/S0962492902000077}.
%Type = Article
\bibitem[{Roma et~al.(1999)Roma, Peskin, and Berger}]{roma1999}
\bibinfo{author}{A.~M. Roma}, \bibinfo{author}{C.~S. Peskin},
  \bibinfo{author}{M.~J. Berger},
\newblock \bibinfo{title}{An adaptive version of the immersed boundary method},
\newblock \bibinfo{journal}{J. Comput. Phys.} \bibinfo{volume}{153}
  (\bibinfo{year}{1999}) \bibinfo{pages}{509--534}.
  \DOIprefix\doi{10.1006/jcph.1999.6293}.
%Type = Article
\bibitem[{Sotiropoulos and Yang(2014)}]{sotiropoulos2014}
\bibinfo{author}{F.~Sotiropoulos}, \bibinfo{author}{X.~Yang},
\newblock \bibinfo{title}{Immersed boundary methods for simulating
  fluid–structure interaction},
\newblock \bibinfo{journal}{Prog. Aerosp. Sci.} \bibinfo{volume}{65}
  (\bibinfo{year}{2014}) \bibinfo{pages}{1--21}.
  \DOIprefix\doi{10.1016/j.paerosci.2013.09.003}.
%Type = Article
\bibitem[{Moriche et~al.(2016)Moriche, Flores, and
  Garc{\'{\i}}a-Villalba}]{moriche2016}
\bibinfo{author}{M.~Moriche}, \bibinfo{author}{O.~Flores},
  \bibinfo{author}{M.~Garc{\'{\i}}a-Villalba},
\newblock \bibinfo{title}{Three-dimensional instabilities in the wake of a
  flapping wing at low {R}eynolds number},
\newblock \bibinfo{journal}{Int. J. Heat Fluid Flow} \bibinfo{volume}{62A}
  (\bibinfo{year}{2016}) \bibinfo{pages}{44--55}.
  \DOIprefix\doi{10.1016/j.ijheatfluidflow.2016.06.015}.
%Type = Article
\bibitem[{Moriche et~al.(2017)Moriche, Flores, and
  Garc{\'{\i}}a-Villalba}]{moriche2017}
\bibinfo{author}{M.~Moriche}, \bibinfo{author}{O.~Flores},
  \bibinfo{author}{M.~Garc{\'{\i}}a-Villalba},
\newblock \bibinfo{title}{On the aerodynamic forces on heaving and pitching
  airfoils at low {R}eynolds number},
\newblock \bibinfo{journal}{J. Fluid Mech.} \bibinfo{volume}{828}
  (\bibinfo{year}{2017}) \bibinfo{pages}{395--423}.
  \DOIprefix\doi{10.1017/jfm.2017.508}.
%Type = Article
\bibitem[{Gonzalo et~al.(2018)Gonzalo, Arranz, Moriche, Garc\'{i}a-Villalba,
  and Flores}]{gonzalo2018}
\bibinfo{author}{A.~Gonzalo}, \bibinfo{author}{G.~Arranz},
  \bibinfo{author}{M.~Moriche}, \bibinfo{author}{M.~Garc\'{i}a-Villalba},
  \bibinfo{author}{O.~Flores},
\newblock \bibinfo{title}{From flapping to heaving: {A} numerical study of
  wings in forward flight},
\newblock \bibinfo{journal}{J. Fluids Struct.} \bibinfo{volume}{83}
  (\bibinfo{year}{2018}) \bibinfo{pages}{293--309}.
  \DOIprefix\doi{10.1016/j.jfluidstructs.2018.09.006}.
%Type = Article
\bibitem[{Arranz et~al.(2020)Arranz, Flores, and
  Garc\'{i}a-Villalba}]{arranz2020}
\bibinfo{author}{G.~Arranz}, \bibinfo{author}{O.~Flores},
  \bibinfo{author}{M.~Garc\'{i}a-Villalba},
\newblock \bibinfo{title}{Three-dimensional effects on the aerodynamic
  performance of flapping wings in tandem configuration},
\newblock \bibinfo{journal}{J. Fluids Struct.} \bibinfo{volume}{94}
  (\bibinfo{year}{2020}) \bibinfo{pages}{102893}.
  \DOIprefix\doi{10.1016/j.jfluidstructs.2020.102893}.
%Type = Article
\bibitem[{Moriche et~al.(2021{\natexlab{a}})Moriche, Gonzalo, Flores, and
  Garcia-Villalba}]{moriche2021a}
\bibinfo{author}{M.~Moriche}, \bibinfo{author}{A.~Gonzalo},
  \bibinfo{author}{O.~Flores}, \bibinfo{author}{M.~Garcia-Villalba},
\newblock \bibinfo{title}{Three-dimensional effects on plunging airfoils at low
  {R}eynolds numbers},
\newblock \bibinfo{journal}{AIAA J.} \bibinfo{volume}{59}
  (\bibinfo{year}{2021}{\natexlab{a}}) \bibinfo{pages}{65--74}.
  \DOIprefix\doi{10.2514/1.J058569}.
%Type = Article
\bibitem[{Moriche et~al.(2021{\natexlab{b}})Moriche, Sedky, Jones, Flores, and
  Garcia-Villalba}]{moriche2021b}
\bibinfo{author}{M.~Moriche}, \bibinfo{author}{G.~Sedky},
  \bibinfo{author}{A.~R. Jones}, \bibinfo{author}{O.~Flores},
  \bibinfo{author}{M.~Garcia-Villalba},
\newblock \bibinfo{title}{Characterization of aerodynamic forces on wings in
  plunge maneuvers},
\newblock \bibinfo{journal}{AIAA J.} \bibinfo{volume}{59}
  (\bibinfo{year}{2021}{\natexlab{b}}) \bibinfo{pages}{751--762}.
  \DOIprefix\doi{10.2514/1.J059689}.
%Type = Article
\bibitem[{Arranz et~al.(2018{\natexlab{a}})Arranz, Moriche, Uhlmann, Flores,
  and Garc{\'\i}a-Villalba}]{arranz2018a}
\bibinfo{author}{G.~Arranz}, \bibinfo{author}{M.~Moriche},
  \bibinfo{author}{M.~Uhlmann}, \bibinfo{author}{O.~Flores},
  \bibinfo{author}{M.~Garc{\'\i}a-Villalba},
\newblock \bibinfo{title}{Kinematics and dynamics of the auto-rotation of a
  model winged seed},
\newblock \bibinfo{journal}{Bioinspir. Biomim.} \bibinfo{volume}{13}
  (\bibinfo{year}{2018}{\natexlab{a}}) \bibinfo{pages}{036011}.
  \DOIprefix\doi{10.1088/1748-3190/aab144}.
%Type = Article
\bibitem[{Arranz et~al.(2018{\natexlab{b}})Arranz, Gonzalo, Uhlmann, Flores,
  and Garc{\'{\i}}a-Villalba}]{arranz2018b}
\bibinfo{author}{G.~Arranz}, \bibinfo{author}{A.~Gonzalo},
  \bibinfo{author}{M.~Uhlmann}, \bibinfo{author}{O.~Flores},
  \bibinfo{author}{M.~Garc{\'{\i}}a-Villalba},
\newblock \bibinfo{title}{A numerical study of the flow around a model winged
  seed in auto-rotation},
\newblock \bibinfo{journal}{Flow Turbul. Combust.} \bibinfo{volume}{101}
  (\bibinfo{year}{2018}{\natexlab{b}}) \bibinfo{pages}{477--497}.
  \DOIprefix\doi{10.1007/s10494-018-9945-z}.
%Type = Article
\bibitem[{Tschisgale et~al.(2017)Tschisgale, Kempe, and
  Fröhlich}]{tschisgale2017}
\bibinfo{author}{S.~Tschisgale}, \bibinfo{author}{T.~Kempe},
  \bibinfo{author}{J.~Fröhlich},
\newblock \bibinfo{title}{A non-iterative immersed boundary method for
  spherical particles of arbitrary density ratio},
\newblock \bibinfo{journal}{J. Comput. Phys.} \bibinfo{volume}{339}
  (\bibinfo{year}{2017}) \bibinfo{pages}{432 -- 452}.
  \DOIprefix\doi{10.1016/j.jcp.2017.03.026}.
%Type = Article
\bibitem[{Toomey and Eldredge(2008)}]{toomey2008}
\bibinfo{author}{J.~Toomey}, \bibinfo{author}{J.~D. Eldredge},
\newblock \bibinfo{title}{Numerical and experimental study of the fluid
  dynamics of a flapping wing with low order flexibility},
\newblock \bibinfo{journal}{Phys. Fluids} \bibinfo{volume}{20}
  (\bibinfo{year}{2008}) \bibinfo{pages}{073603}.
  \DOIprefix\doi{10.1063/1.2956372}.
%Type = Article
\bibitem[{Lee and Choi(2015)}]{lee2015}
\bibinfo{author}{I.~Lee}, \bibinfo{author}{H.~Choi},
\newblock \bibinfo{title}{A discrete-forcing immersed boundary method for the
  fluid–structure interaction of an elastic slender body},
\newblock \bibinfo{journal}{J. Comput. Phys.} \bibinfo{volume}{280}
  (\bibinfo{year}{2015}) \bibinfo{pages}{529--546}.
  \DOIprefix\doi{10.1016/j.jcp.2014.09.028}.
%Type = Article
\bibitem[{Yeh and Alexeev(2016)}]{yeh2016}
\bibinfo{author}{P.~D. Yeh}, \bibinfo{author}{A.~Alexeev},
\newblock \bibinfo{title}{Effect of aspect ratio in free-swimming plunging
  flexible plates},
\newblock \bibinfo{journal}{Comput. Fluids} \bibinfo{volume}{124}
  (\bibinfo{year}{2016}) \bibinfo{pages}{220 -- 225}.
  \DOIprefix\doi{10.1016/j.compfluid.2015.07.009}.
%Type = Article
\bibitem[{Yeh and Alexeev(2014)}]{yeh2014}
\bibinfo{author}{P.~D. Yeh}, \bibinfo{author}{A.~Alexeev},
\newblock \bibinfo{title}{Free swimming of an elastic plate plunging at low
  {R}eynolds number},
\newblock \bibinfo{journal}{Phys. Fluids} \bibinfo{volume}{26}
  (\bibinfo{year}{2014}) \bibinfo{pages}{053604}.
  \DOIprefix\doi{10.1063/1.4876231}.
%Type = Article
\bibitem[{Hunt et~al.(1988)Hunt, Wray, and Moin}]{hunt1988}
\bibinfo{author}{J.~C.~R. Hunt}, \bibinfo{author}{A.~A. Wray},
  \bibinfo{author}{P.~Moin},
\newblock \bibinfo{title}{Eddies, stream, and convergence zones in turbulent
  flows},
\newblock \bibinfo{journal}{Center For Turbulence Research}
  \bibinfo{volume}{Report CTR-S88} (\bibinfo{year}{1988}).
%Type = Article
\bibitem[{Raspa et~al.(2014)Raspa, Ramananarivo, Thiria, and
  Godoy-Diana}]{raspa2014}
\bibinfo{author}{V.~Raspa}, \bibinfo{author}{S.~Ramananarivo},
  \bibinfo{author}{B.~Thiria}, \bibinfo{author}{R.~Godoy-Diana},
\newblock \bibinfo{title}{Vortex-induced drag and the role of aspect ratio in
  undulatory swimmers},
\newblock \bibinfo{journal}{Phys. Fluids} \bibinfo{volume}{26}
  (\bibinfo{year}{2014}) \bibinfo{pages}{041701}.
  \DOIprefix\doi{10.1063/1.4870254}.
%Type = Article
\bibitem[{Humphrey(1987)}]{humphrey1987}
\bibinfo{author}{J.~A.~C. Humphrey},
\newblock \bibinfo{title}{Fluid mechanic constraints on spider ballooning},
\newblock \bibinfo{journal}{Oecologia} \bibinfo{volume}{73}
  (\bibinfo{year}{1987}) \bibinfo{pages}{469--477}.
  \DOIprefix\doi{10.1007/BF00385267}.
%Type = Inproceedings
\bibitem[{Zhao et~al.(2017)Zhao, Panayotova, Chuang, Sheldon, Bourouiba, and
  Miller}]{zhao2017}
\bibinfo{author}{L.~Zhao}, \bibinfo{author}{I.~N. Panayotova},
  \bibinfo{author}{A.~Chuang}, \bibinfo{author}{K.~S. Sheldon},
  \bibinfo{author}{L.~Bourouiba}, \bibinfo{author}{L.~A. Miller},
\newblock \bibinfo{title}{Flying spiders: Simulating and modeling the dynamics
  of ballooning},
\newblock in: \bibinfo{editor}{A.~T. Layton}, \bibinfo{editor}{L.~A. Miller}
  (Eds.), \bibinfo{booktitle}{Women in Mathematical Biology},
  \bibinfo{year}{2017}, pp. \bibinfo{pages}{179--210}.
  \DOIprefix\doi{10.1007/978-3-319-60304-9\_10}.
%Type = Article
\bibitem[{Cho et~al.(2018)Cho, Neubauer, Fahrenson, and Rechenberg}]{cho2018}
\bibinfo{author}{M.~Cho}, \bibinfo{author}{P.~Neubauer},
  \bibinfo{author}{C.~Fahrenson}, \bibinfo{author}{I.~Rechenberg},
\newblock \bibinfo{title}{An observational study of ballooning in large
  spiders: {N}anoscale multifibers enable large spiders' soaring flight},
\newblock \bibinfo{journal}{PLOS Biol.} \bibinfo{volume}{16}
  (\bibinfo{year}{2018}) \bibinfo{pages}{1--27}.
  \DOIprefix\doi{10.1371/journal.pbio.2004405}.
%Type = Article
\bibitem[{Suter(1991)}]{suter1991}
\bibinfo{author}{R.~B. Suter},
\newblock \bibinfo{title}{Ballooning in spiders: results of wind tunnel
  experiments},
\newblock \bibinfo{journal}{Ethol. Ecol. Evol.} \bibinfo{volume}{3}
  (\bibinfo{year}{1991}) \bibinfo{pages}{13--25}.
  \DOIprefix\doi{10.1080/08927014.1991.9525385}.
%Type = Misc
\bibitem[{Courtney et~al.(2020)Courtney, Stevens, Zhang, and
  Zhao}]{courtney2020}
\bibinfo{author}{R.~J. Courtney}, \bibinfo{author}{T.~Stevens},
  \bibinfo{author}{W.~Zhang}, \bibinfo{author}{L.~Zhao}, \bibinfo{title}{Flying
  spiders: {W}hat is the drag acting on a spider-dragline in free-fall?},
  \bibinfo{howpublished}{AIAA Scitech 2020 Forum. AIAA 2020-1539},
  \bibinfo{year}{2020}. \DOIprefix\doi{10.2514/6.2020-1539}.
%Type = Article
\bibitem[{Reynolds et~al.(2006)Reynolds, Bohan, and Bell}]{reynolds2006}
\bibinfo{author}{A.~M. Reynolds}, \bibinfo{author}{D.~A. Bohan},
  \bibinfo{author}{J.~R. Bell},
\newblock \bibinfo{title}{Ballooning dispersal in arthropod taxa with
  convergent behaviours: dynamic properties of ballooning silk in turbulent
  flows},
\newblock \bibinfo{journal}{Biol. Lett.} \bibinfo{volume}{2}
  (\bibinfo{year}{2006}) \bibinfo{pages}{371--373}.
  \DOIprefix\doi{10.1098/rsbl.2006.0486}.
%Type = Book
\bibitem[{Paidoussis(1998)}]{paidoussis1998}
\bibinfo{author}{M.~P. Paidoussis}, \bibinfo{title}{Fluid-structure
  interactions: slender structures and axial flow}, volume~\bibinfo{volume}{1},
  \bibinfo{publisher}{Academic press}, \bibinfo{year}{1998}.
%Type = Article
\bibitem[{Yu et~al.(2019)Yu, Liu, and Amandolese}]{yu2019}
\bibinfo{author}{Y.~Yu}, \bibinfo{author}{Y.~Liu},
  \bibinfo{author}{X.~Amandolese},
\newblock \bibinfo{title}{A review on fluid-induced flag vibrations},
\newblock \bibinfo{journal}{Appl. Mech. Rev.} \bibinfo{volume}{71}
  (\bibinfo{year}{2019}). \DOIprefix\doi{10.1115/1.4042446}.
%Type = Article
\bibitem[{Bouchet et~al.(2006)Bouchet, Mebarek, and Du{\v{s}}ek}]{bouchet2006}
\bibinfo{author}{G.~Bouchet}, \bibinfo{author}{M.~Mebarek},
  \bibinfo{author}{J.~Du{\v{s}}ek},
\newblock \bibinfo{title}{Hydrodynamic forces acting on a rigid fixed sphere in
  early transitional regimes},
\newblock \bibinfo{journal}{Eur. J. Mech.-B/Fluids} \bibinfo{volume}{25}
  (\bibinfo{year}{2006}) \bibinfo{pages}{321--336}.
  \DOIprefix\doi{10.1016/j.euromechflu.2005.10.001}.
%Type = Inproceedings
\bibitem[{Tomboulides et~al.(1993)Tomboulides, Orszag, and
  Karniadakis}]{tomboulides1993}
\bibinfo{author}{A.~Tomboulides}, \bibinfo{author}{S.~Orszag},
  \bibinfo{author}{G.~Karniadakis},
\newblock \bibinfo{title}{Direct and large-eddy simulations of axisymmetric
  wakes},
\newblock in: \bibinfo{booktitle}{31st Aerospace Sciences Meeting},
  \bibinfo{year}{1993}, p. \bibinfo{pages}{546}.
  \DOIprefix\doi{10.2514/6.1993-546}.
%Type = Article
\bibitem[{Johnson and Patel(1999)}]{johnson1999}
\bibinfo{author}{T.~A. Johnson}, \bibinfo{author}{V.~C. Patel},
\newblock \bibinfo{title}{Flow past a sphere up to a {R}eynolds number of 300},
\newblock \bibinfo{journal}{J. Fluid Mech.} \bibinfo{volume}{378}
  (\bibinfo{year}{1999}) \bibinfo{pages}{19–70}.
  \DOIprefix\doi{10.1017/S0022112098003206}.

\end{thebibliography}


\end{document}